\documentstyle[11pt,epsfig]{article}

\def\laq{\raise 0.4ex\hbox{$<$}\kern -0.8em\lower 0.62
ex\hbox{$\sim$}}
\def\gaq{\raise 0.4ex\hbox{$>$}\kern -0.7em\lower 0.62
ex\hbox{$\sim$}}

\begin{document}

\begin{titlepage}
\begin{flushright}
CERN-PH-TH/2006-056
\end{flushright}

\vspace{0.8cm}

\begin{center}

\huge{Entropy perturbations and large-scale magnetic fields}

\vspace{0.8cm}

\large{Massimo Giovannini \footnote{e-mail address: massimo.giovannini@cern.ch}}

\normalsize

\normalsize
\vspace{0.3cm}
{{\sl Centro ``Enrico Fermi", Compendio del Viminale, Via 
Panisperna 89/A, 00184 Rome, Italy}}\\
\vspace{0.3cm}
{{\sl Department of Physics, Theory Division, CERN, 1211 Geneva 23, Switzerland}}
\vspace*{2cm}

\begin{abstract}
\noindent
An appropriate gauge-invariant framework for  the treatment of magnetized curvature and entropy modes is developed.  It is shown that large-scale  magnetic fields, present after neutrino decoupling, affect curvature and  
entropy perturbations. The evolution of different magnetized modes is then studied across the matter-radiation transition both analytically and numerically.  From the observation that, after equality (but before decoupling) the (scalar) 
Sachs-Wolfe contribution must be (predominantly) adiabatic, constraints on the magnetic power spectra are deduced. The present results motivate the experimental analysis of more general initial conditions of CMB anisotropies  (i.e. mixtures of magnetized adiabatic and isocurvature modes during the pre-decoupling phase). The role of the possible correlations between the different components of the fluctuations is partially discussed.
\end{abstract}
\end{center}

\end{titlepage}
\newpage
\renewcommand{\theequation}{1.\arabic{equation}}
\section{Introduction}
\setcounter{equation}{0}

The observational evidence inferred from the spatial 
variations of the Cosmic Microwave Background (CMB) temperature
suggests that the curvature perturbations that are present 
prior to decoupling (but after equality) must be predominantly 
adiabatic and Gaussian with an (almost) scale-invariant spectrum 
\cite{map1,map2,map3}. 
A mode is defined to be (predominantly) adiabatic if the fluctuations 
of the entropy density  are subleading  after matter-radiation equality
 (but prior to decoupling).  
An exactly adiabatic mode of curvature fluctuations 
may be produced, for instance, by single-field inflationary models. If this is the case, the scalar fluctuations of the inflaton field (with wavelength larger than the Hubble radius) are transferred to curvature fluctuations whose value stays (almost) frozen from the early stages of the evolution of the Universe until the pre-decoupling epoch when the minute spatial variations of the geometry are imprinted, via the Sachs-Wolfe effect, on large-scale temperature inhomogeneities.

In addition to adiabatic fluctuations there may be one 
or more non-adiabatic modes sometimes named (with a somehow confusing 
terminology) isocurvature modes. The specific number of (regular) isocurvature modes (up to four) is determined by the physical properties of the plasma
where curvature fluctuations evolve. A mode, in the present 
terminology, simply means a consistent solution of the governing 
equations of the metric and plasma fluctuations, i.e. 
a consistent solution of the perturbed Einstein equations and of the 
lower multipoles of the Boltzmann hierarchy. 

Consider as an example the physical system formed by two  fluids, i.e.  radiation and pressureless matter.  This is a pretty good approximation in the description of pre-decoupling 
physics if the radiation is identified with the sum of photons and neutrinos and the pressureless fluid accounts for the cold dark matter (CDM) component. 
If this is the case,  the plasma suports  not only an adiabatic mode, but also a
non-adiabatic mode for which the entropy fluctuations of the CDM-radiation 
system do not vanish \cite{bardeen,ks}. It is therefore possible 
 that the pre-decoupling curvature fluctuations have a dominant (adiabatic) contribution and a subdominant (entropic) correction.
Sticking to the simplified situation described above, 
the large-scale  contribution to the angular power spectrum would be 
determined by three pre-decoupling power spectra, i.e. the spectrum of the adiabatic mode, the spectrum of the entropic mode and the spectrum of their cross-correlation.

This possibility leads to a rich parameter space and 
to an interesting phenomenological framework (see, for instance, 
\cite{hannu1,juss1,hannu2}). As an example in Ref. \cite{hannu2} a 
recent likelihood analysis determined the isocurvature fraction allowed 
by WMAP data (as well as by large scale structure data) by taking into account both the adiabatic and the entropic 
contribution of the CDM-radiation system.  

The study of isocurvature and entropy modes is often conducted 
within a two-fluid system because of the inherent mathematical complications related with the possible proliferation of non-adiabatic modes. It will be demonstrated here that the presence 
of a large-scale magnetic fields during the pre-decoupling epoch 
affects the whole system of curvature and entropy perturbations both qualitatively and quantitatively. 

If the origin of large-scale magnetic fields is primordial 
(as opposed to astrophysical) it is plausible to expect that the pre-decoupling plasma is effectively magnetized. 
Different studies have been devoted, in recent years, to constrain a hypothetical primeval magnetic field from CMB studies (see \cite{kan1} for a short 
review and \cite{g1} for a more extended discussion). 
Indeed, large-scale magnetic fields that are present in the pre-decoupling 
epoch may affect the polarization power spectra (both the EE, BB and EB 
angular power spectra) either directly \cite{BB1,BB2,BB3} or through the Faraday effect \cite{FAR1,FAR2,FAR3,FAR4,FAR5,FAR6} that may rotate an initial E mode (produced thanks to the polarized nature of Thompson scattering) into a B mode.  Barrow and Subramanian, in a series of papers \cite{bs1,bs2,bs3,bs4}, analyzed the 
vector modes (and the polarization power spectra) induced by large-scale magnetic fields. Recently \cite{BB2}
a numerical approach has been developed to analyze vector and tensor 
modes. Vector, as well as tensor modes have been discussed also in \cite{BB1}. 
While the effect of vector modes will be crucial at small scales (i.e. $\ell\geq 1000$)
the contribution of tensor modes may be relevant also for larger scales (i.e. $\ell < 100$).

Large-scale magnetic fields induce vector and tensor modes but also various scalar modes whose classification and  analysis will be one of the themes of the present investigation. 
More specifically, large-scale magnetic fields affect the usual adiabatic and entropic modes \cite{cl,tur}  that may be  present in the pre-decoupling epoch.  
In fact,  the divergence of the magnetohydrodynamical Lorentz force influences the tight-coupling evolution of photons and baryons. 
Moreover, the magnetic fields gravitate. Therefore, they contribute to the Hamiltonian constraint and (in a weaker fashion) to the momentum constraint. In \cite{ts2}, 
an interesting covariant approach for the analysis of cosmological perturbations induced by large-scale magnetic fields has been proposed and further developed in \cite{ts3}.  The approach used in \cite{ts3} was to assume a background 
magnetic field which uniform but, at the same time, sufficiently 
weak not to break the isotropy of the background geometry.  The present approach is more similar to the one of Ref. \cite{g3} where the magnetic field has been taken fully inhomogeneous. Moreover, the 
problems related to the present analysis concern the specific modifications induced by large-scale magnetic fields on the 
adiabatic and non-adiabatic modes that are normally used 
to set initial conditions for CMB anisotropies. 
This goal requires a full description of the magnetized photon-baryon 
system in the presence of neutrinos and CDM particles both before 
and after decoupling. 
The strategy would then be to study how large-scale magnetic fields affect the system of equations describing adiabatic and non-adiabatic modes before the time of radiation-matter equality, i.e. deep in the radiation-dominated epoch. The magnetized solutions will be then evolved both analytically and numerically across the radiation-matter transition and up to the time of photon decoupling. 

As mentioned, large-scale magnetic fields are assumed here 
to be fully inhomogeneous and characterized by an appropriate 
two-point function. In this way the spatial isotropy of the background geometryis not broken and the line element  can then be written, in the spatially flat case, as  
\begin{equation}
ds^2 = g_{\mu\nu} dx^{\mu} d x^{\nu} = a^2(\tau)[ d\tau^2 - d \vec{x}^2],
\label{metric}
\end{equation}
where $\tau$ is the conformal time coordinate and $a(\tau)$ is the scale factor.
Defining the appropriately rescaled magnetic fields as $\vec{B} = a^2 
\vec{{\cal B}}$ (see section2), the relevant two-point function can be written (see the appendix
for a summary on the conventions) as 
\begin{equation}
\langle B_{i}(\vec{k},\tau) B^{j}(\vec{p},\tau) \rangle =  \frac{2\pi^2}{k^3} \,P_{i}^{j}(k)\, P_{\rm B}(k,\tau)\, \delta^{(3)}(\vec{k} + \vec{p}),
\label{Bcorr0}
\end{equation}
where $P_{\rm B}(k)$ is the magnetic field power spectrum and $P_{i}^{j}(k) = 
(\delta_{i}^{j} - k_{i}k^{j}/k^2)$.
Large-scale magnetic fields 
will be treated here in a magnetohydrodynamical (MHD) approximation  
 \cite{pl1} in curved 
space-time. This description is justified by different considerations.
In fact, before decoupling, not only the plasma is globally neutral but 
the value of the Debye scale much smaller than the Hubble radius 
and it is also much smaller than the electron and photon 
mean free path \cite{g1}. As far as the magnetic properties of the 
system are concerned, the plasma behaves, in this situation, as 
a single (intrinsically charged but globally neutral) fluid where 
the total Ohmic current is solenoidal and the Ohmic electric fields 
are suppressed by the finite value of the conductivity. The 
same kind of MHD description has been used 
also in the analysis  of the vector and tensor modes of the geometry 
(see, for instance, \cite{BB1,BB2} and \cite{bs1,bs2}). 
It should be borne in mind, 
however, that it would be incorrect to use the same kind of MHD 
approximation in the analysis of Faraday effect (as sometimes 
assumed).  Faraday effect concerns the propagation of electromagnetic 
waves in a cold (or warm) plasma and cannot be treated within MHD 
which is valid for typical frequencies much smaller than the plasma 
frequency and typical scales much larger than the Debye scale. So in full
analogy with the discussion of Faraday effect in the laboratory (see 
for instance  \cite{plb1,plb2}) a two-fluid description is mandatory \cite{FAR4}.

A direct consequence of the MHD approach is that magnetic flux and 
magnetic helicity will be, to a good approximation, conserved 
in the limit of large conductivity \cite{pl1}. 
In the Friedmann-Robertson-Walker background (\ref{metric}), flux conservation implies that 
the magnetic field strength increases as $a^{-2}$ as we go back in time.
At the present time magnetic fields are observed, at the $\mu$ G level, in galaxies \cite{beck}, clusters \cite{fer,setti} and in some superclusters \cite{kronberg}. 
Even if the fields have different properties in diverse classes 
of structures and even if the determination of magnetic fields in superclusters is still in progress, it is rather plausible to admit 
that we live in a sort of magnetized Universe \cite{magn1,magn2}.
In the case of galaxies, if the magnetic field present prior to gravitational 
collapse was of the order of $0.1$ nG, compressional amplification alone 
could have produced the observed magnetic field even without the need 
of a dynamo action. In recent years interesting progresses have been 
made in the analysis of the dynamo problem especially in connection
with the role of global conservation laws of magnetized plasmas 
\cite{sub,laz}. 
Therefore, 
the protogalactic field could be much smaller than $0.1$ nG and still 
be explain the observed magnetic fields in galaxies, while in this framework cluster magnetic fields are related to the way dynamo saturates. Aside from this consideration,
if we blue-shift a protogalactic magnetic field (coherent over a typical 
Mpc scale) of $0.1$ n G at the epoch of photon decoupling we will end up 
with a magnetic field of mG strength that can potentially affect 
CMB anisotropies. The mechanisms that 
may produce magnetic fields will not be specifically examined and 
we refer the reader to a recent revies \cite{magn2}. However, 
 most of the mechanisms 
predict a stochastically distributed magnetic field that does not break spatial isotropy.  It will then  be interesting 
to scrutinize the implications of a pre-decoupling field on the scalar 
CMB anisotropies. In this sense the ambition of the present approach is 
a model-indepedent approach that is
 similar to recent attempts of constraining 
mixtures of adiabatic and non-adiabatic modes in spite of their specific 
origin (see \cite{hannu1,juss1,hannu2} and references therein). 

To be more specific, the strategy will be to treat 
the magnetic fields as a further element of the primordial plasma 
that may contain both adiabatic (or entropic) and non-adiabatic 
modes. In doing  a gauge invariant description will be adopted
since, as it will be illustrated, a simpler description 
of the matter-radiation transition can be achieved. 
To say anything concerning CMB anisotropies we have to estimate what 
is the value of curvature perturbations after equality (but before 
decoupling) and in the long wavelength limit \footnote{In the present 
paper we will often refer to the ``long wavelength limit" of some expressions 
or equation. By this we simply mean that the relevant wavelengths are the ones whose typical size is larger than the Hubble radius at the corresponding 
epoch. If we define the comoving wave-number $k$, the long wavelengths, in the 
present terminology, 
are the ones satisfying $k \tau < 1$.}. Such an estimate, is, unfortunately,
not available in the present literature.
It is then appropriate to introduce the description developed in this paper 
by defining the density contrast on uniform curvature hypersurfaces, conventionally denoted by $\zeta$ \cite{ks,bard2,bard3} 
\begin{equation}
\zeta = - {\cal H} \frac{\delta \rho_{\rm t}}{\rho'_{\rm t}},\qquad \rho_{\rm t} = 
\sum_{\rm i} \rho_{\rm i},\qquad \rho_{\rm t}'= - 3{\cal H} (\rho_{\rm t} + p_{\rm t}),
\label{zetadef}
\end{equation}
where the prime denotes a derivation with respect to $\tau$ and ${\cal H} = 
(\ln{a})'$; $\rho_{\rm t}$ is the (total) background fluid density while 
$p_{\rm t}$ is the pressure. Similarly, in Eq. (\ref{zetadef}),
 $\delta \rho_{\rm t} = \delta \rho  + \delta\rho_{\rm B}$, $\delta \rho$ 
is the fluid density 
fluctuation in the uniform curvature gauge \cite{HW,OD} and 
$\delta\rho_{B}$ is the fully inhomogeneous magnetic 
energy density which is gauge-invariant and quadratic in the magnetic field 
intensities. If the  fluid has different consituents 
(for instance, in one of the simplest cases, dusty matter and radiation), then 
 the total density contrast on uniform curvature hypersurfaces, i.e. $\delta \rho_{\rm t}$  can be written as the weighted sum of the individual fluid 
perturbations and of the magnetic field density contrast, i.e. 
\begin{equation}
\zeta = \sum_{\rm i} \frac{\rho_{\rm i}'}{\rho_{\rm t}'} \zeta_{\rm i} + \zeta_{B},
\qquad  \zeta_{\rm i} = - {\cal H} \frac{\delta\rho_{\rm i}}{\rho_{\rm i}'},\qquad 
\zeta_{\rm B} = - {\cal H} \frac{\delta\rho_{\rm B}}{\rho_{\rm t}'} .
\label{zetaidef1}
\end{equation}
 Given the form of Eqs. (\ref{zetaidef1})
the relative fluctuations in the entropy density $\varsigma$ can be defined as 
\begin{equation}
{\cal S}_{\rm i\,j} = \frac{\delta \varsigma_{\rm i\,j}}{\varsigma_{\rm i\,j}}= 
- 3 (\zeta_{\rm i} - \zeta_{\rm j}), \qquad 
{\cal S}_{{\rm i}\,{\rm B}} = \frac{\delta \varsigma_{{\rm i}\,B}}{\varsigma_{{\rm i}\,B}}= - 3 (\zeta_{\rm i} - \zeta_{\rm B}). 
\label{SiB}
\end{equation}
In this language, the adiabatic mode corresponds to the situation where 
$\zeta_{{\rm i\,j}}=0$ and $\zeta_{{\rm i}\, B}=0$ for each fluid of the mixture.
The other important variable treated here will be gauge-invariant curvature fluctuation
\footnote{In the present paper curvature 
fluctuations are identified with the fluctuations of the spatial curvature in the comoving orthogonal gauge 
\cite{press,lyth}. The gauge-invariant generalization of this quantity is usually denoted by ${\cal R}$. Clearly, in different 
gauges, the physical meaning of ${\cal R}$ may be different.} usually denoted 
by  ${\cal R}$, and  related to $\zeta$ as  
\begin{equation}
\zeta = {\cal R} + \frac{\nabla^2 \Psi}{12\,\pi\, G \,a^2 (p_{\rm t} + \rho_{\rm t})},
\label{zetatoR}
\end{equation}
where $\Psi$ is the (gauge-invariant) Bardeen potential. 
In the long wavelength limit, the second term at right hand side of Eq. (\ref{zetatoR}) can be neglected and $\zeta$ is approximately equal to ${\cal R}$.
When the plasma is constituted by a mixture of relativistic fluids, the evolution of $\zeta$ can be written, in the long wavelength limit , as 
\begin{equation}
\zeta' = - \frac{{\cal H}}{p_{\rm t} + \rho_{\rm t}} \delta p_{\rm nad} + \frac{{\cal H}}{p_{\rm t} +\rho_{\rm t}} \biggl( c_{\rm s}^2 - \frac{1}{3} \biggr) \delta\rho_{\rm B},
\label{zetacons}
\end{equation}
where $\delta p_{\rm nad}$ is the non-adiabatic pressure density variation 
accounting for the contribution of entropy fluctuations (see Eq. \ref{dpnad}). 
 Equation (\ref{zetacons}) contains, at the right hand side, not only the contribution of $\delta p_{\rm nad}$ but also the 
effect of the inhomogeneous magnetic field.

The logic followed in the present investigation will then be to provide a fully gauge-invariant 
system of equations where the evolution of  magnetized curvature and entropy modes can be characterized 
as regular solutions deep in the radiation-dominated regime. Then the various modes will be followed
across the radiation-transition both analytically and numerically. 

In Section 2 a gauge-invariant system 
of governing equations will be proposed for the simultaneous description of magnetized adiabatic and entropy modes.  In Section 3 the  analysis of various magnetized modes will be presented in the realistic 
situation where the plasma is constituted by a mixture of baryons, neutrinos, CDM and photons.The fate of the regular magnetized modes across the radiation-matter equality will be discussed 
in Section 4.  The precise amplitude of the curvature perturbations before decoupling will then 
be computed for the wavelengths that are larger than the Hubble radius after equality. In Section 5 some more phenomenological considerations on the properties of the Sachs-Wolfe 
plateau will be discussed with the aim of providing a possible starting 
point for more specific numerical discussions. Finally, Section 6 contains the concluding discussion and a summary of the possible developments of the present investigation. A series of technical results have been collected in the Appendix in connection with the proper parametrization of the various power 
spectra arising in the problem.

\renewcommand{\theequation}{2.\arabic{equation}}
\section{Magnetized curvature fluctuations}
\setcounter{equation}{0}
\subsection{General considerations}
The system of magnetized curvature perturbations can be 
obtained by considering the case of a gravitating charged 
fluid which is, however, globally neutral for scales larger 
than the Debye length. Around the time of decoupling, for a typical temperature 
of $0.3$ eV and for an electron density $n_{\rm e} = 10^{3}\,{\rm cm}^{-3}$ the Debye length is  of the order of $10\, {\rm cm}$, 
i.e. much smaller than the Hubble radius at the corresponding epoch.
The background equations (and the background metric (\ref{metric}))
are not affected by the presence of magnetic fields
and have the standard form in terms of the (total) 
energy and pressure densities, i.e. $\rho_{\rm t}$ and $p_{\rm t}$:
\begin{eqnarray}
 {\cal H}^2 = \frac{8\pi G}{3} a^2 \rho_{\rm t},\qquad {\cal H}^2 - {\cal H}' = 4\pi G a^2 ( \rho_{\rm t} + p_{\rm t}), \qquad \rho_{\rm t}' + 3 {\cal H} (\rho_{\rm t} + p_{\rm t}) =0.
\label{FL}
\end{eqnarray}
Denoting with $ \delta_{\rm s}$ the first-order (scalar) fluctuation of the corresponding quantity, the perturbed  Einstein equations can be written 
as 
\begin{equation}
\delta_{\rm s} {\cal G}_{\mu}^{\nu} = 8\pi G ( \delta_{\rm s} T_{\mu}^{\nu} + {\cal T}_{\mu}^{\nu}),
\label{p1}
\end{equation}
where ${\cal G}_{\mu}^{\nu}$ is the Einstein tensor 
while $\delta_{\rm s} T_{\mu}^{\nu}$ 
and ${\cal T}_{\mu}^{\nu}$ are, respectively, the energy-momentum tensors 
of the fluid perturbations and of the electromagnetic sources.  
The perturbed covariant conservation equations and by the equations for the Maxwell field strength can instead be written as
\begin{eqnarray}
&& \delta_{\rm s} (\nabla_{\mu} T^{\mu}_{\nu}) + \nabla_{\mu} {\cal T}^{\mu}_{\nu}
=0
\label{p2}\\
&& \nabla_{\mu} F^{\mu\nu} = 4\pi j^{\nu}.
\label{p3}
\end{eqnarray}
In Eqs. (\ref{p1})--(\ref{p3}), $\nabla_{\mu}$ denotes the covariant derivative 
defined from the metric $g_{\mu\nu}$ defined in Eq. (\ref{metric}).  In Eq. (\ref{p3}) $j^{\nu} =\sigma_{c} F^{\mu\nu}u_{\mu} $                                                                                                                                                                                                                                                                                                                                                                                                                                            
 is the Ohmic current. The (total) velocity field, $u_{\mu}$, obeys, as usual 
 $g^{\mu\nu} u_{\mu}u_{\nu} =1$. Recalling that  $F_{0i} = a^2 {\cal E}_{i}$ and 
$F_{ij} = - a^2 \epsilon_{ijk} {\cal B}^{k}$ the evolution equations for the 
rescaled electric and magnetic fields become:
\begin{eqnarray}
&& \vec{\nabla} \cdot \vec{E} =0, \qquad \vec{\nabla} \cdot \vec{B}=0,
\label{sol}\\
&& \frac{\partial \vec{B}}{\partial \tau} = - \vec{\nabla}\times \vec{E},
\label{e1}\\
&& \vec{\nabla}\times \vec{B}  = 4\pi \vec{J} + \frac{\partial \vec{E}}{\partial \tau},
\label{e2}\\
&& \vec{J} = \sigma( \vec{E} + \vec{v} \times \vec{B}),
\label{e3}
\end{eqnarray}
where 
\begin{equation}
\vec{E} = a^2 \vec{{\cal E}}, \qquad \vec{B} = a^2 \vec{{\cal B}}, \qquad \vec{J} = a^3 \vec{j}.
\end{equation}
The one-fluid MHD
description adopted here \cite{pl1} amounts then to the requirement  that
both the Maxwell fields and the induced Ohmic current are solenoidal
\begin{equation}
\vec{\nabla}\cdot \vec{E}=0,\qquad \vec{\nabla} \cdot \vec{B}=0,
\qquad \vec{\nabla} \cdot \vec{J}=0,
\label{sol2}
\end{equation}
which implies that 
\begin{equation}
\vec{J} = \frac{\vec{\nabla}\times \vec{B}}{4\pi}, \qquad \vec{E} = 
 \frac{\vec{\nabla}\times \vec{B}}{4\pi \sigma} - \vec{v} \times \vec{B}.
 \label{el1}
 \end{equation}
Since we will be interested in effects occurring over length-scales much larger 
than the Debye scale, the displacement current can be safely 
neglected in Eq. (\ref{e3}).  Note 
that $\vec{v}$ appearing in Eq. (\ref{el1}) is the bulk velocity of the plasma.
In MHD, the bulk velocity of the plasma is essentially the center of mass 
velocity of the baryon-electron system. Therefore, $\vec{v}$ coincides, in practice 
with the baryon velocity field.   In Eq. (\ref{el1}), 
$\sigma$ is the conductivity. In a relativistic plasma 
$\sigma \simeq T/\alpha_{\rm em}$ while in a non-relativistic 
plasma $\sigma \simeq (T/\alpha_{\rm em}) (T/m_{\rm e})^{1/2}$ where 
$m_{\rm e}$ is the electron mass and $T$ is common temperature of electrons 
and baryons. 

The evolution equations for the coupled system defined by 
Eqs. (\ref{p1}), (\ref{p2}) and (\ref{p3}) can then be derived.  
For the present purposes it is useful to employ the uniform 
curvature gauge \cite{HW,OD} 
where the perturbed elements of the metric tensor are 
\begin{equation}
\delta_{\rm s} g_{00} = 2 a^2 \phi, \qquad \delta_{\rm s} g_{i0} = - a^2 \partial_{i} {\cal A}.
\end{equation}
The covariant conservation of the total energy-momentum tensor, i.e.
Eq.  (\ref{p3}) implies then 
two separate equations one for the total velocity field and the 
other for the density fluctuation. These two equations can be written as 
\begin{eqnarray}
&& \delta_{\rm s} \rho' + (p_{\rm t} +  \rho_{\rm t}) \theta + 
3 {\cal H} (\delta_{\rm s} \rho + \delta_{\rm s} p) = \frac{ \vec{E} \cdot\vec{J}}{a^4},
\label{con1}\\
&& (\theta + \nabla^2 {\cal A})' + {\cal H} ( 1 - 3 c_{\rm s}^2) (\theta + \nabla^2 {\cal A}) + \frac{\nabla^2 \delta_{\rm s} p}{p_{\rm t} + \rho_{\rm t}} - \frac{\partial_{j} 
\partial^{i} \Pi_{i}^{j}}{p_{\rm t} + \rho_{\rm t}}  + \nabla^2 \phi = 
\frac{ \vec{\nabla} \cdot( \vec{J}\times \vec{B})}{a^4 (p_{\rm t} + \rho_{\rm t})},
\label{con2}
\end{eqnarray}
where $\theta = \partial_{i} v^{i}$ is the divergence of the (total) velocity field and 
where  $c_{\rm s}^2 = p_{\rm t}' /\rho_{\rm t}'$ is the (total) sound speed.  To derive Eqs. (\ref{con1}) and 
(\ref{con2}) from Eq. (\ref{p2}), the perturbed components of the energy-momentum 
tensor of the fluid have been parametrized as
\begin{eqnarray}
&& \delta_{\rm s} T_{0}^{0} = \delta_{\rm s} \rho, \qquad \delta T_{i}^{j} = - 
\delta_{\rm s} p + \Pi_{i}^{j},
\label{enp}\\
&& \delta_{\rm s} T_{0}^{i} = ( p_{\rm t} + \rho_{\rm t}) v^{i},\qquad 
\delta_{\rm s} T_{i}^{0}= - (p_{\rm t} + \rho_{\rm t}) (v_{i} + \partial_{i} {\cal A}).
\label{velp}
\end{eqnarray}
In Eq. (\ref{velp}) the term $\Pi_{i}^{j}$ represents the contribution of the 
anisotropic stress.  The energy-momentum tensor of the electromagnetic sources is, instead, 
\begin{equation}
{\cal T}_{\mu}^{\nu} = \frac{1}{4\pi} \biggl( -F_{\mu\alpha} F^{\nu\alpha} 
+ \frac{1}{4} \delta_{\mu}^{\nu} F_{\alpha\beta} F^{\alpha\beta}\biggr).
\label{elm}
\end{equation}
Equation (\ref{p1}) implies two constraints (stemming from the 
($00$) and ($0i$) components). The Hamiltonian and momentum 
constraints are then, in the off-diagonal gauge, 
\begin{eqnarray}
&& {\cal H} \nabla^2 {\cal A} + 3 {\cal H}^2 \phi = - 4\pi G a^2 ( \delta_{\rm s}\rho + \delta \rho_{\rm B}),
\label{HAM0}\\
&& \nabla^2 [ {\cal H} \phi + ({\cal H}^2 - {\cal H}') {\cal A}] = - 
4\pi G a^2 [ (p_{\rm t} + \rho_{\rm t}) \theta + \delta {\cal V}_{\rm B}],
\label{MOM0}
\end{eqnarray}
where 
\begin{equation}
\delta \rho_{\rm B} = \frac{ \vec{B}^2}{8\pi a^4},\qquad \delta {\cal V}_{\rm B} = \frac{ \vec{\nabla}\cdot( \vec{E} \times \vec{B})}{4\pi a^4},
\label{comp1}
\end{equation}
are, respectively, the magnetic energy density and the Poynting vector.
Notice that $\delta {\cal V}_{\rm B}$, in the one-fluid MHD approximation, 
contains two terms: the first is suppressed by the conductivity, the second is of higher-order. In fact, using the second relation of Eq. (\ref{el1}), 
\begin{equation}
\delta {\cal V}_{\rm B} = \frac{\vec{\nabla}\cdot(\vec{J}\times \vec{B})}{4\pi \sigma a^{4}  } - \frac{\vec{\nabla}\cdot[ (\vec{v} \times \vec{B})\times \vec{B}]}{ 4\pi a^4}.
\label{int1}
\end{equation}
The $(ij)$ component of Eq. (\ref{p1}) can be written as 
\begin{eqnarray}
&&\biggl[ ({\cal H}^2 + 2 {\cal H}') \phi + {\cal H} \phi' + 
\frac{1}{2} \nabla^2 ( {\cal A}' + 2 {\cal H} {\cal A} + \phi)\biggr]\delta_{i}^{j} - \frac{1}{2} \partial_{i}\partial^{j}[ {\cal A}'  + 2 {\cal H} {\cal A} +\phi]
\nonumber\\
&& = 4\pi G a^2 [ (\delta_{\rm s} p + \delta p_{\rm B}) \delta_{i}^{j} - 
\Pi_{i}^{j} - {\tilde{\Pi}}_{i}^{j}],
\label{ijeq1}
\end{eqnarray}
where 
\begin{equation}
\delta p_{\rm  B} = \frac{\delta \rho_{\rm B}}{3},\qquad \tilde{\Pi}_{i}^{j}
= \frac{1}{4\pi a^4}\biggl[ B_{i}B^{j} - \frac{1}{3} \vec{B}^2 \delta_{i}^{j}\biggr].
\label{ansB}
\end{equation}

The electromagnetic contribution 
to the Hamiltonian constraint (\ref{HAM0}) contains also a purely electric term.
However,  by virtue of Eq. (\ref{el1}) the electric contribution is suppressed with the respect to the magnetic one.  Furthermore, terms containing the product of a quadratic (magnetic) contribution and of a metric (or velocity)  
fluctuation are of higher order and will then  be neglected.

\subsection{Gauge-invariant governing equations}
The system formed by Eqs.
(\ref{con1}), (\ref{con2}) and (\ref{ijeq1}) (supplemented by the 
constraints (\ref{HAM0}) and (\ref{MOM0})) mixes the 
four variables $ \phi$, ${\cal A}$, $\delta_{\rm s}\rho$ and $\theta$.
The gauge-invariant generalization of the total density contrast on uniform 
curvature hypersurfaces, i.e.  $\delta_{\rm s}\rho$ is given by  $\zeta$, i.e. 
\begin{equation}
\zeta = - \frac{\delta_{\rm s} \rho_{\rm t}}{\rho_{\rm t}'}{\cal H},
\label{DEF0}
\end{equation}
where $\delta_{\rm s} \rho_{\rm t} = \delta_{\rm s} \rho + \delta \rho_{\rm B}$ .
On the other hand, the gauge-invriant generlization of  $\phi$ is related with the curvature fluctuations on comoving orthogonal hypersurfaces, 
denoted by ${\cal R}$:
\begin{equation}
{\cal R} = - \frac{{\cal H}^2}{{\cal H}^2 - {\cal H}'} \phi.
\label{DEF1}
\end{equation}
Finally, the variable ${\cal A}$ is related with one of the two Bardeen 
potentials, usually denoted by $\Psi$:
\begin{equation}
\Psi = - {\cal H} {\cal A},
\label{DEF2}
\end{equation}
The variables $\zeta$, ${\cal R}$ and $\Psi$ are all gauge-invariant and while 
they have been introduced, for simplicity, in the uniform curvature 
gauge, they have interesting physical interpretations also in different 
gauges. For instance, $\zeta$ can be interpreted as the curvature 
perturbation on constant density hypersurfaces. This interpretation 
is clear by looking at the gauge-invariant form of the Hamiltonian 
constraint. Inserting Eqs.  (\ref{DEF0}), (\ref{DEF1}) and (\ref{DEF2}) 
into Eqs. (\ref{HAM0}) and (\ref{MOM0}) the gauge-invariant forms of the 
Hamiltonian and momentum constraints can be obtained:
\begin{eqnarray}
&& \zeta = {\cal R} + \frac{\nabla^2 \Psi}{ 12 \pi G a^2 (\rho_{\rm t} +
p_{\rm t})},
\label{HAM1}\\
&& \nabla^2 ( {\cal R} + \Psi) = {\cal H} \theta + \frac{{\cal H}}{p_{\rm t} 
+ \rho_{\rm t}} \delta_{\rm s} {\cal V}_{\rm B}.
\label{MOM1}
\end{eqnarray}
In the gauge-invariant approach presented here, the Hamiltonian constraint 
(\ref{HAM1}) fixes the relation between $\zeta$ and ${\cal R}$. 
If we are interested in the evolution 
of the system for length-scales much larger than the Hubble radius we will have that the term containing $\nabla^2\Psi$ is sub-dominant and the constraint is satisfied if $\zeta \simeq {\cal R}$. The momentum 
constraint (\ref{MOM1}) can be used to compute the divergence 
of the total velocity field $\theta$ once ${\cal R}$ and $\Psi$ are known.
Using Eqs. (\ref{DEF0}), (\ref{DEF1}) and (\ref{DEF2}) 
into Eq. (\ref{con1}),  and recalling Eqs. (\ref{FL}), 
the following  (gauge-invariant) 
evolution equation for $\zeta$ can then be obtained
\begin{equation}
\zeta'  = - \frac{{\cal H}}{p_{\rm t} + \rho_{\rm t}} \biggl[ \delta p_{\rm nad} - \biggl( c_{\rm s}^2 - \frac{1}{3} \biggr)  \delta \rho_{\rm B} \biggr] - \frac{\theta}{3} + \frac{\vec{E} \cdot \vec{J}}{3 a^4 ( p_{\rm t} + \rho_{\rm t})}.
\label{zetamas}
\end{equation}
In Eq. (\ref{zetamas}), the pressure fluctuation of the fluid component 
has been split as $\delta_{\rm s} p = c_{\em s}^2 \delta_{\rm s} \rho + \delta p_{\rm nad}$,
where $\delta p_{\rm nad}$ is the term proportional to the intrinsic entropy fluctuations of the fluid and 
$c_{\rm s}^2 = p_{\rm t}'/\rho_{\rm t}'$ is the total sound speed. This 
decomposition is usual when dealing with mixtures of relativistic 
fluids even in the absence of magnetic fields.
The explicit form of $\delta p_{\rm nad}$ can be  evaluated once the chemical 
composition of the fluid is specified. By introducing the individual 
sound speeds of the various fluids 
\footnote{Clearly, by definition of total sound speed 
$c_{\rm s}^2 = \sum_{\rm i}(\rho_{\rm i}' /\rho_{\rm t}')c_{\rm s\,i}^2$.}
and the relative fluctuations of the specific entropy in a generic pair 
of fluids, i.e. 
\begin{equation}
c_{\rm s\,i}^2 = \frac{p_{\rm i}'}{\rho_{\rm i}'}, \qquad  {\cal S}_{\rm i\,j} = - 3 ( \zeta_{\rm i} - \zeta_{\rm j}),
\label{defcsi}
\end{equation}
the general form of $\delta p_{\rm nad}$ can be written as 
\begin{equation}
\delta p_{\rm nad} = \frac{1}{6 {\cal H} \rho_{\rm t}'} \sum_{{\rm i},{\rm j}} \rho_{\rm i}' \rho_{\rm j}' ( c_{\rm s\, i}^2 - 
c_{\rm s\, j}^2 ) {\cal S}_{\rm i \, j}.
\label{dpnad}
\end{equation}
The indices appearing 
in Eq. (\ref{dpnad}) (not to be confused with covariant spatial indices) label the different components of the fluid and, by definition, ${\cal S}_{i\,j} = - {\cal S}_{\rm j\, i}$ (see Eq. (\ref{defcsi}).
The gauge-invariant density contrasts of each fluid, i.e.  $\zeta_{\rm i}$, can be 
evaluated, for instance, in the uniform curvature gauge where 
they correspond to 
\begin{equation}
\zeta_{\rm i} = - {\cal H}\frac{\delta \rho_{\rm i}}{\rho_{\rm i}'},\qquad 
\rho_{\rm t} = \sum_{\rm i} \rho_{\rm i}
\label{zetai}
\end{equation}
as already introduced in Eq. (\ref{zetaidef1}).
The variable $\zeta$ that appears in the Hamiltonian 
constraint (\ref{HAM1}) can be also written in terms 
of the individual $\zeta_{\rm i}$ supplemented by 
tha magnetic contribution $\zeta_{\rm B}$:
\begin{equation}
\zeta = \sum_{\rm i} \frac{\rho_{\rm i}'}{\rho_{\rm t}'} \zeta_{\rm i} + \zeta_{B}, 
\qquad \zeta_{B} = - {\cal H} \frac{\delta\rho_{B}}{\rho_{\rm t}'}.
\label{zetasum}
\end{equation}
In connection with Eq. (\ref{zetamas}) the following  remarks are in order:
\begin{itemize} 
\item{} even if the intrinsic (non-adiabatic) contribution vanishes, the 
evolution of $\zeta$ may change over scales larger than the Hubble
radius since its evolution is directly sourced by the magnetic energy 
density;
\item{} this new (entropic) contribution vanishes when the Universe 
is dominated by radiation  but becomes relevant as soon as 
the speed of sound $c_{\rm s}^2 \neq 1/3$ so that, for instance, it is relevant 
in the context of the radiation-matter transition;
\item{} by virtue of the momentum constraint (\ref{MOM1}) and 
of the Hamiltonian constraint (\ref{HAM1}), the contribution 
containing $\theta$ at the left hand side of Eq. (\ref{zetamas})
is proportional, to leading order, to $\nabla^2 \zeta$ and to $\nabla^4\Psi$ (both contributions are then small on super-Hubble scales);
\item{} as previously remarked in deriving Eq. (\ref{MOM1}) the 
term $\vec{E}\cdot\vec{J}$ appearing in Eq. (\ref{zetamas})
 is strictly zero in the superconducting limit and negligible for finite 
 conduvctivity.
 \end{itemize}
 The last statement can be easily verified since, in the MHD approximation, 
 the relations of Eq. (\ref{el1}) imply the following chain of equalities
 \begin{equation}
  \vec{E} \cdot \vec{J} = \frac{\vec{J}\cdot \vec{\nabla}\times \vec{B}}{\sigma} - (\vec{v} \times \vec{B})\cdot\vec{J}= \frac{(\vec{\nabla}\times\vec{B})^2}{4\pi \sigma} - \frac{(\vec{v} \times \vec{B})\cdot(\vec{\nabla}\times \vec{B})}{4\pi}.
\label{int2}
\end{equation}
Now, the first term in the second equality of Eq. (\ref{int2}) vanishes 
in the limit of large conductivities or, more precisely, in the limit $\sigma/H \gg 1$ where $H = a {\cal H}$ is the Hubble parameter. The second 
term in the last equality of Eq. (\ref{int2}) is of higher order. 
To complete the discussion we have now to study the relations 
implied by Eq. (\ref{ijeq1}) in terms of $\zeta$, ${\cal R}$ and $\Psi$. 
Using Eqs. (\ref{DEF0}), (\ref{DEF1}) and (\ref{DEF2}) into Eq. (\ref{ijeq1}) we  can extract the trace-free (gauge-invariant) contribution and apply the differential operator $\partial_{i}\partial^{j}$:
\begin{equation}
\nabla^4\biggl[ {\cal R} + \Psi + \frac{2 \rho_{\rm t}}{3 ( \rho_{\rm t} 
+ p_{\rm t})} \biggl( \frac{\Psi'}{{\cal H}} + \Psi \biggr)\biggr]  = 
- \frac{16\pi G a^2 \rho_{\rm t}}{3 (\rho_{\rm t} + p_{\rm t})}[ \nabla^4 \Pi 
+ \nabla^4 \tilde{\Pi}].
\label{tfree}
\end{equation}
On the other hand, the  trace-full contribution stemming from Eq. (\ref{ijeq1}) simply leads to the gauge-invariant evolution of ${\cal R}$, i.e. 
\begin{equation}
{\cal R}'  = - \frac{{\cal H}}{p_{\rm t} + \rho_{\rm t}}\biggl[ \delta p_{\rm nad} - \biggl(c_{\rm s}^2 - \frac{1}{3} \biggr)\delta \rho_{\rm B}\biggr]
- \frac{ c_{\rm s}^2 {\cal H} \nabla^2\Psi}{4\pi G a^2 (p_{\rm t} + \rho_{\rm t})} + \frac{2}{3} \frac{{\cal H}}{p_{\rm t} + \rho_{\rm t}} (\nabla^2 \Pi + \nabla^2 \tilde{\Pi}).
\label{Rmas}
\end{equation}
Equation (\ref{zetamas}) has the same physical content 
of Eq. (\ref{Rmas}). The reason is that for typical wavelengths 
larger than the Hubble radius, Eq. (\ref{HAM1}) implies that 
$\zeta \simeq {\cal R}$. This statement can be verified 
by showing that, for instance, Eq. (\ref{Rmas}) 
reduces to Eq. (\ref{zetamas}) when the constraints 
(\ref{HAM1}) and (\ref{MOM1}) are taken into account.
In fact, using Eq. (\ref{HAM1}) to express ${\cal R}$ in terms of $\zeta$ 
and taking the first time derivative we can substitute ${\cal R}'$ in 
Eq. (\ref{Rmas}). 
This manipulation will give rise to a bunch of terms: the first one will be essentially $\zeta'$ while the other contain the laplacian of $\Psi$ and its derivatives. Then we can use Eqs. (\ref{MOM1}) and (\ref{tfree}) 
to get rid of $\Psi'$.  The resulting equation will exactly be Eq. (\ref{zetamas}) which was obtained from the covariant conservation equation.

Equations (\ref{zetamas})  and (\ref{Rmas}) suggest that the magnetic field acts, effectively, as
a non-adiabatic variation of the pressure density. Therefore it is useful, sometimes, to think that 
\begin{equation}
\delta {\cal P}_{\rm nad} = \delta p_{\rm nad}  - ( 3 c_{\rm s}^2 -1) \zeta_{\rm B}, \qquad 
\zeta_{\rm B} = \frac{\delta\rho_{\rm B}}{ 3 (\rho_{\rm t} + p_{\rm t})},
\end{equation}
where $\zeta_{\rm B}$ now obeys 
\begin{equation}
\zeta_{\rm B}' +  {\cal H} ( 1 - 3 c_{\rm s}^2) \zeta_{\rm B} =0. 
\label{zetaB}
\end{equation}

\renewcommand{\theequation}{3.\arabic{equation}}
\section{Magnetized curvature perturbations before equality}
\setcounter{equation}{0}
\subsection{Physical scales}
In the absence of  magnetic fields, the commonly employed strategy 
is to use  Eqs. (\ref{zetamas}) and (\ref{Rmas}) for the estimate of the 
values of  curvature perturbations in the long wavelength limit after the equality (denoted, in what follows, 
by $\tau_{\rm eq}$) but before  decoupling  (denoted by $\tau_{\rm dec}$). The magnetic contribution 
to Eqs. (\ref{zetamas}) and (\ref{Rmas}) change the evolution 
of $\zeta$ and ${\cal R}$, as discussed in the previous section. 
Moreover, the contribution to the anisotropic stress arises 
both from the collisionless neutrinos and from the magnetic fields. 
In the absence of collisionless neutrinos the magnetic fields can only induce, during 
radiation, a singular mode since the magnetic anisotropic stress 
cannot be counterbalanced by the anisotropic stress of the fluid.
The solutions of the 
gauge-invariant system of section 2 will now be studied 
 in the limit $\tau \ll \tau_{\rm eq}$. Then 
the different magnetized modes (both adiabatic and non-adiabatic) will be followed through the matter radiation transition with the purpose 
of computing in detail the value of the curvature perturbations before 
decoupling and for wavelengths larger than the Hubble radius at the 
corresponding epoch. The radiation-matter transition will be modeled by the interpolating scale factor
\begin{equation}
a(\tau) = a_{\rm eq} \biggl[ \biggl(\frac{\tau}{\tau_{1}}\biggr)^2 + 2 \biggl( \frac{\tau}{\tau_{1}}\biggr)\biggr],
\label{aint}
\end{equation}
which is a solution of Eqs. (\ref{FL})  when 
the plasma contains both pressureless matter and radiation. 
In Eq. (\ref{aint}), $a_{\rm eq}$ is the value of the scale factor at equality. The values of $a_{\rm eq}$ and $\tau_{1}$ are fixed by the critical fractions of
matter and radiation. In particular
 \footnote{The scale factor will be normalized in such a way that $a_{0}=1$ and the factor $h$, i.e. the 
indetermination in the Hubble constant, will be taken $0.73$ for numerical estimates.}, 
\begin{equation}
\frac{a_{\rm eq}}{a_{0}} = \frac{\overline{\Omega}_{\rm r}}{\overline{\Omega}_{\rm m}} \simeq \frac{4.15 \times 10^{-5}}{h^2 \overline{\Omega}_{\rm m}}, 
\qquad \tau_{1} = \frac{2}{H_{0}} \sqrt{\frac{a_{\rm eq}}{\overline{\Omega}_{\rm m}}} \simeq 287.48\, 
\biggl(\frac{h^2 \overline{\Omega}_{\rm m}}{0.134} \biggr)^{-1}. 
\label{par1}
\end{equation}
and $\overline{\Omega}_{\rm m}$ and $\overline{\Omega}_{\rm r}$ 
are the present values of the critical fractions of matter and radiation, 
\begin{equation}
\overline{\Omega}_{\rm m} =  
\overline{\Omega}_{\rm c} + \overline{\Omega}_{\rm b},\qquad 
\overline{\Omega}_{\rm r} = 
\overline{\Omega}_{\gamma} + \overline{\Omega}_{\nu}.
\label{omegamr}
\end{equation}
The redshift of decoupling will be taken to be  $1+ z_{\rm dec}= a_{\rm dec}^{-1}  \simeq 1100$, implying  from Eq. (\ref{par1}) that 
$\tau_{\rm dec}\simeq 2.36 \, \tau_{\rm eq}$ (where $\tau_{\rm eq} = (\sqrt{2} -1) \tau_{1}$).   The numerical values of the matter and radiation density 
fractions will be taken to be   
\begin{equation}
h^2 \overline{\Omega}_{\rm m} \simeq 0.134,
\qquad h^2 \overline{\Omega}_{\rm r} \simeq 4.15 \times 10^{-5},
\label{ommr}
\end{equation}
with 
\begin{equation}
h^2 \overline{\Omega}_{\rm c} \simeq 0.111, \qquad 
h^2 \overline{\Omega}_{\rm b} \simeq 0.023, \qquad 
h^2 \overline{\Omega}_{\gamma} \simeq 
2.47 \times 10^{-5}, \qquad 
h^2 \overline{\Omega}_{\nu} \simeq 1.68 \times 10^{-5}.
\label{parameters}
\end{equation}
Concerning the set of parameters adopted in Eqs. (\ref{ommr}) and 
(\ref{parameters}) few comments are in order. The recent WMAP three year 
data \cite{map3}, when combined with ``gold" sample of SNIa \cite{riess}  give values compatible with the ones adopted here.  However, if WMAP data alone 
are used we would have, instead, a slightly smaller value for the
critical fraction of matter, i.e. $h^2 \Omega_{\rm m} \simeq 1.268$.
The values of the parameters chosen here are just illustrative 
in order to model, in a realistic fashion, the radiation-matter 
transition. Therefore, any difference in $h^2 \Omega_{\rm m}$ will
just affect the absolute value of $\tau_{\rm 1}$ (and, consequently, 
of $\tau_{\rm eq}$) but will not change the features of the forthcoming 
 numerical analysis whose purpose is just to corroborate the analytical 
 results. A final comment on neutrino masses is appropriate. If neutrino 
masses are in the range of the ${\rm meV}$ (i.e. smaller 
than the temperature at $z_{\rm dec}$), they will  
be non-relativistic today. However, they will be relativistic 
around $z_{\rm dec}$ and will then contribute
to the radiation fraction $\Omega_{\rm r}$ 
(and not to the matter fraction). In the 
following (when writing the evolution equations of the neutrino component)
neutrino masses will then be neglected since our analysis applies for 
$\tau \leq \tau_{\rm dec}$. In setting the initial conditions 
(i.e. for $\tau \ll \tau_{\rm eq}$) the neutrinos are effectively massless
since the temperature of the radiation plasma is higher than $T_{\rm dec}$.
\subsection{Mixtures of four fluids}
For $\tau \ll \tau_{\rm eq}$ 
(but always after neutrino decoupling, i.e. $T< 1 \, {\rm MeV}$)
 the radiation component dominates and,  within the gauge-invariant approach presented in section 2,
it is particularly simple to set adiabatic or isocurvature initial conditions 
and to solve the equations deep inside 
the radiation-dominated phase. Few relevant cases will now be illustrated 
in view of the numerical analysis reported in section 3.
When the plasma is formed 
by baryons, photons, CDM particles and massless neutrinos, 
the gauge-invariant density perturbations of each fluid obey
\begin{eqnarray}
&& \zeta_{\rm b}' = - \frac{\theta_{\rm b}}{3},\qquad \zeta_{\gamma}'= - \frac{\theta_{\gamma}}{3},
\label{bgamma}\\
&& \zeta_{\rm c}' = - \frac{\theta_{\rm c}}{3}, \qquad \zeta_{\nu}'= - \frac{\theta_{\nu}}{3},
\label{cnu}
\end{eqnarray}
where the subscripts ${\rm b}$, $\gamma$, ${\rm c}$ and $\nu$ refer, respectively, to the baryon, photon, CDM and 
neutrino components of the plasma.  
Since the divergences of the peculiar velocities carry a spatial 
gradient, they are small for modes that are larger than the Hubble radius prior to equality.  Adiabatic initial conditions amount then to the choice that all the density contrasts are equal and constant, i.e. 
\begin{equation}
\zeta_{\rm b} = \zeta_{\rm c} = \zeta_{\gamma} = \zeta_{\nu} = {\rm constant}.
\label{adincon}
\end{equation}
Equation (\ref{adincon})
implies that all the non adiabatic pressure density variation of the fluid vanishes exactly by virtue of Eq. (\ref{dpnad}). 
If the initial conditions for the gauge-invariant density contrasts are different 
from Eq. (\ref{adincon}), then the corresponding solution is said to be non-adiabatic. In fact, any deviation from the initial conditions (\ref{adincon}) 
necessarily leads to a non-vanishing entropy perturbation. For instance, if $\zeta_{\gamma} \neq \zeta_{\nu}$, ${\cal S}_{\gamma\,\nu}\neq 0$. This is the situation of the so-called neutrino isocurvature mode. Yet another example could be the one of the CDM-radiation mode where ${\cal S}_{\gamma\,{\rm c}}\neq 0$ and ${\cal S}_{\nu\,{\rm c}}\neq 0$. 
The evolution equations of the velocity fields of the CDM and of the neutrinos are
that appear at the right hand side of Eqs. (\ref{bgamma}) and (\ref{cnu}) are:
\begin{eqnarray}
&& \theta_{\rm c}' + {\cal H} \theta_{\rm c} -\nabla^2 \biggl[ \frac{\Psi'}{{\cal H}} + \frac{{\cal H}^2 - {\cal H}'}{{\cal H}^2} 
( \Psi + {\cal R}) \biggr] =0,
\label{cvel}\\
&& \theta_{\nu}'   - \nabla^2 \biggl[ \biggl(\frac{\Psi}{{\cal H}}\biggr)'  
+ \frac{{\cal H}^2 - {\cal H}'}{{\cal H}^2 }
 {\cal R}\biggr]  + \nabla^2 \zeta_{\nu} - \nabla^2 \sigma_{\nu}=0.
\label{nuvel}
\end{eqnarray}
The quantity $\sigma_{\nu}$ appearing in Eq. (\ref{nuvel}) 
is  the neutrino anisotropic stress. In fact, since the neutrinos 
are collisionless particles, their anisotropic stress (connected to the 
quadrupole moment of the neutrino phase space distribution as $\sigma_{\nu}
= {\cal F}_{\nu\,2}/2$) is not erased after neutrino decoupling occurring, approximately, when the plasma was dominated by radiation and for 
temperatures of the order of $1$ MeV. 
The total anisotropic stress introduced in Eq. (\ref{tfree})  
will then be written, for convenience, as
\begin{equation}
\partial_{i}\partial^{j} \Pi_{j}^{i} + \partial_{i}\partial^{j} \tilde{\Pi}_{j}^{i} = 
(p_{\nu} + \rho_{\nu}) \nabla^2 \sigma_{\nu} + (p_{\gamma} + \rho_{\gamma}) 
\nabla^2 \sigma_{\rm B},
\label{anis1}
\end{equation}
 where $\sigma_{\rm B}$ is the anisotropic stress of the magnetic fields.
Recalling the explicit form of $\tilde{\Pi}_{i}^{j}$ given in Eq. (\ref{ansB}), 
it is easy to deduce that 
 \begin{equation}
\nabla^2\sigma_{\rm B} = \frac{3}{4} \frac{\vec{\nabla}\cdot[\vec{J}\times
\vec{B}]}{a^4 \rho_{\gamma}} + \frac{1}{4} \nabla^2 \Omega_{\rm B},
\label{anis2}
\end{equation}
where it has been used that $\vec{\nabla} \cdot \vec{J}= 0$  and that 
$4 \pi \vec{J} = \vec{\nabla}\times \vec{B}$. In Eq. (\ref{anis2}) 
the ratio bewteen the magnetic energy density 
and the photon energy density:
\begin{equation}
\Omega_{\rm B}(\vec{x}) = \frac{\delta\rho_{\rm B}(\vec{x})}{\rho_{\gamma}} = \frac{B^2(\vec{x})}{8\pi a^4 \rho_{\gamma}}.
\label{OMB}
\end{equation} 
Since $\rho_{\gamma} \simeq a^{-4}$, Eq. (\ref{OMB}) implies that  $\Omega_{\rm B}(\vec{x})$ is constant in time. 
Therefore, according to Eqs. (\ref{anis2}) and (\ref{OMB}),
 the contribution of the 
magnetohydrodynamical energy-momentum tensor to the scalar fluctuations 
reduces, effectively,  to the magnetic energy (and pressure) densities (i.e. $\delta\rho_{\rm B}$ 
and $\delta\rho_{\rm B}/3$) and to the divergence of the Lorentz force.
The collisionless nature of the neutrinos implies 
that not only the quadrupole moment (i.e. the anisotropic stress) of the 
neutrino phase space distribution contributes to the fluid evolution but also
the higher moments should be taken into account. In particular, the following gauge-invariant equations have to be introduced
\begin{eqnarray}
&& \sigma_{\nu}' = \frac{4}{15} \theta_{\nu} - \frac{3}{10} k {\cal F}_{\nu\,3},
\label{snuprime}\\
&& {\cal F}_{\nu\,\ell}' = \frac{k}{2 \ell + 1} [ \ell {\cal F}_{\nu\,(\ell -1)} - (\ell + 1) {\cal F}_{\nu\,(\ell + 1)}].
\label{hierarchy}
\end{eqnarray} 
Equation (\ref{hierarchy}) applies for multipoles larger than the quadrupole.
To specify initial conditions for neutrinos would then require, 
in principle,  all the multipole moments 
of the (perturbed) neutrino phase space distribution. In practice, it will  
 be sufficient to analyse the initial conditions of the lower 
moltipoles i.e. monopole (density constrast), dipole (peculiar velocity)
and quadrupole (ansotropic stress). 

Unlike the CDM and neutrinos, the evolution of the baryons and of the photons is coupled by Thompson scattering:
\begin{eqnarray}
&& \theta_{\gamma}' = \nabla^2 \biggl[ \biggl( \frac{\Psi}{{\cal H}}\biggr)' + \frac{{\cal H}^2 - {\cal H}'}{{\cal H}^2} {\cal R} 
\biggr] - \nabla^2 \zeta_{\gamma} + a n_{\rm e} x_{\rm e} \sigma_{\rm T} ( \theta_{\rm b} - \theta_{\gamma}),
\label{photv}\\
&& \theta_{\rm b}' + {\cal H} \theta_{\rm b} = 
\nabla^2 \biggl[ \frac{\Psi'}{{\cal H}} + \frac{{\cal H}^2  - {\cal H}'}{{\cal H}^2} ({\cal R} + \Psi)\biggr] 
+ \frac{\vec{\nabla} \cdot [\vec{J} \times \vec{B}]}{\rho_{\rm b} a^4}
+ \frac{4}{3} \frac{\rho_{\gamma}}{\rho_{\rm b}} a n_{\rm e} x_{\rm e} \sigma_{\rm T} (\theta_{\gamma} - 
\theta_{\rm b}),
\label{barv}
\end{eqnarray}
where $\sigma_{\rm T}$ is the Thompson scattering cross section, $n_{\rm e}$ is the electron density and $x_{\rm e}$ is the ionization fraction.
By taking the difference of Eqs. (\ref{photv}) and (\ref{barv}) 
the evolution of  the combination $(\theta_{\gamma} - \theta_{\rm b})$ 
can be obtained with the result that 
\begin{equation}
(\theta_{\gamma} - \theta_{\rm b})' + a n_{\rm e} x_{\rm e} \sigma_{\rm T} \biggl[ 1 + \frac{4}{3} \frac{\rho_{\gamma}}{\rho_{\rm b}}\biggr] (\theta_{\gamma} - \theta_{\rm b})= {\cal H} \theta_{\rm b}- \nabla^2(\zeta_{\gamma}+\Psi) - \frac{\vec{\nabla} \cdot [\vec{J} \times \vec{B}]}{\rho_{\rm b} a^4}.
\label{TC1}
\end{equation}
For $\tau \ll \tau_{\rm eq}$ the characteristic time scale for the synchronization of photons and baryons is 
set by $(n_{\rm e} \sigma_{\rm T})^{-1}$  which is small in comparison both with the 
expansion time, i.e. $a \tau$,  and with the typical oscillation time, i.e. $a\tau/k$ where $k$ is the Fourier wavenumber. 
Consequently, any deviation of $(\theta_{\gamma} - \theta_{\rm b})$ 
rapidly decays away. This aspect can be appreciated from Eq. (\ref{TC1}) by taking the limit $\sigma_{\rm T}\to \infty$. No matter how strong the forcing term is, 
$(\theta_{\gamma} -\theta_{\rm b}) \to 0$. Because 
of the tight coupling between photons and baryons, 
the baryon-photon velocity as a single physical quantity, i.e. $\theta_{\gamma{\rm b}}$ whose evolution equation can be obtained 
by multiplying Eq. (\ref{photv}) by $(4/3) \rho_{\gamma}$ and by adding it to Eq. (\ref{barv}) (multiplied by $\rho_{\rm b}$).
The scattering terms then cancel and the evolution of $\theta_{\gamma{\rm b}}$, 
becomes:
\begin{eqnarray}
&&  \zeta_{\gamma}' = - \frac{\theta_{\gamma{\rm b}}}{3},\qquad 
\zeta_{\rm b}' = - \frac{\theta_{\gamma{\rm b}}}{3},
\nonumber\\
&& \theta_{\gamma{\rm b}}' + \frac{R_{\rm b} {\cal H}}{R_{\rm b} + 1}  \theta_{\gamma{\rm b}} = \nabla^2\biggl[ \frac{\Psi'}{{\cal H}} + \frac{{\cal H}^2 - {\cal H}'}{{\cal H}^2} ({\cal R} +\Psi) \biggr]- \frac{\nabla^2(\Psi+ \zeta_{\gamma})}{1 + R_{\rm b}} + \frac{ 4 \nabla^2 \sigma_{\rm B} - \nabla^2\Omega_{\rm B}}{ 4 ( 1 + R_{\rm b} )}.
\label{TC}
\end{eqnarray}
To derive the last term in Eq. (\ref{TC}), 
Eq. (\ref{anis2}) has been used to express 
the Lorentz force in terms of the anisotropic stress and of the magnetic energy density in units of the photon energy density.  In Eq. (\ref{TC}) the baryon to photon ratio has been also introduced as 
\begin{equation}
R_{\rm b} = \frac{3}{4} \frac{\rho_{\rm b}}{\rho_{\gamma}} = 
\biggl( \frac{698}{z + 1}\biggr) \biggl( \frac{h^2 \overline{\Omega}_{\rm b}}{ 0.023}\biggr) \biggl( \frac{h^2 \overline{\Omega}_{\gamma}}{2.47 \times 10^{-5}}\biggr)^{-1}
\end{equation}
where $ z + 1= 1/a(\tau)$ is the redshift with the conventions adopted 
in Eq. (\ref{aint}).
Since Eq. (\ref{TC}) involves only gauge-invariant quantities, 
it is simple to obtain each corresponding equation in a specific gauge.
For instance, the conformally Newtonian gauge is often employed for numerical
discussions of the tight-coupling approximation. In this gauge 
the perturbed metric components as simply $ \delta_{\rm s} g_{00} = 2 a^2 \phi$ and $\delta_{\rm s} g_{i j} = 2 a^2 \psi \delta_{ij}$. The 
gauge-invariant quantities appearing in Eq. (\ref{TC}) read, in the longitudinal gauge,
 \begin{eqnarray}
 && \zeta_{\gamma} \to - \psi + \frac{\delta_{\gamma}}{4}, \qquad 
 \zeta_{\rm b} \to - \psi + \frac{\delta_{\rm b}}{3},\qquad \theta_{\gamma{\rm b}}\to 
 \theta_{\gamma{\rm b}},
 \label{T1}\\
 && \Psi \to \psi,\qquad {\cal R} \to - \psi - 
 \frac{{\cal H}^2}{{\cal H}^2 - {\cal H}'}\biggl( \phi + \frac{\psi'}{{\cal H}}\biggr)
 \label{T2}
 \end{eqnarray}
where $\delta_{\gamma}$ and $\delta_{\rm b}$ are the density contrasts 
computed in the conformally Newtonian gauge.  Equation (\ref{TC}) 
becomes, after easy algebra, 
\begin{eqnarray}
&& \delta_{\gamma}' = 4 \psi' - \frac{4}{3} \theta_{\gamma{\rm b}}, \qquad 
\delta_{\rm b}' = 3 \psi' - \theta_{\gamma{\rm b}},
\nonumber\\
&& \theta_{\gamma{\rm b}}' + \frac{R_{\rm b} {\cal H}}{R_{\rm b} + 1}  \theta_{\gamma{\rm b}} = -\frac{\nabla^2 \delta_{\gamma}}{4(1 + R_{\rm b})} - \nabla^2 \phi 
+ \frac{4 \nabla^2\sigma_{\rm B} - 
\nabla^2\Omega_{\rm B}}{ 4 ( 1 + R_{\rm b}) } ,
\label{TC2}
\end{eqnarray}
which is the standard form to the tight coupling equations in the conformally 
Newtonian gauge if $\sigma_{\rm B}= \Omega_{\rm B} =0$.
It is  useful to mention that the total $\zeta$ (see Eq. (\ref{zetasum})) is given by 
\begin{equation}
\zeta = \frac{\rho_\gamma'}{\rho_{\rm t}'} \zeta_{\rm \gamma} +
\frac{\rho_{\nu}'}{\rho_{\rm t}'} \zeta_{\nu} + 
\frac{\rho_{\rm c}'}{\rho_{\rm t}'} \zeta_{\rm c} + 
\frac{\rho_{\rm b}'}{\rho_{\rm t}'} \zeta_{\rm b}
+ \frac{\delta\rho_{\rm B}}{3 (\rho_{\rm t} + p_{\rm t})},
\label{zetatt}
\end{equation}
where $\rho_{\nu}$, $\rho_{\gamma}$, $\rho_{\rm b}$ and $\rho_{\rm c}$ are,
respectively,  the unperturbed energy densities of neutrinos, photons, baryons 
and CDM particles. This observation is relevant for the actual solution of the Hamiltonian constraint (\ref{HAM1}).
Finally, to solve the momentum constraint (\ref{MOM0}), we recall that the total 
peculiar velocity $\theta$ is given, in terms of the peculiar velocities 
of each fluid, by the following expression
\begin{equation}
(p_{\rm t} + \rho_{\rm t}) \theta = (p_{\nu} + \rho_{\nu}) \theta_{\nu} + 
(p_{\gamma}+ \rho_{\gamma}) \theta_{\gamma} + (p_{\rm c} + \rho_{\rm c}) 
\theta_{\rm c} + (p_{\rm b} + \rho_{\rm b}) \theta_{\rm b}.
\label{sumvel}
\end{equation}
With the same notations of Eq. (\ref{zetatt}), $p_{\gamma} = \rho_{\gamma}/3$, 
$p_{\nu}= \rho_{\nu}/3$ while $p_{\rm b} = p_{\rm c} =0$.
\subsection{The adiabatic mode}
By going to Fourier space,  Eqs. (\ref{bgamma}) and 
(\ref{cnu}) can be easily solved, to lowest order in $k\tau < 1$ and for $\tau \ll \tau_{\rm eq}$. Denoting 
with ${\cal R}_{*}$ the curvature perturbation on super-Hubble scales 
and with $\Psi_{*}$ value of the Bardeen potential in the same epoch we will have 
that  the Hamiltonian and momentum constraints imply  
\begin{eqnarray}
&& {\cal R}= {\cal R}_{*}, 
\label{ad1}\\
&& \zeta = {\cal R}_{*} - \frac{k^2 \tau^2}{6} \Psi_{*},
\label{ad2}\\
&& \Psi_{*} = - 2 \frac{ 5 + 2 R_{\nu}}{15 + 4 R_{\nu}} {\cal R}_{*} 
+ \frac{R_{\gamma} ( 4 \sigma_{\rm B} - R_{\nu} \Omega_{\rm B})}{ 15 + 4 R_{\nu}}.
\label{ad3}
\end{eqnarray}
In Eq. (\ref{ad3})  the fractions of neutrinos and photons 
present in the radiation-dominated plasma have been introduced 
according to the following conventions:
\begin{equation}
R_{\gamma} = 1 - R_{\nu}, \qquad R_{\nu} = \frac{r}{1 + r},\qquad r= \frac{7}{8} N_{\nu} \biggl(\frac{4}{11}\biggr)^{4/3};
\end{equation}
with $N_{\nu} =3$,  $r = 0.681$. 
The gauge-invariant density contrasts will then be determined to be 
\begin{equation}
\zeta_{\gamma} = \zeta_{\nu} = \zeta_{\rm b} = \zeta_{\rm c} = {\cal R}_{*} - 
\frac{\Omega_{\rm B}}{4} R_{\gamma}.
\label{ad4}
\end{equation} 
By explicit integration of the corresponding equations, the expressions 
of the peculiar velocities are  
\begin{eqnarray}
&& \theta_{\rm c} = - k^2 \tau ( {\cal R}_{*} + \Psi_{*}),\qquad 
\theta_{\nu} = - k^2 \tau \biggl[( \Psi_{*} + {\cal R}_{*}) + 
\frac{\Omega_{\rm B}}{4} R_{\gamma} 
- \frac{R_{\gamma}}{R_{\nu}} \sigma_{\rm B} 
\biggr], 
\label{ad5}\\
&& \theta_{\gamma\,{\rm b}} = - k^2 \tau \biggl[(\Psi_{*} + {\cal R}_{*})   
+ \sigma_{\rm B} - \frac{R_{\nu}}{4}\Omega_{\rm B} \biggr].
\label{ad6}
\end{eqnarray}
Finally, the expression of the neutrino anisotropic stress turns out to be: 
\begin{equation}
\sigma_{\nu} = - \frac{R_{\gamma}}{R_{\nu}}\sigma_{\rm B} + 
\frac{k^2 \tau^2}{3 R_{\nu}} \biggl( {\cal R}_{*} + \frac{3}{2} \Psi_{*}\biggr).
\label{ad7}
\end{equation}
From Eq. (\ref{ad3}), $\Psi_{*}$ is determined from ${\cal R}_{*}$ 
and from the parameters of the magnetized plasma, therefore the 
$\Psi_{*}$ appearing in the other formulae have been used just 
to avoid too cumbersome expressions. 
In the limit $\Omega_{\rm B} \to 0$ and $\sigma_{\rm B}\to 0$ the solution reproduces the standard adiabatic mode. Using Eqs. (\ref{T1}) and 
(\ref{T2}) relating the gauge-invariant quantities with the conformally 
Newtonian fluctuations, we obtain, for instance
\begin{equation}
\psi_{*} = \phi_{*} \biggl(1 + \frac{2}{5} R_{\nu}\biggr) + \frac{R_{\gamma}}{5} ( 4 \sigma_{\rm B} - R_{\nu} \Omega_{\rm B}), \qquad 
{\cal R}_{*} = - \frac{\phi_{*}}{10} ( 15 + 4 R_{\nu}) + \frac{R_{\gamma}}{5} ( 4 \sigma_{\rm B} - R_{\nu} \Omega_{\rm B}).
\label{ad8}
\end{equation}
If $\Omega_{\rm B} \to 0$ and $\sigma_{\rm B}\to 0$, Eq. reproduces the well 
known result with the difference of the Newtonian potentials being 
constant (to zeroth order in $k\tau$) and proportional to the fraction of neutrinos \cite{MAB}. 

\subsection{The non-adiabatic modes}

The discussion presented in the case of the adiabatic mode can be 
repeated, with the due differences, in the case of the isocurvature modes.  
Here only few examples will be given with particular attention to those features that will be useful for the forthcoming numerical aspects of the analysis. 
In the case of the 
neutrino isocurvature density mode, to lowest order 
in $k\tau$, the Bardeen potential is given by the following expression\footnote{As in the case of the magnetized adiabatic mode, the only free parameter of the 
solution is ${\cal S}_{*}$; the value of $\Psi_{*}$ is determined by 
${\cal S}_{*}$ according to Eq. (\ref{nu1}). For practical purposes, however, we will 
give some of the quantities as a function of both ${\cal S}_{*}$ and $\Psi_{*}$.} 
\begin{equation}
\Psi = \Psi_{*} = \frac{4\, R_{\gamma} \,R_{\nu}}{3 (4 R_{\nu} + 15)}  \biggl[ 
{\cal S}_{*} - \frac{3}{R_{\nu}} \sigma_{\rm B} + \frac{3}{4} \Omega_{\rm B}\biggr],
\label{nu1}
\end{equation}
where the term ${\cal S}_{*}$ is the specific entropy of the photon-neutrino 
system. The term ${\cal S}_{*}$ is given, in the present approach, exactly 
by the mismatch between the density contrasts of the photons and of the 
neutrinos, i.e. ${\cal S}_{\nu\gamma} = - 3 (\zeta_{\nu} - \zeta_{\gamma})= S_{*}$.
The density contrasts of each individual species are then:
\begin{equation}
 \zeta_{\nu} = - \frac{R_{\gamma}}{3}
 \biggl( {\cal S}_{*} +\frac{3}{4} \Omega_{\rm B}\biggr),
\qquad \zeta_{\gamma} = \frac{R_{\nu}}{3} S_{*} - \frac{R_{\gamma}}{4} \Omega_{\rm B},
\qquad \zeta_{\rm c} = \zeta_{\rm b} = - \frac{R_{\gamma}}{4} \Omega_{\rm B}.
\label{nu4}
\end{equation}
Again, integrating the evolution equations of the peculiar velocities the following results can be obtained:
\begin{eqnarray}
&& \theta_{\rm c} = - k^2 \tau \Psi_{*},\qquad \theta_{\nu} = -k^2 \tau \biggl[ 
\frac{R_{\gamma}}{3} {\cal S}_{*} + \Psi_{*}   + \frac{R_{\gamma}}{R_{\nu}} \sigma_{\rm B} - \frac{\Omega_{\rm B}}{4} \biggr],
\label{nu5}\\
&& \theta_{\gamma{\rm b}} = - k^2 \tau \biggl[ 2 \Psi_{*} + \frac{R_{\nu}}{3} {\cal S}_{*} 
- \frac{R_{\gamma} + 1}{4} \Omega_{\rm B} + \sigma_{\rm B} \biggr].
\label{nu6}
\end{eqnarray}
A feature of the neutrino isocurvature  density mode is that 
the total density constrast, $\zeta$ vanishes to zeroth order in $k\tau \ll 1$. Consequently also the total curvature fluctuation, i.e. ${\cal R}$, vanishes
in for typical wavelengths larger than the Hubble radius.  The neutrino 
anisotropic stress is given, in this case, by 
\begin{equation}
\sigma_{\nu} = - \frac{R_{\gamma}}{R_{\nu}} \sigma_{\rm B} + \frac{k^2 \tau^2}{2 R_{\nu}} \Psi_{*}.
\label{nu7}
\end{equation}
It is useful to make contact with the conformally Newtonian gauge where, 
according to Eqs. (\ref{T1}) and (\ref{T2}), the two Newtonian potentials are 
\begin{equation}
\phi_{*} = \frac{ 8 R_{\gamma} R_{\nu}}{3 ( 4 R_{\nu} + 15)} \biggl[ S_{*} - \frac{3}{R_{\nu}} \sigma_{\rm B} + \frac{3}{4} \Omega_{\rm B} \biggr],\qquad 
\psi_{*} = - \frac{ 4 R_{\gamma} R_{\nu}}{3 ( 4 R_{\nu} + 15)} \biggl[ S_{*} - \frac{3}{R_{\nu}} \sigma_{\rm B} + \frac{3}{4} \Omega_{\rm B} \biggr].
\label{nu8}
\end{equation}
Notice that $2 \psi_{*} + \phi_{*} =0$. This implies, in the conformally Newtonian 
gauge, that ${\cal R} = - (\psi_{*} + \phi_{*}/2)$ vanishes, as  already pointed out 
by looking directly at the gauge-invariant fluctuations $\zeta$ and ${\cal R}$.

Consider finally the situation where the non vanishing fluctuations in the 
specific entropy are given by 
${\cal S}_{{\rm c}\,\gamma}$ and ${\cal S}_{{\rm c}\,\nu}$ and define, conventionally, 
\begin{equation}
{\cal S}_{{\rm c}\,\gamma} = {\cal S}_{{\rm c}\,\nu} = S_{*}.
\label{cdm1}
\end{equation}
This is the case of the CDM-radiation isocurvature mode.
Then, from Eq. (\ref{dpnad}), by making the sum over the fluid components 
explicit we do obtain  
\begin{equation}
\delta p_{\rm nad} = \frac{\rho_{\rm c}' (\rho_{\gamma}' + \rho_{\nu}')}{9 {\cal H} \rho_{\rm t}'} {\cal S}_{*}, 
\label{cdm2}
\end{equation}
where we used that ${\cal S}_{\gamma\,{\rm c}} = {\cal S}_{\nu\,{\rm c}} = - {\cal S}_{*}$. Now defining $\rho_{\rm r} = \rho_{\gamma} + \rho_{\nu}$, Eq. (\ref{cdm2}) 
can also be written as 
\begin{equation}
\delta p_{\rm nad} = \rho_{\rm c}\, c_{\rm s}^2\, {\cal S}_{*},\qquad c_{s}^2 = 
\frac{4}{3} \frac{\rho_{\rm r}}{3 \rho_{\rm c} + 4 \rho_{\rm r}} =  
\frac{4 a_{\rm eq} }{3( 3 a + 4 a_{\rm eq})},
\label{cdm3}
\end{equation}
where the second equality in the definition of the total sound speed $c_{\rm s}^2$
follows from the conventions illustrated in connection with Eq. (\ref{aint}).
Therefore, to lowest order in $k\tau <1$ and for $\tau < \tau_{\rm eq}$
we have that 
\begin{eqnarray}
&&{\cal R} \simeq \zeta = - \frac{ 4 {\cal S}_{*} + 3 \Omega_{\rm B} R_{\gamma}}{16} 
\biggl(\frac{a}{a_{\rm eq}}\biggr), 
\label{cdm4}\\
&& \Psi = \frac{ 15 + 4 R_{\nu}}{8 ( 15 + 2 R_{\nu})} \biggl( {\cal S}_{*} + \frac{3}{4}
\Omega_{\rm B} R_{\gamma}\biggr)
\biggl(\frac{a}{a_{\rm eq}}\biggr),
\label{cdm5}
\end{eqnarray}
where we used that $\overline{\Omega}_{\rm b} \ll \overline{\Omega}_{\rm c}$ (so that $a_{\rm eq} \simeq \overline{\Omega}_{\rm r}/\overline{\Omega_{\rm c}}$).  
In the limit 
$\Omega_{\rm B}\to 0$, Eq. (\ref{cdm4}) implies that 
${\cal R} \simeq - {\cal S}_{*} (a/a_{\rm eq})/4$ which reproduces the 
standard result obtainable for the CDM-radiation isocurvature mode 
in the absence of magnetic fields. In the same limit of vanishing magnetic contribution (and with vanishing 
fractional contribution of neutrinos, i.e. $R_{\nu}\to 0$)  $\Psi \to {\cal S}_{*}
(a/a_{\rm eq})/8 $ which is also well known from the analysis of the two fluid 
system in the absence of neutrinos. In this case the relevant density 
contrasts are given by 
\begin{equation}
\zeta_{\gamma} = \zeta_{\nu} = - \frac{\Omega_{\rm B}}{4} R_{\gamma}, \qquad 
\zeta_{\rm c} = - {\cal S}_{*} - \frac{\Omega_{\rm B}}{4} R_{\gamma},
\label{cdm6} 
\end{equation}
as it follows from the Hamiltonian constraint and from the definition 
of $\zeta$ in terms of the density contrasts of each fluid.
The peculiar velocities can then be directly obtained to lowest order. In particular 
it is interesting to comment about the neutrino sector where 
the solution for the lowest multipoles can  be written as 
\begin{eqnarray}
&& \theta_{\nu} = \frac{15}{8} \frac{\overline{\Omega}_{\rm c}}{ 15 + 2 R_{\nu}} \biggl[ S_{*} + \frac{3}{4} \Omega_{\rm B} R_{\gamma} \biggr] k^2 \tau^2 + 
k\tau \biggl[ \frac{R_{\gamma}}{R_{\nu}} \sigma_{\rm B} - \frac{\Omega_{\rm B}}{4} R_{\gamma} \biggr],
\nonumber\\
&& \sigma_{\nu} = - \frac{R_{\gamma}}{R_{\nu}} \sigma_{\rm B} + 
\frac{k^2 \tau^3}{6} \frac{\overline{\Omega}_{\rm c}}{15 + 2 R_{\nu}} \biggl[ S_{*} + \frac{3}{4} \Omega_{\rm B} R_{\gamma} \biggr],
\nonumber\\
&& {\cal F}_{\nu\,3} = \frac{8}{15} \biggl[ \frac{R_{\gamma}}{R_{\nu}} \sigma_{\rm B} - \frac{\Omega_{\rm B}}{4}\biggr] k\tau.
\label{cdm7}
\end{eqnarray}
Notice that to set correctly initial conditions in the magnetized case the 
octupole is important. However, in the limit of vanishing magnetic inhomogeneities, the octupole can be simply set to zero as all the remaining terms 
in the hierarchy. Finally, in the case of the CDM-radiation mode, using Eqs. (\ref{T1}) and (\ref{T2}), the conformally Newtonian potentials are: 
\begin{equation}
\psi = \frac{ 15 + 4 R_{\nu}}{8 ( 15 + 2 R_{\nu})} \biggl( {\cal S}_{*} + \frac{3}{4}
\Omega_{\rm B} R_{\gamma}\biggr)
\biggl(\frac{a}{a_{\rm eq}}\biggr),\qquad \phi = \frac{ 15 - 4 R_{\nu}}{8 ( 15 + 2 R_{\nu})} \biggl( {\cal S}_{*} + \frac{3}{4}
\Omega_{\rm B} R_{\gamma}\biggr)
\biggl(\frac{a}{a_{\rm eq}}\biggr).
\label{cdm9}
\end{equation}
The two longitudinal potentials $\psi$ and $\phi$ are only 
equal in the limit $R_{\nu} \to 0$. In the two-fluid treatment of the problem this would actually be the case since no neutrinos are allowed.

\renewcommand{\theequation}{4.\arabic{equation}}
\section{Magnetized curvature fluctuations after equality }
\setcounter{equation}{0}
The regular solutions describing both adiabatic and entropy modes will now be followed through the transition between the radiation-dominated stage of expansion and the matter-dominated epoch. 
In fact, as already mentioned, the decoupling time takes place after $\tau_{\rm eq}$  (i.e. $\tau_{\rm dec} = 2.36 \,\tau_{\rm eq}$ for $h^2 \overline{\Omega}_{\rm m} = 0.134$).

\subsection{The adiabatic mode}

In the absence of magnetic fields and in the absence of non-adiabatic 
pressure density fluctuations, Eqs. (\ref{HAM1}) and (\ref{zetamas}) 
imply that the total 
curvature perturbation ${\cal R}$, i.e. the curvature fluctuation on comoving 
orthogonal hypersurfaces, is conserved for wavelengths much larger than the 
Hubble radius, i.e. for $k\tau \ll 1$.  Conversely, the 
Bardeen potential $\Psi$ diminishes, in the transition from radiation to matter 
of a factor $9/10$ (assuming vanishing neutrino anisotropic stress). 
In the absence of magnetic fields, the effect of the neutrino anisotropic 
stress is often ignored for order of magnitude estimates. 
However, if the plasma is magnetized, this approximation is no longer 
justified as clearly demonstrated by working out the solutions presented 
in section 3. In fact, the anisotropic 
stress of the neutrinos  is balanced by the magnetic anisotropic 
stress and the subsequent evolution of the system will tell what happens 
once the initial conditions are accurately set. 
Adiabatic (magnetized) initial conditions are 
imposed according to Eqs. (\ref{ad1}), (\ref{ad2}) and (\ref{ad3}). The initial 
conditions for the density contrasts can be found in Eq. (\ref{ad4}). The velocity 
fields and the neutrino anisotropic stress are then the ones 
derived, respectively, in Eqs. (\ref{ad5})--(\ref{ad6}) and (\ref{ad7}). 
By setting $\Omega_{\rm B} = \sigma_{\rm B} =0$ 
in the initial conditions the standard adiabatic mode (with canonical 
dependence upon the fractional contribution of massless neutrinos)  is reproduced.

\begin{figure}
\begin{center}
\begin{tabular}{|c|c|}
      \hline
      \hbox{\epsfxsize = 6.5 cm  \epsffile{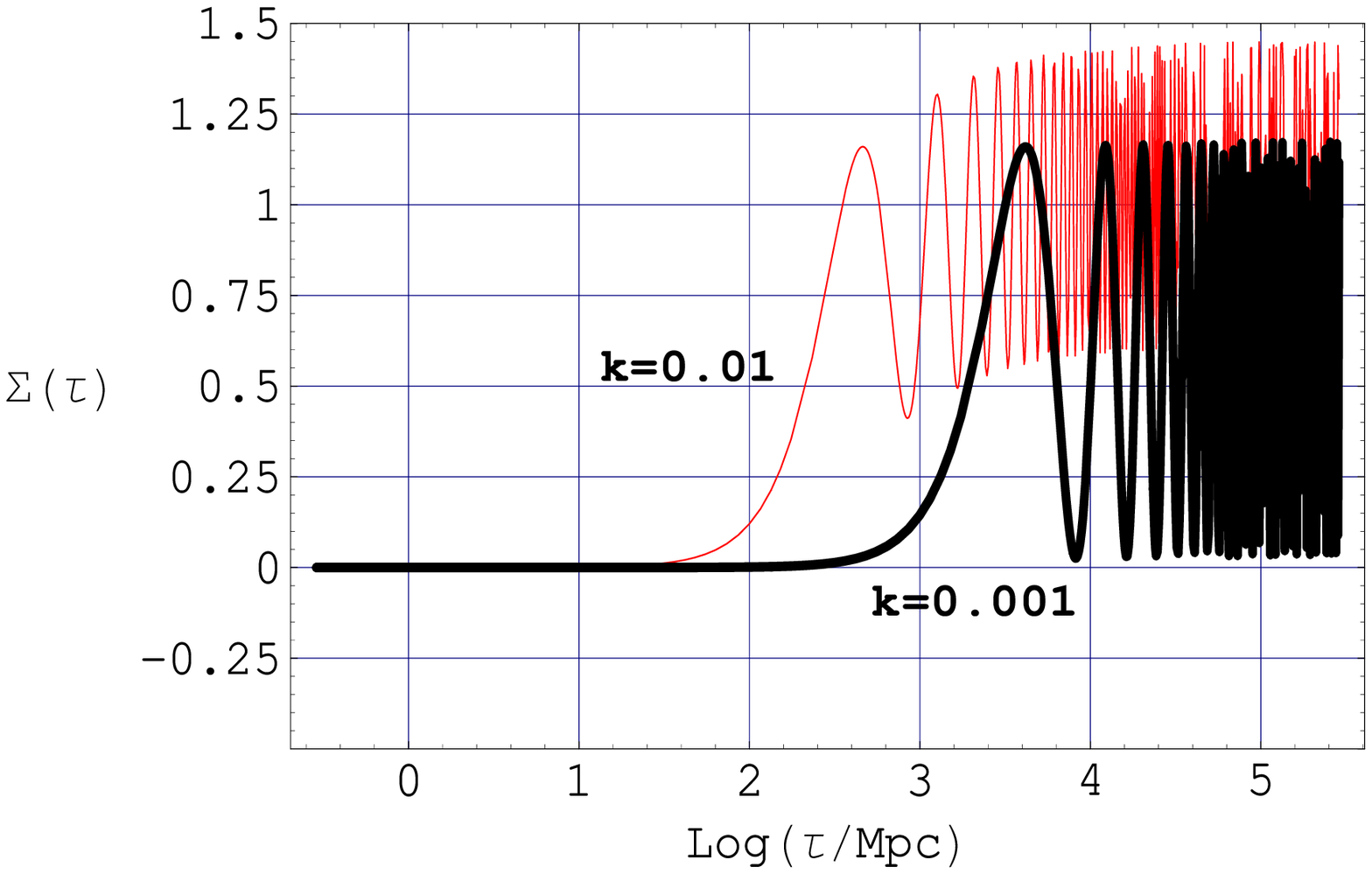}} &
      \hbox{\epsfxsize = 7 cm  \epsffile{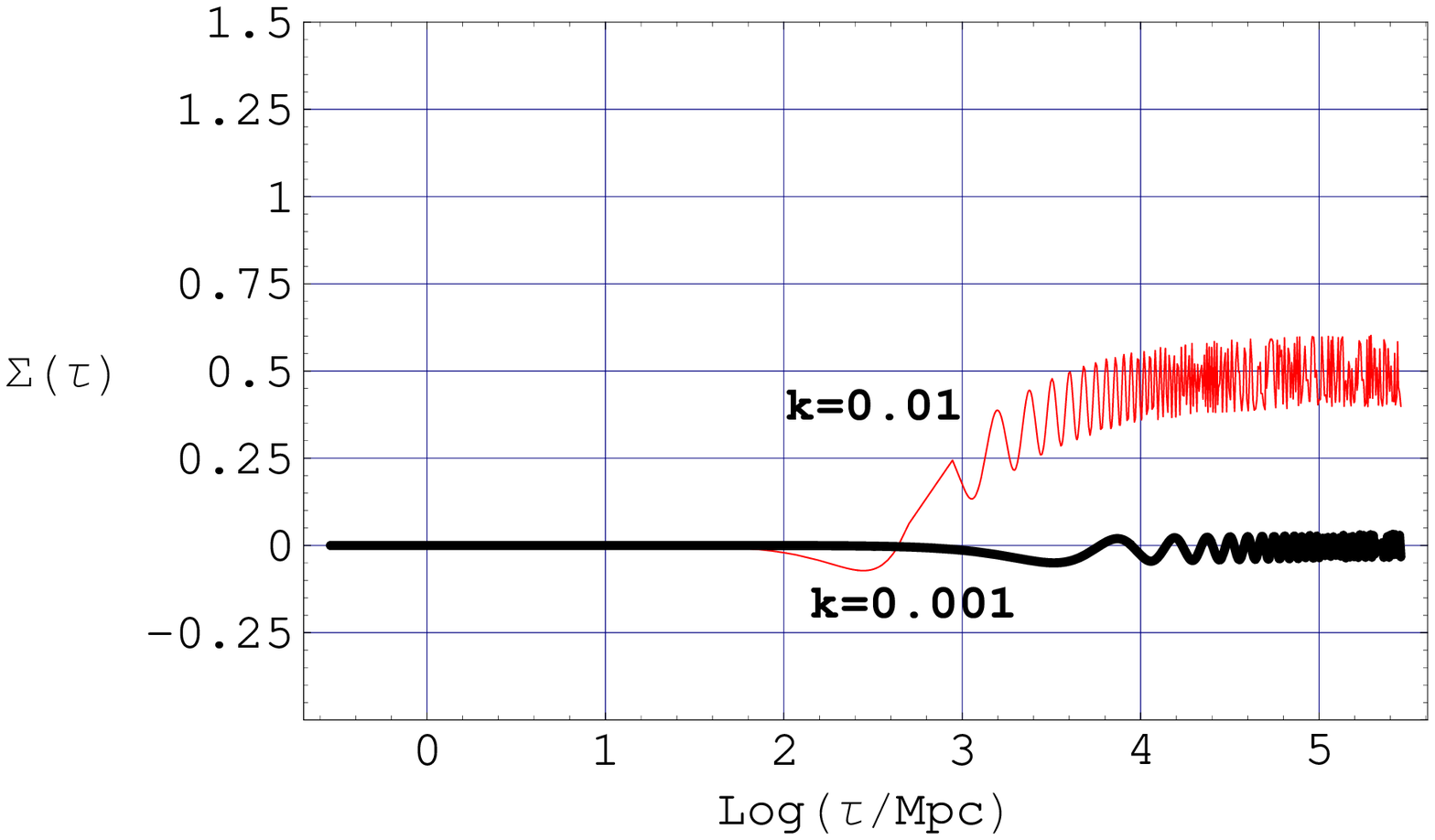}}\\
      \hline
\end{tabular}
\end{center}
\caption[a]{The time evolution of the total anisotropic stress for two different 
amplitudes of the magnetic field intensity. The comoving wave-numbers are measured in ${\rm Mpc}^{-1}$. In the plot at the left $\Omega_{\rm B} = {\cal R}_{*}$. In the plot at the right $\Omega_{\rm B} = 0.1 {\cal R}_{*}$.}
\label{F1}
\end{figure}

\begin{figure}
\begin{center}
\begin{tabular}{|c|c|}
      \hline
      \hbox{\epsfxsize = 7 cm  \epsffile{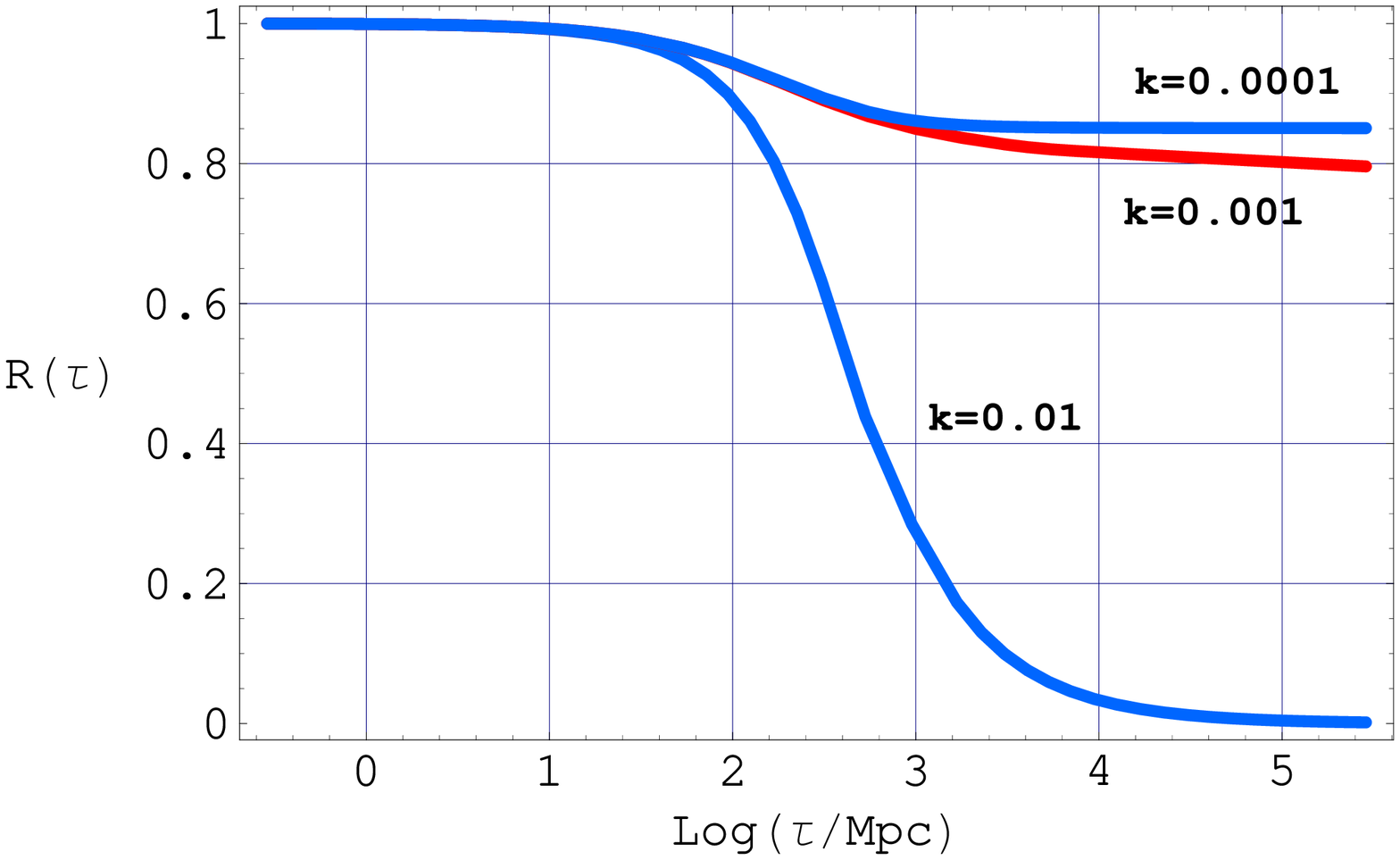}} &
      \hbox{\epsfxsize = 6.6 cm  \epsffile{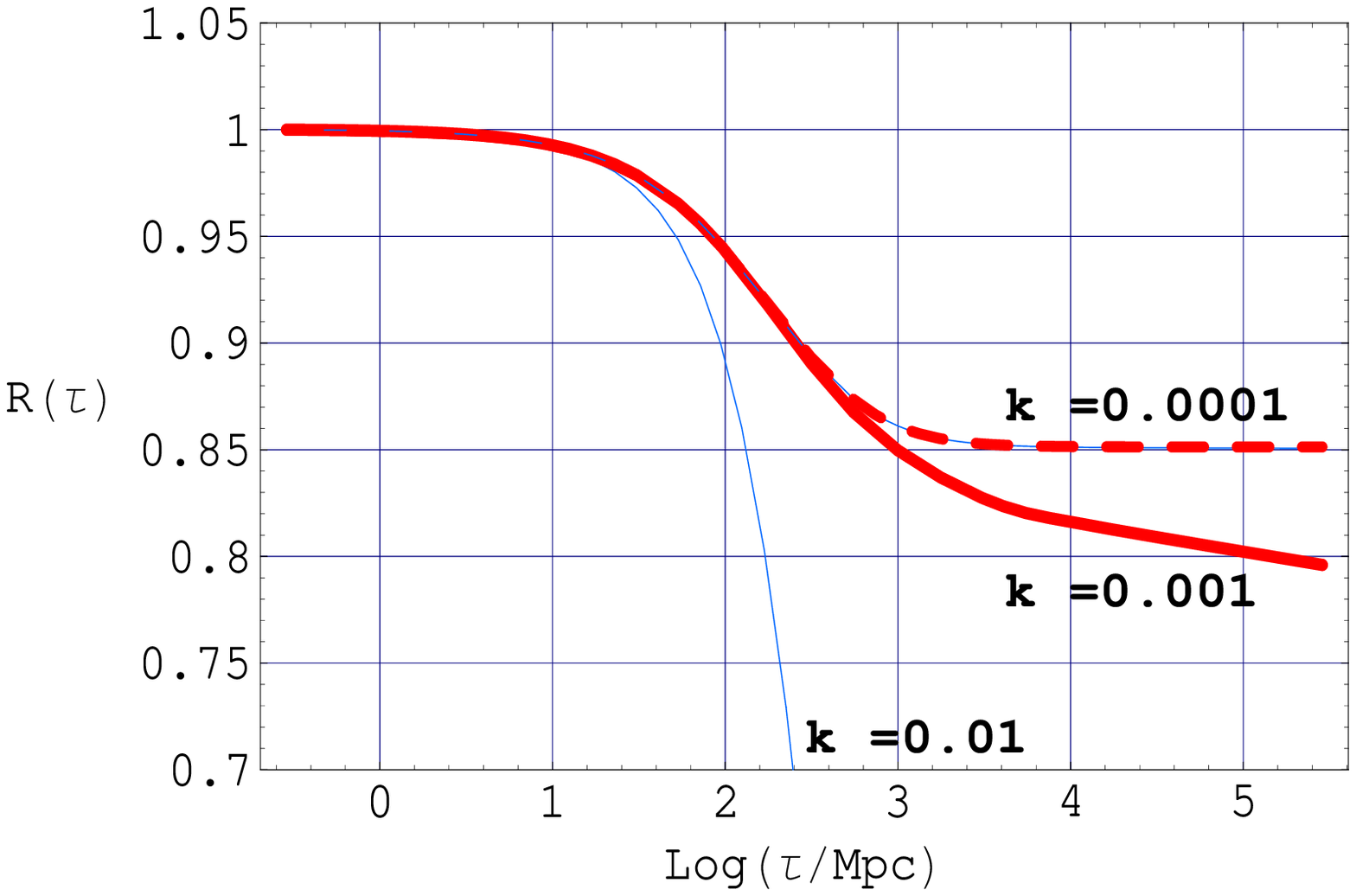}}\\
      \hline
\end{tabular}
\end{center}
\caption[a]{The time evolution of curvature perturbations for $\Omega_{\rm B}= 
{\cal R}_{*}$. The dashed line in the right plot is the 
analytical result valid for wavelengths larger than the Hubble radius. 
In both plots ${\cal R}$ is rescaled by its initial value, i.e. ${\cal R}_{*}$.}
\label{F2}
\end{figure}

In the realistic situation the adiabatic mode will have a nearly scale-invariant 
spectrum (with red tilt) while $\Omega_{\rm B}$ will have a nearly scale-invariant 
spectrum (with blue tilt). For simplicity the relevant results concerning 
the power spectra are collected in the appendix here we just recall 
that the spectra of ${\cal R}_{*}$ and $\Omega_{\rm B}$  and $\sigma_{\rm B}$ can 
be written, according with the notations of the appendix, as 
\begin{equation}
{\cal P}_{{\cal R}}(k) = {\cal A}_{{\cal R}} \biggl(\frac{k}{k_{\rm p}}\biggr)^{n_{r} -1}, 
\label{sp0}
\end{equation}
and as 
\begin{equation}
{\cal P}_{\Omega}(k) = {\cal F}(\varepsilon) 
\overline{\Omega}^2_{{\rm B}L} \biggr(\frac{k}{k_{L}}\biggl)^{2 \varepsilon}, \qquad P_{\sigma} (k) = \frac{( 3 - 2\varepsilon) (188 -4 \varepsilon^2 - 66 \varepsilon )}{3 (6 - \varepsilon) (3 -\varepsilon  ) ( 2 \varepsilon + 1)} {\cal P}_{\Omega}(k),
\label{sp1}
\end{equation}
where $k_{\rm p}= 0.05 \,{\rm Mpc}^{-1}$ is a conventional pivot scale 
of the scalar adiabatic modes and $k_{L} = 1 \, {\rm Mpc}^{-1}$ is the 
pivot scale usually employed to parametrize the magnetic power 
spectra. The quantity $\varepsilon$ is the spectral tilt 
of the magnetic energy spectrum i.e. $ 0 < \varepsilon <1$.
Through the years it emerged that a nearly scale invariant 
magnetic energy spectrum is probably the most interesting 
to discuss in CMB studies since it is not strongly affected by diffusive 
processes (see, for instance, \cite{g1,BB2}). 
The other numerical quantities appearing in Eq. (\ref{sp1}) are defined in Eqs. 
(\ref{Feps}) and (\ref{OMC2}) of the appendix.
If the magnetic fields are absent and $n_{r} = 1$, the Sachs-Wolfe 
plateau (see section 5) corresponds to 
\begin{equation}
\frac{\ell (\ell + 1)}{2\pi} C_{\ell} = \frac{{\cal A}_{{\cal R}}}{25},\qquad
{\cal A}_{\cal R} = 2.65 \times 10^{-9}. 
\end{equation} 
The first interesting exercise (even if not realistic as we shall see later on) is to assume that the magnetic field 
is exactly comparable in magnitude with the curvature 
perturbations, i.e. ${\cal R}_{*} = \Omega_{\rm B}$. In particular without loss of generality, we will assume $n_{r} =1$ and $\varepsilon =0.01$. 
In this situation, as dictated by Eq. (\ref{ad7}) total anisotropic stress
$\Sigma(\tau) = \sigma_{\nu} + R_{\gamma}/R_{\nu} \sigma_{\rm B}$
 is initially vanishing for typical wavelengths larger than the Hubble radius. 
Figure \ref{F1} (plot at the left hand side) 
shows that the total anisotropic stress remains vanishing as long as the mode is outside the horizon and it
oscillates as soon as the mode reenters. In Fig. \ref{F1} the time around 
the radiation-matter transition is measured in ${\rm Mpc}$ while 
the comoving wave-numbers are measured in ${\rm Mpc}^{-1}$.  
In the right plot of Fig. \ref{F1} we use the same initial conditions 
but with a value of $\Omega_{\rm B}$ that 
is one tenth of the amplitude of the initial curvature perturbations 
${\cal R}_{*}$: the amplitude of the  oscillations of the total anisotropic stress 
diminishes. 
The evolution of the curvature perturbation ${\cal R}$ presents 
some interesting aspects that allows a comparison with 
the analytical results. In Fig. \ref{F2} (left plot), the 
evolution of ${\cal R}$ is reported for different wave-numbers and for the same 
range of parameters chosen in the case of Fig. \ref{F1}(left plot), i.e. 
$\Omega_{\rm B} = {\cal R}_{*}$.  When the given wavelength 
is larger than the Hubble radius during the radiation epoch(i.e. $\tau \ll \tau_{\rm eq}$), $\zeta' \simeq {\cal R}' \simeq 0$. This conclusion can be derived, for instance, from Eq. (\ref{zetamas})) and from the Hamiltonian constraint (\ref{HAM1}). Equation 
(\ref{zetamas}) tells that when $c_{\rm s}^2 \simeq 1/3$ (and when $\delta p_{\rm nad}=0$, as in the case of the adiabatic mode), $\zeta$ (and therefore ${\cal R}$, by virtue of Eq. (\ref{HAM1}) ) are conserved for $k \tau \ll 1$.  As soon as the 
given wavenumber reenters the Hubble radius the conservation of ${\cal R}$ no longer holds and this can be appreciated from Fig. \ref{F2} (left plot). The 
interesting situation occurs when the given wavelength is larger than the Hubble
radius after equality (but before decoupling). These are the modes 
determining the large-scale contribution to the (scalar) CMB anisotropies and, in 
particular to the SW plateau. From Fig. \ref{F2} it is clear that the magnetic 
field drives a growth of ${\cal R}$ when the given mode is still outside 
the horizon. In the right plot of Fig. \ref{F2} the evolution of ${\cal R}$ 
is compared with the case where the magnetic field is absent 
(the numerical values of the other initial conditions are the same). 
The horizontal line with amplitude $1$ is the value of ${\cal R}$ in units 
of ${\cal R}_{*}$ when the magnetic contribution is absent. The middle curve 
represents the contribution of the magnetic field when $\Omega_{\rm B} = \sigma_{\rm B} = {\cal R}_{*}$. The dashed line in the right plot of Fig. \ref{F2} is the analytical 
result  obtained by solving Eqs. (\ref{zetamas}) and (\ref{HAM1}) for $k\tau \ll 1$. 
The analytical result reproduces very well with the numerical estimate.
Consider, indeed, Eq. (\ref{zetamas}) recalling that within the notations 
introduced so far 
\begin{equation}
c_{\rm s}^2 =  \frac{4 a_{\rm eq} }{3( 3 a + 4 a_{\rm eq})},\qquad w(a) = \frac{a_{\rm eq} }{3( a + a_{\rm eq})}.
\label{cw}
\end{equation}
Then Eq. (\ref{zetamas}) becomes, in the limit $k \tau \ll 1$
\begin{equation}
\frac{ d \zeta}{d a} = - 
\frac{ R_{\gamma} \Omega_{\rm B} a_{\rm eq} }{ ( 3 a + 4 a_{\rm eq})^2}. 
\label{zetaex}
\end{equation}
The initial conditions deep in the radiation epoch (i.e. $a/a_{\rm eq} \to 0$) are such that the Hamiltonian constraint must be satisfied, i.e. $ \zeta_{*} \simeq {\cal R}_{*} + {\cal O}( k^2 \tau^2)$. Consequently, the solution of Eq. (\ref{zetaex})  and (\ref{HAM1}) will be 
\begin{equation}
{\cal R}\simeq \zeta(a) = {\cal R}_{*} - \frac{3 a  R_{\gamma} 
\Omega_{\rm B}}{4 ( 3 a + 4 a_{\rm eq})}.
\label{zetasol}
\end{equation}
In the limit $a/a_{\rm eq} > 1$ the above expression leads to 
\begin{equation}
{\cal R} \simeq {\cal R}_{*} - \frac{R_{\gamma}}{4} \Omega_{\rm B} + {\cal O}(k^2 \tau^2),
\end{equation}
For $\Omega_{\rm B} \simeq {\cal R}_{*}$ we get that ${\cal R}/{\cal R}_{*} \simeq ( 1 - R_{\gamma}/4) \simeq 0.85$ which is the asymptotic value reported, with the 
dashed line, in Fig. \ref{F2}.
The same result can be readily obtained from Eq. (\ref{Rmas}) since the neutrino anisotropic stress counterbalances  
the magnetic stress with the net result, as verified numerically in Fig. \ref{F1}, that 
the total stress can indeed be neglected for  $k \tau \ll 1$. 
Once ${\cal R}$ and $\zeta$ have been computed, $\Psi$ can be derived 
from Eq. (\ref{tfree}) whose explicit form can be written as 
\begin{equation}
\frac{ d \Psi}{ d \alpha } + \frac{5 \alpha + 6}{2 \alpha (\alpha + 1)}\Psi = 
\frac{  3 \overline{\Omega}_{\rm B}}{ 8 ( \alpha + 1)} - 
{\cal R}_{*} \frac{3 \alpha + 4 }{ 2 \alpha ( \alpha + 1)},
\label{psieq}
\end{equation}
where the variable $\alpha = a/a_{\rm eq}$ has been introduced.
Recalling now that 
\begin{equation}
\frac{ d \ln{f(\alpha)}}{d \alpha} = \frac{5 \alpha + 6}{2 \alpha (\alpha + 1)}, \qquad 
f(a) = \frac{\alpha^3}{\sqrt{\alpha +1} }, 
\label{fa}
\end{equation}
 Eq. (\ref{psieq})  can be written as 
\begin{equation}
\frac{d }{d \alpha} [ f(\alpha) \Psi(\alpha)] = - {\cal R}_{*} \frac{\alpha^2 (3 \alpha + 4)}{2 (\alpha + 1)^{3/2}} + \frac{ 3 \overline{\Omega}_{B} \alpha^3}{ 8 ( \alpha + 1)^{3/2}}.
\label{inteq1}
\end{equation}
Direct integration of Eq. (\ref{inteq1}) between $0$ and $a$ implies 
\begin{eqnarray}
f(\alpha) \Psi(\alpha) - f(0) \Psi(0) &=& - {\cal R}_{*} \frac{ 16( \sqrt{ \alpha + 1} -1) + 9 \alpha^3 + 2 \alpha^2 - 8 \alpha}{15 \sqrt{\alpha + 1}}
\nonumber\\
&+& \frac{3 R_{\gamma}\Omega_{B}}{40}
\frac{32 ( 1 - \sqrt{\alpha + 1})+ 2 \alpha^3 - 4\alpha^2 - 16 \alpha}{\sqrt{\alpha + 1}}.
\label{INT1}
\end{eqnarray}
Now, since $f(0)$ goes to zero quite fast for $a\to 0$, the term
$f(0)\Psi(0)\to 0$ provided $\Psi(0)$ does not diverge\footnote{This is not only the case for the magnetized adiabatic mode but also for the majority of the isocurvature mode with the exception of the neutrino isocurvature velocity 
mode. In fact, in the case of the adiabatic mode $\Psi(0)$ is constant but 
$f(0)\to 0$ (see Eq. (\ref{fa})). In the case of the isocurvature modes 
$\Psi(0)\to 0$. } in $\alpha\to 0$.
Therefore Eq. (\ref{INT1}) implies
\begin{eqnarray}
\Psi(\alpha) &=& - {\cal R}_{*} \frac{ 16 ( \sqrt{a+ 1} -1) + 9 \alpha^3 + 2 \alpha^2 - 8 \alpha}{ 15 \alpha^3} 
\nonumber\\
&+& \frac{3 R_{\gamma} \overline{\Omega}_{B}}{40} \frac{32 ( 1 - \sqrt{\alpha + 1}) + 2 \alpha [ 8 + \alpha ( \alpha -2)]}{\alpha^3}.
\label{Psif}
\end{eqnarray}
In the limit $a > a_{\rm eq}$ (i.e. $\alpha > 1$), Eq. (\ref{Psif}) implies 
that $\Psi(\alpha)$ goes to a constant (in time) and that, in particular,
\begin{equation}
\lim_{\alpha\gg 1} \Psi(\alpha) \simeq - \frac{3}{5} {\cal R}_{*} + \frac{3\, R_{\gamma}}{20} \Omega_{\rm B}.
\end{equation}
The reported results are gauge-independent, this means that if we would 
perform the numerical integration in a specific gauge (like, for 
instance, the conformally Newtonian gauge) exactly the same results 
for ${\cal R}$ and $\zeta$ will be obtained. This exercise has been 
done. We integrated explicitly the evolution equations in the longitudinal 
gauge and obtained exactly the same results derived by numerical integration 
of the gauge-invariant system of equations. 

As an example, we can illustrate the equivalence of these two approaches 
by presenting the numerical evaluation of the Sachs-Wolfe contribution  
in the sudden decoupling limit where the visibility function is approximated by a Dirac delta functions
$\delta (\tau - \tau_{\rm dec})$ and the optical depth is simply its integral, i.e. 
a step function $\theta(\tau - \tau_{\rm dec})$.  In this (naive but physically 
useful) limit the brightness perturbation (related to the Stokes parameter 
I) can be written as 
\begin{equation}
\Delta_{{\rm I}}(\vec{k}, \tau_{0} ) = \int_{\tau_{\rm dec}}^{\tau_{0}} e^{i k \mu (\tau - \tau_0)}[ \psi' + \phi'] d\tau + 
e^{i k \mu(\tau_{\rm dec} - \tau_{0})} [ \Delta_{{\rm I}\,0} + \phi + \mu V]_{\tau_{\rm dec}},
 \label{SWBZ2}
 \end{equation}
where $\mu = \hat{k}\cdot \hat{n}$ is the angle between the direction 
defined by the momentum of the photon and the direction defined by the 
Fourier mode $\vec{k}$.  The term $\Delta_{{\rm I}\,0}$ is exactly 
$\delta_{\gamma}/4$ while 
\begin{equation}
V = \frac{\theta_{\rm b}}{i k} 
= \frac{\theta_{\gamma}}{ik}.
\label{redef}
\end{equation}
The first equality in Eq. (\ref{redef}) is just a redefinition; the second 
equality holds in the tight coupling approximation.
\begin{figure}
\begin{center}
\begin{tabular}{|c|c|}
      \hline
      \hbox{\epsfxsize = 7 cm  \epsffile{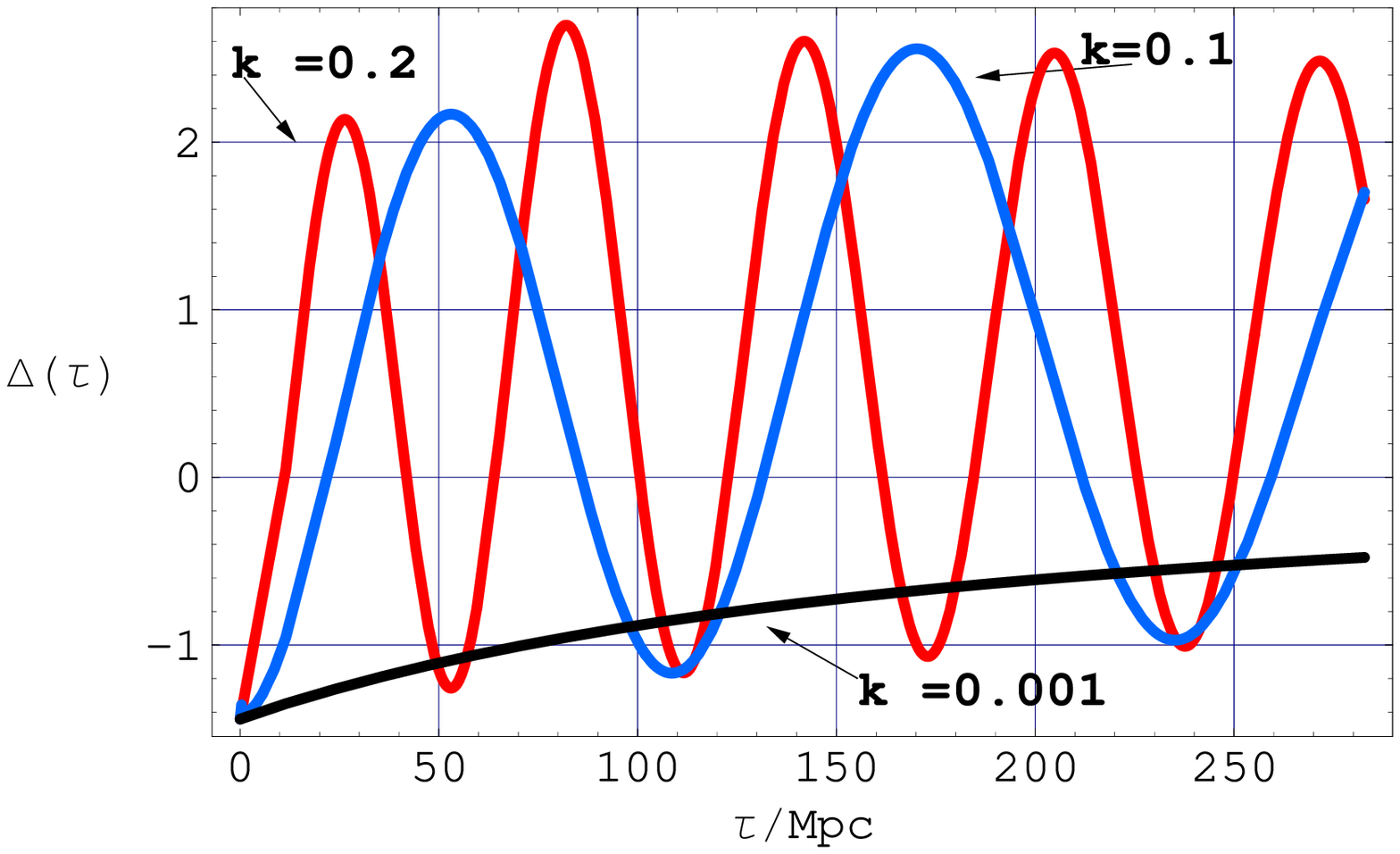}} &
      \hbox{\epsfxsize = 6.6 cm  \epsffile{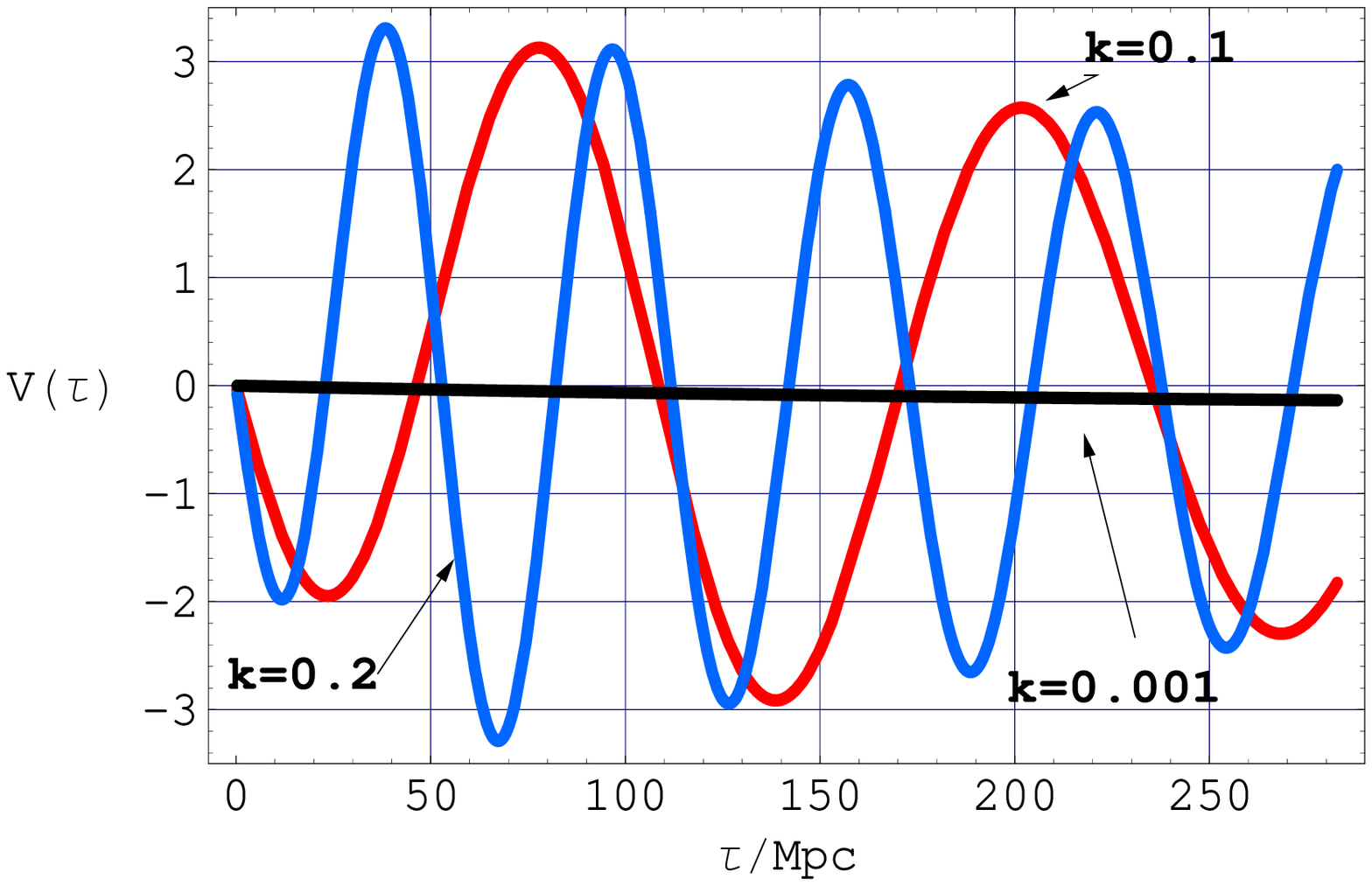}}\\
      \hline
\end{tabular}
\end{center}
\caption[a]{The evolution of the Sachs-Wolfe terms for the case $\Omega_{\rm B} = {\cal R}_{*}$ and for three typical wave-numbers in units of ${\rm Mpc}^{-1}$. }
\label{F3}
\end{figure}
In Fig. \ref{F3} the result of the numerical integration is 
illustrated for the case where $\Omega_{\rm B} = {\cal R}_{*}$. The final 
time in the linear scale appearing in both plots of Fig. \ref{F3} is 
$\tau_{\rm dec}$ (recall 
that $\tau_{\rm dec} = 282.81\, {\rm Mpc}$ and $\tau_{\rm eq}=
119.07$ for the typical values of the parameters discussed in Eqs. (\ref{par1}), 
(\ref{ommr}) and (\ref{parameters})). In the left plot of Fig. \ref{F3}
 we illustrate the combination
$\Delta(\tau) = \delta_{\gamma}/4 + \phi$ (recall Eqs. (\ref{T1}) and (\ref{T2})
for the definitions of the conformally  Newtonian fluctuations). In the 
plot at the right of Fig. \ref{F3} we illustrate $V(\tau)$.

\begin{figure}
\begin{center}
\begin{tabular}{|c|c|}
      \hline
      \hbox{\epsfxsize = 7.1 cm  \epsffile{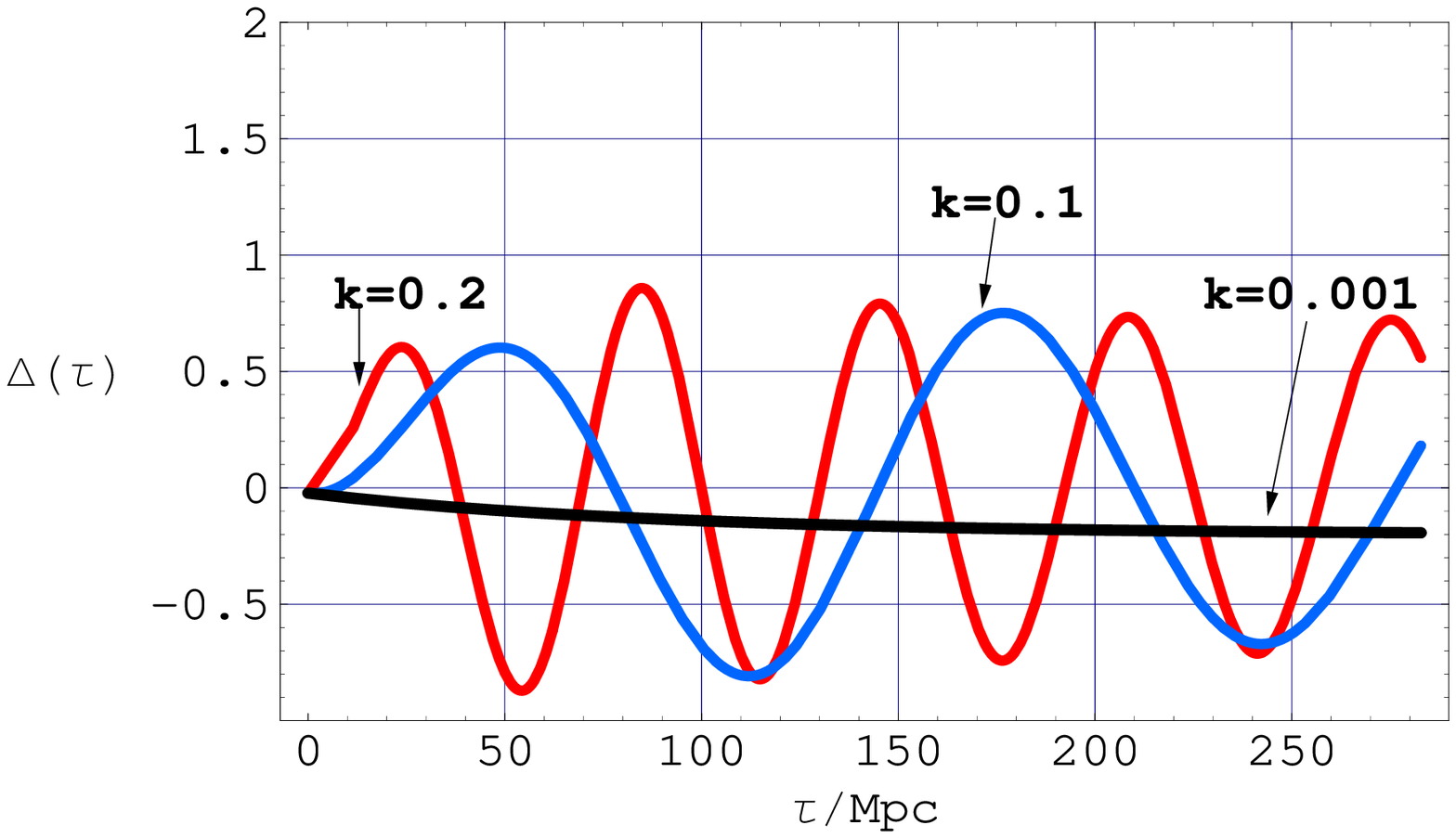}} &
      \hbox{\epsfxsize = 6.6 cm  \epsffile{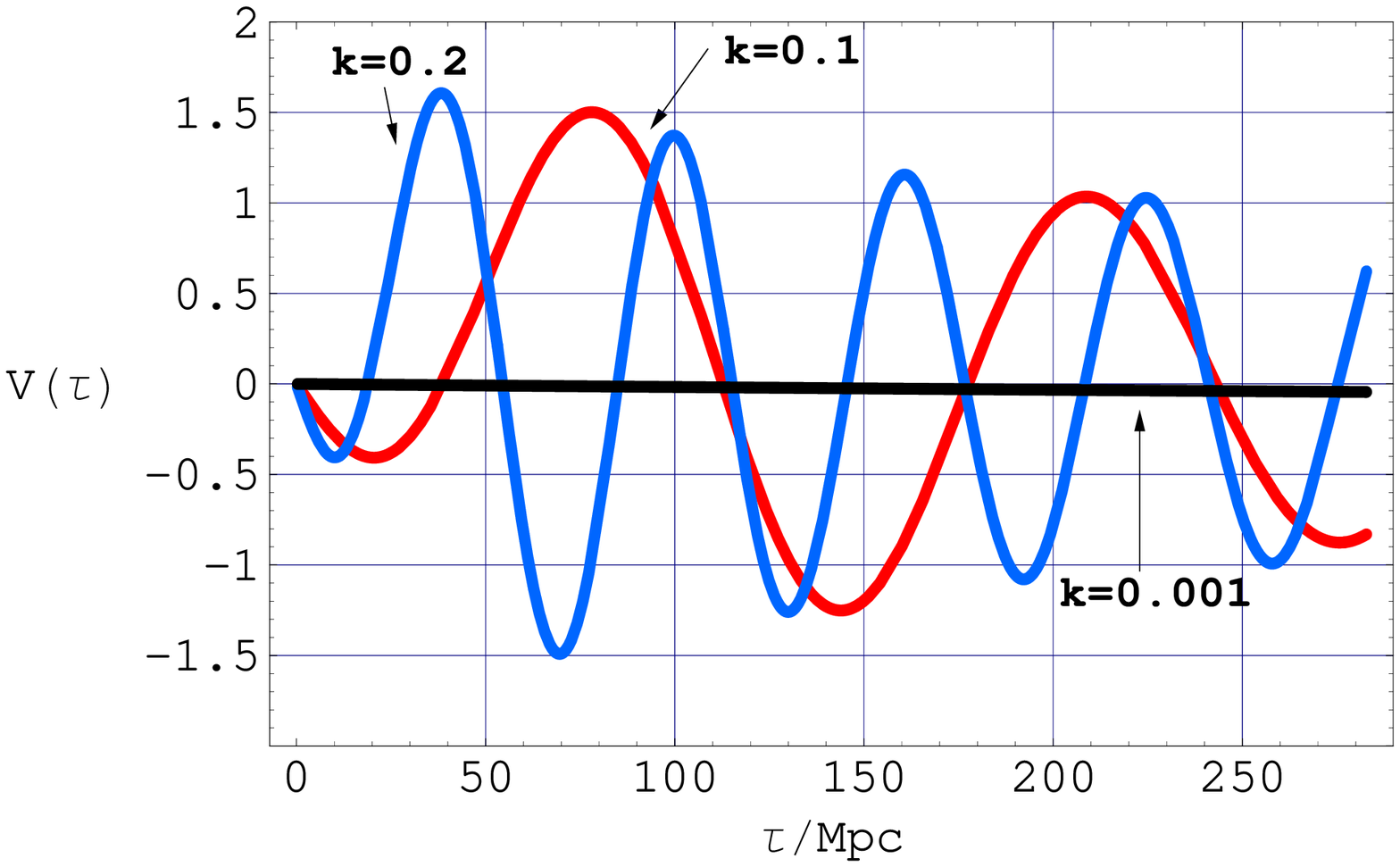}}\\
      \hline
\end{tabular}
\end{center}
\caption[a]{The same quantities illustrated in Fig. \ref{F3} but now for $\Omega_{\rm B} =0.001 {\cal R}_{*}$.}
\label{F4}
\end{figure}

The results of Fig. \ref{F3} should be compared with the ones 
of Fig. \ref{F4} where the same quantities are illustrated but for 
 $\Omega_{\rm B} =0.001\, {\cal R}_{*}$. 
The comparison of the two plots shows that the introduction of magnetized 
initial conditions affects the amplitude of the oscillations and, less 
strongly, the phases. For $\Omega_{\rm B} =0 $ 
the results coincide with the ones of Fig. \ref{F4}. This means that in order not to affect the dominant adiabatic mode, $\Omega_{\rm B}$ should not exceed 
$10^{-3} {\cal R}_{*}$. We will get back to this point in section 5.
It is appropriate to note here that the oscillations appearing in Figs. \ref{F3} and \ref{F4} are the Sakharov oscillations that will determine the Doppler 
peak structure. The requirement that the dominant adiabatic mode 
is not affected by the presence of the magnetic field intensity 
implies then a standard Doppler peak structure.

\subsection{The isocurvature mode}

\begin{figure}
\begin{center}
\begin{tabular}{|c|c|}
      \hline
      \hbox{\epsfxsize = 7 cm  \epsffile{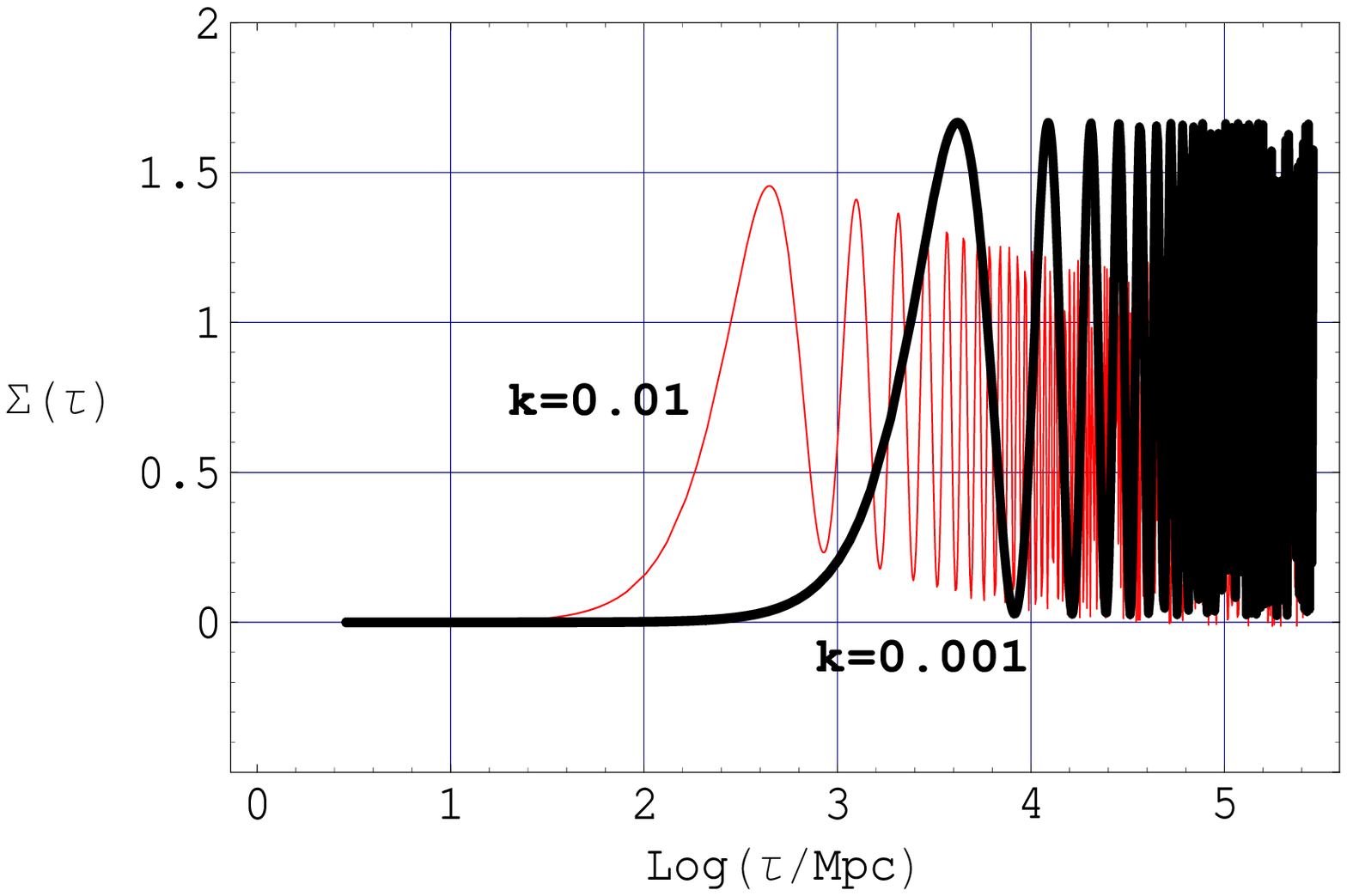}} &
      \hbox{\epsfxsize = 6.7 cm  \epsffile{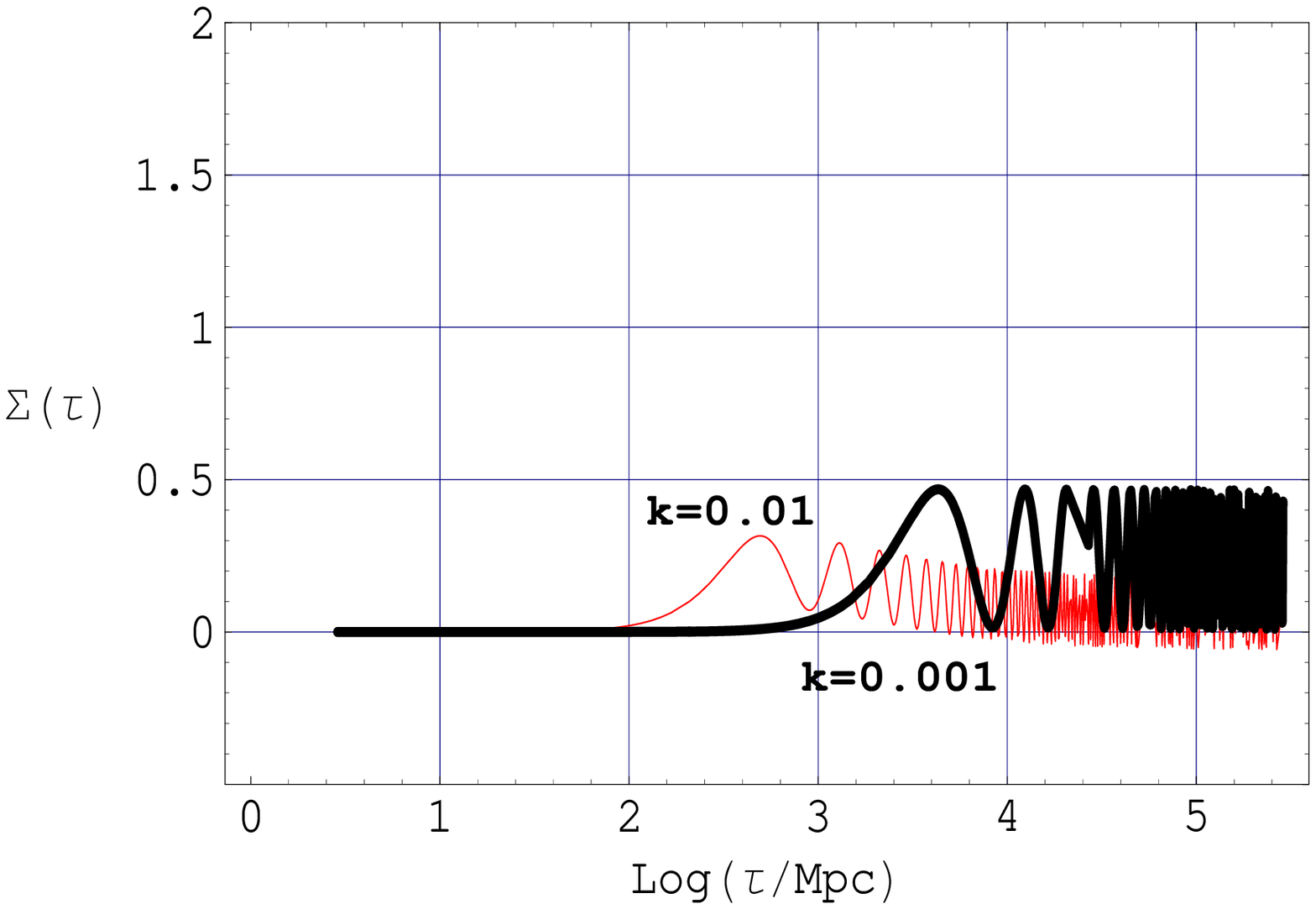}}\\
      \hline
\end{tabular}
\end{center}
\caption[a]{The evolution of the total anisotropic stress  in the case of the 
CDM-radiation (magnetized) non-adiabatic mode. In the plot at the left 
$\Omega_{\rm B} = {\cal S}_{*}$. In the plot at the right $\Omega_{\rm B} = 0.1 {\cal S}_{*}$.}
\label{F5}
\end{figure}

Equations (\ref{zetaex}) and (\ref{zetasol}) suggest that the magnetized contribution can indeed be seen as an isocurvature correction to an observationally dominant adiabatic contribution. This aspect can be 
appreciated by looking at the second term at the right hand side of Eq. 
(\ref{zetasol}): notice that this term goes to zero as $a/a_{\rm eq} \ll 1$. 
It is therefore natural to consider an example of a magnetized isocurvature 
mode, like, for instance, the CDM-radiation isocurvature mode. 
In this case the proper (magnetized)  initial conditions for the numerical integration can be found in Eqs. (\ref{cdm3})--(\ref{cdm7}). Notice, again, that 
by setting $\sigma_{\rm B} \to 0$ and $\Omega_{\rm B} \to 0$ we do recover the 
initial conditions for the standard CDM-radiation isocurvature 
mode with non-vanishing fractional contribution of massless neutrinos.
In the case where $\Omega_{\rm B} = {\cal S}_{*}$ (where ${\cal S}_{*}$ parametrizes 
the primordial entropy fluctuation in the CDM-radiation system) the anisotropic 
stress evolves as in Fig. \ref{F5} (left plot). Conversely, if $\Omega_{\rm B} = 0.1
{\cal S}_{*}$, the numerical results are reported, always in terms of the total 
anisotropic stress, in the right plot of Fig. \ref{F5}. The two plots 
present the same features already discussed in the case of the adiabatic 
mode (see Fig. \ref{F1}). 
\begin{figure}
\begin{center}
\begin{tabular}{|c|c|}
      \hline
      \hbox{\epsfxsize = 7 cm  \epsffile{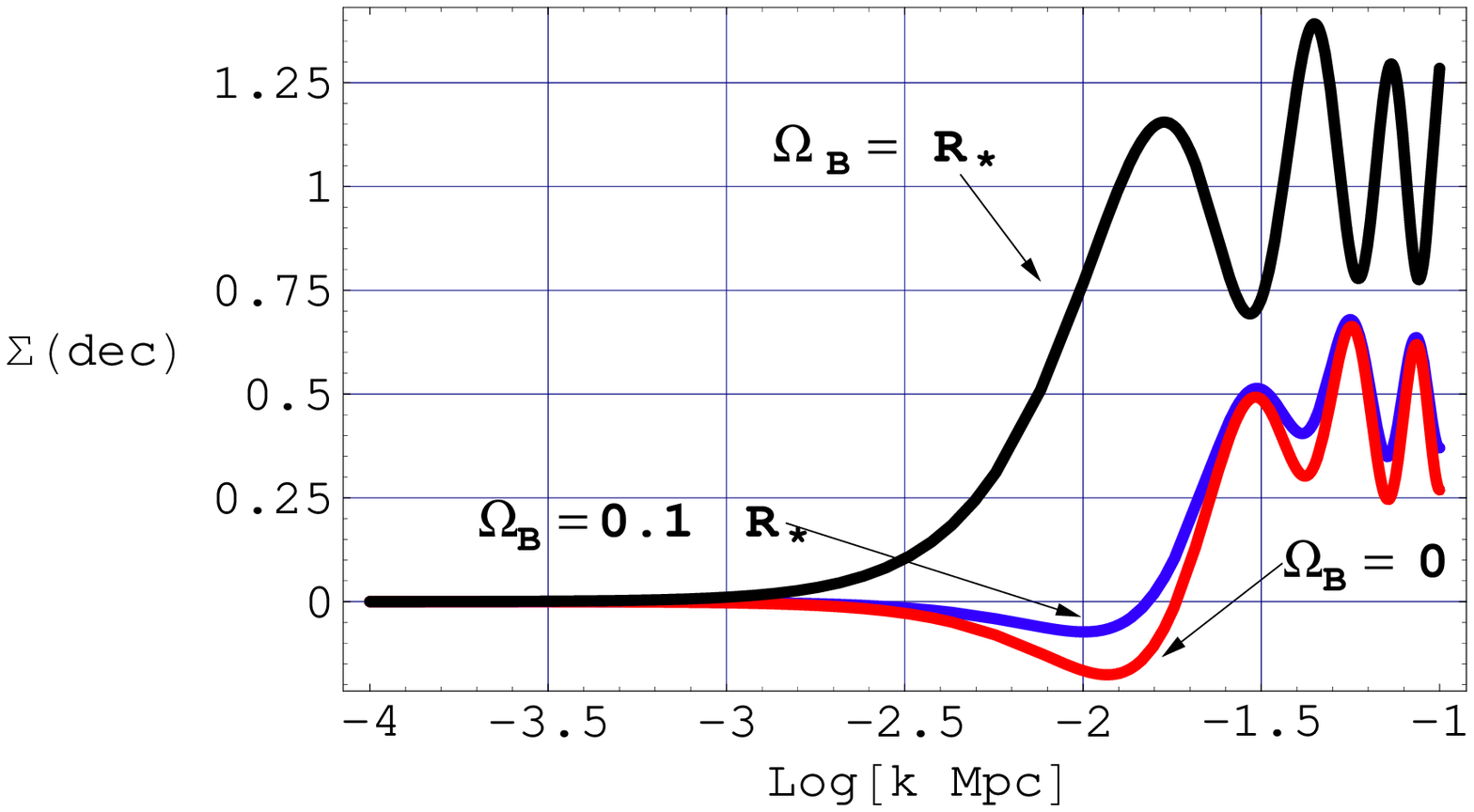}} &
      \hbox{\epsfxsize = 6.8 cm  \epsffile{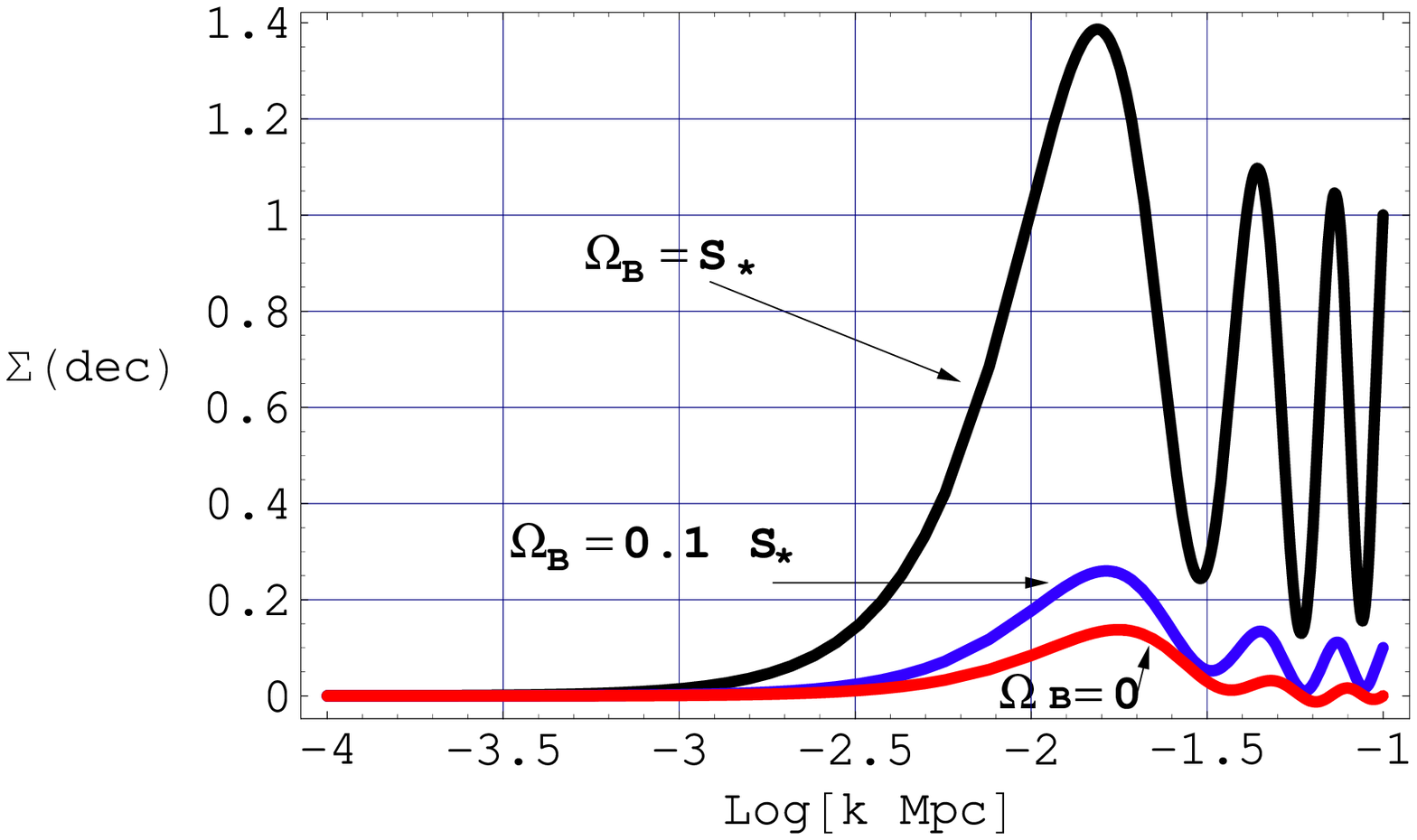}}\\
      \hline
\end{tabular}
\end{center}
\caption[a]{The total anisotropic stress is illustrated at a fixed time 
(coinciding with $\tau_{\rm dec}$) but as a function of the wave-number
measured in ${\rm Mpc}^{-1}$. In the left plot the adiabatic case 
is illustrated. The right plot illustrates instead the CDM-radiation mode.
In both cases different strengths of the magnetic field intensity have been 
reported.}
\label{F5a}
\end{figure}
In Fig. \ref{F5a}, for comparison, the behaviour of $\Sigma(\tau_{\rm dec})$ is illustrated 
as a function of the wave-number in the case of adiabatic initial 
conditions (left plot), and in the case of isocurvature initial conditions 
(right plot). The different values of the magnetic field strengths 
are referred to the amplitude of the initial (adiabatic or non-adiabatic) mode.

The crucial difference between the (magnetized) adiabatic mode 
and the (magnetized) CDM-radiation mode can be illustrated by looking at the 
evolution of ${\cal R}$ and $\Psi$. 
\begin{figure}
\begin{center}
\begin{tabular}{|c|c|}
      \hline
      \hbox{\epsfxsize = 7 cm  \epsffile{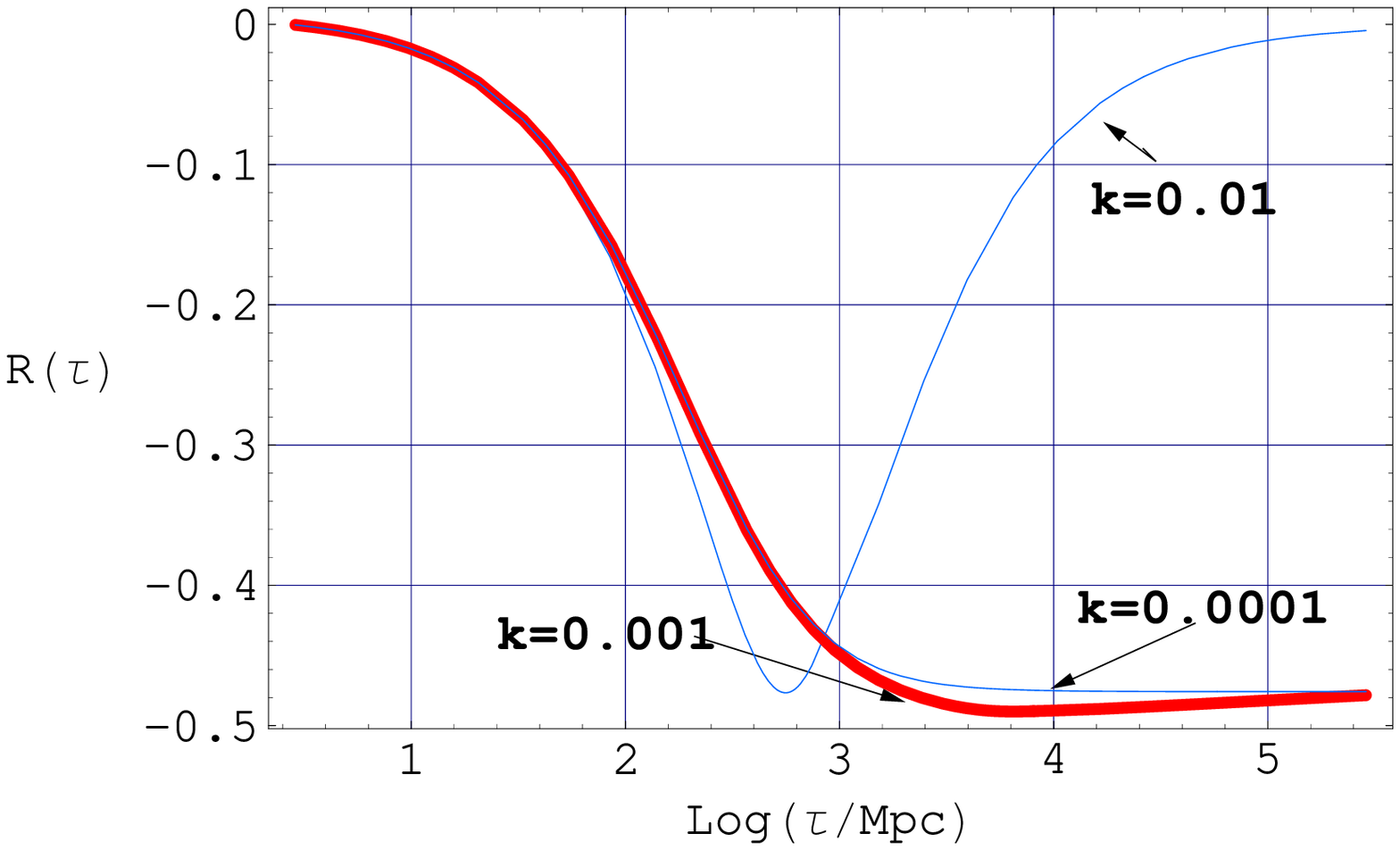}} &
      \hbox{\epsfxsize = 7 cm  \epsffile{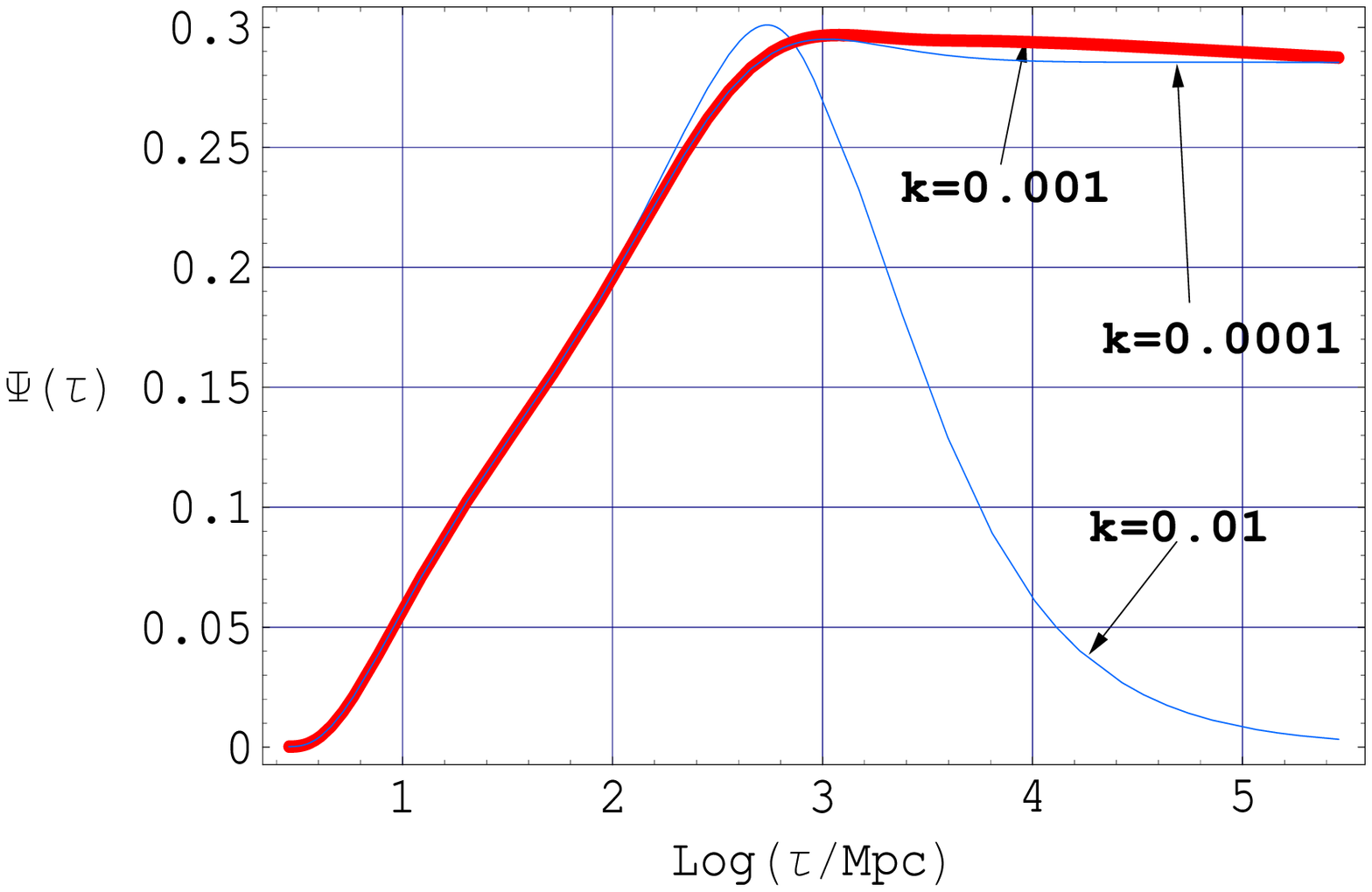}}\\
      \hline
\end{tabular}
\end{center}
\caption[a]{The evolution of the magnetized CDM-radiation 
isocurvature mode is illustrated in terms of ${\cal R}$ and $\Psi$ for 
$\Omega_{\rm B} = {\cal S}_{*}$.}
\label{F6}
\end{figure}
In contrast with the adiabatic result, however, 
both $\Psi$ and ${\cal R}$ vanish for $a/a_{\rm eq} \to 0$. This feature 
is quantitatively illustrated in the plots of Fig. \ref{F6}. In the right plot 
the evolution of $\Psi$ is reported in the asymptotic region where 
the analytical estimates are easily compared with the numerical results.
Both for ${\cal R}$ and $\Psi$ for different sets of magnetized isocurvature 
initial conditions the analytical results reproduce correctly the numerical ones.
The analytical derivation resembles, in this case, the one already performed 
in the adiabatic case with the difference that, now, the non-adiabatic pressure 
density variation is given by
\begin{equation}
- \frac{{\cal H}}{(p_{\rm t} + \rho_{\rm t})} \delta p_{\rm nad}= - \frac{ 4 {\cal H}{\cal  S }a a_{\rm eq}}{( 4 a_{\rm eq} + 3 a)^2}.
\label{nadex}
\end{equation}
By means of this expression, Eqs. (\ref{zetamas}) and (\ref{HAM1}) imply 
that, for $a/a_{\rm eq} \gg1$
\begin{equation}
 {\cal R } \simeq - \frac{{\cal S}_{*}}{3} - \frac{R_{\gamma}}{4} \Omega_{\rm B},
\qquad
 \Psi \simeq \frac{{\cal S}_{*}}{5} + \frac{3 R_{\gamma}}{20} \Omega_{\rm B}.
\label{asym}
\end{equation}

The results obtained so far in the case of the adiabatic and in the case 
of the isocurvature mode can be  summarized by the following two  matrices, i.e. 
\begin{equation}
\pmatrix{
 {\cal R}_{f} \cr
{\cal S}_{f}\cr
\Omega_{{\rm B}\,f}} = 
\pmatrix{{\cal M}_{{\cal R}{\cal R}} & {\cal M}_{{\cal R}{\cal S}} & 
{\cal M}_{{\cal R}{\rm B}}\cr
0 & {\cal M}_{{\cal S}{\cal S}}& {\cal M}_{{\cal S} {\rm B}}\cr
0 & 0& {\cal M}_{{\rm B} {\rm B}} }
\pmatrix{
 {\cal R}_{*}\cr
{\cal S}_{*}\cr
\Omega_{{\rm B}}}.
 \label{MAT1}
\end{equation}
In the case of a mixture of (magnetized) adiabatic and CDM-radiation
modes, we find, for $a > a_{\rm eq}$ 
\begin{equation}
{\cal M}_{{\cal R}{\cal R}} \to 1, \qquad {\cal M}_{{\cal R}{\cal S}} \to - \frac{1}{3},\qquad {\cal M}_{{\cal R}{\rm B}} - \frac{R_{\gamma}}{4},\qquad {\cal M}_{{\cal S}{\cal S}} \to 1,\qquad {\cal M}_{{\cal S}{\rm B}}\to 0,
\label{MAT2}
\end{equation}
and  ${\cal M}_{{\rm B}{\rm B}} \to 1$. The matrix (\ref{MAT1}) gives the values 
of the various fluctuations after equality in terms of the initial conditions 
set for $\tau \ll \tau_{\rm eq}$. In similar terms, for $\Psi$ we will have  
\begin{equation}
\pmatrix{
 \Psi_{f} \cr
{\cal S}_{f}\cr
\Omega_{{\rm B}\,f}} = 
\pmatrix{{\cal M}_{{\Psi}{\Psi}} & {\cal M}_{{\Psi}{\cal S}} & 
{\cal M}_{{\Psi}{\rm B}}\cr
0 & {\cal M}_{{\cal S}{\cal S}}& {\cal M}_{{\cal S} {\rm B}}\cr
0 & 0& {\cal M}_{{\rm B} {\rm B}} }
\pmatrix{
\Psi_{*}\cr
{\cal S}_{*}\cr
\Omega_{{\rm B}}}.
 \label{MAT3}
\end{equation}
where, now, 
\begin{equation}
{\cal M}_{\Psi\Psi} \to \frac{9}{10}, \qquad {\cal M}_{\Psi {\cal S}}\to 
\frac{1}{5}, \qquad {\cal M}_{\Psi {\rm B}}\to \frac{3 R_{\gamma}}{20}
\label{MAT4}
\end{equation}
while the other matrix elements are the same as in the case of Eq. (\ref{MAT1}).
\begin{figure}
\begin{center}
\begin{tabular}{|c|c|}
      \hline
      \hbox{\epsfxsize = 7 cm  \epsffile{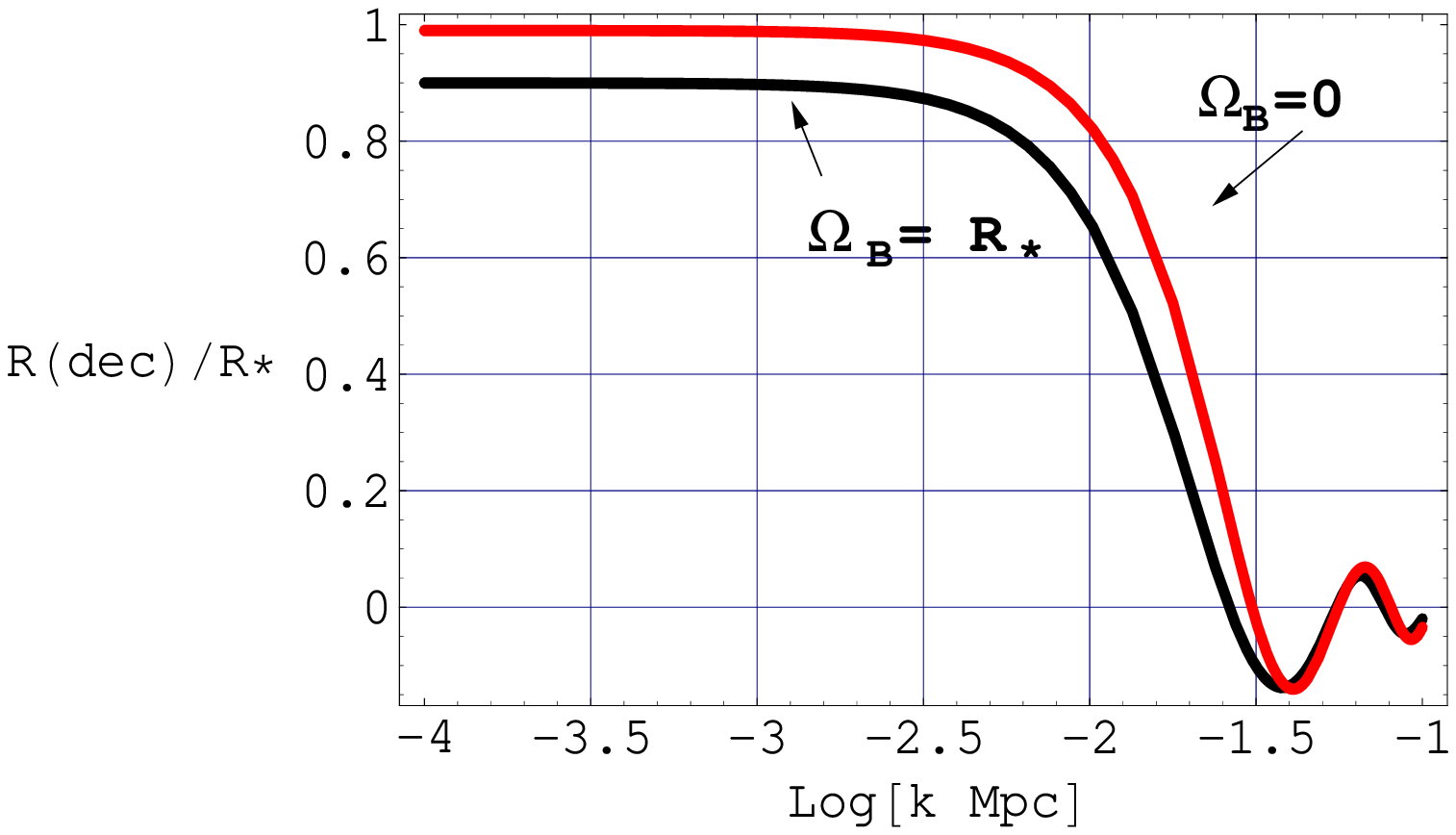}} &
      \hbox{\epsfxsize = 6.8 cm  \epsffile{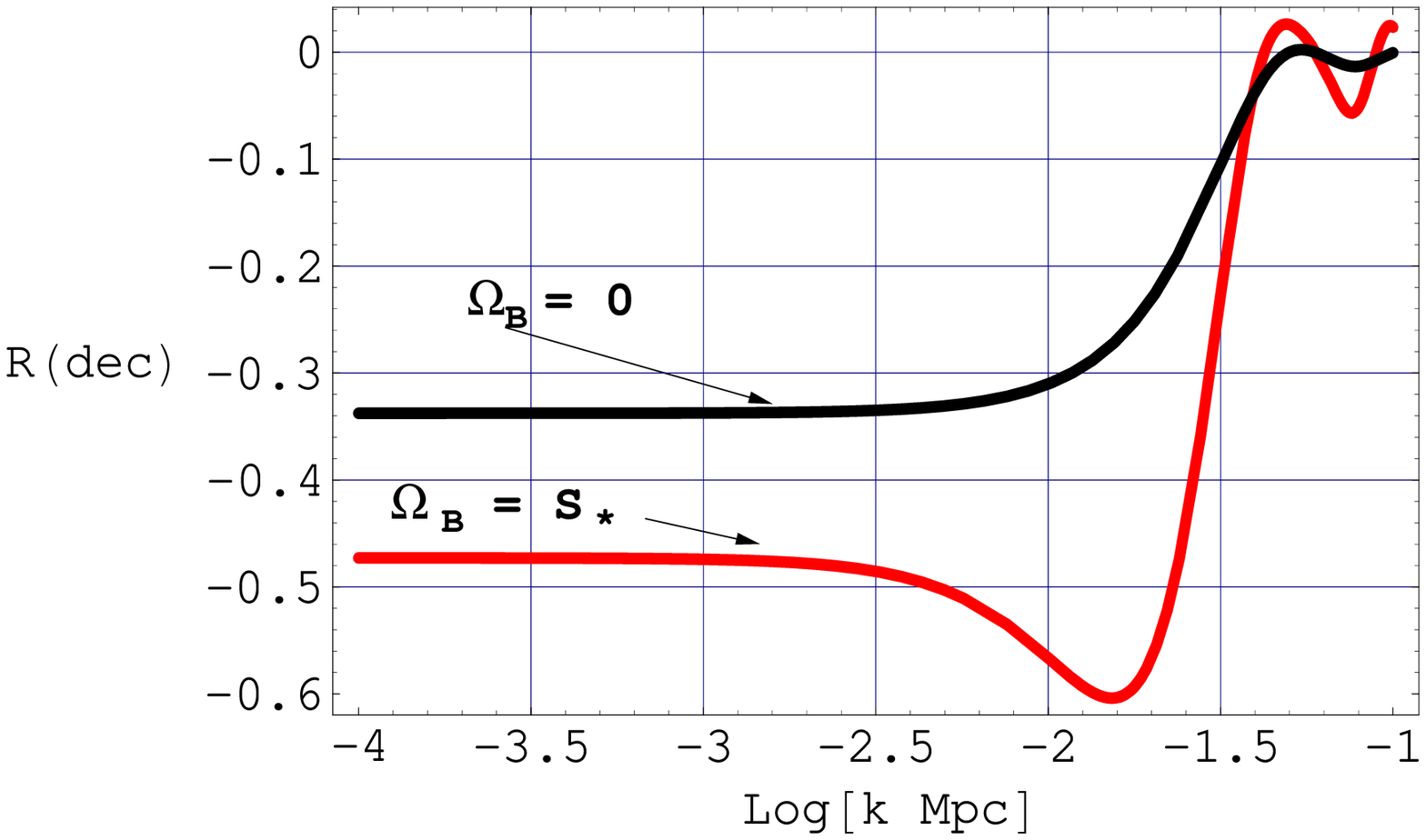}}\\
      \hline
\end{tabular}
\end{center}
\caption[a]{The gauge-invariant curvature perturbations ${\cal R}$ 
are reported at fixed time (coinciding with $\tau_{\rm dec}$) and as a function 
of the comoving wavenumber in units of ${\rm Mpc}^{-1}$. In the left plot the adiabatic case is illustrated.
In the right plot non-adiabatic initial conditions are imposed.}
\label{F6a}
\end{figure}
In Fig. \ref{F6a} the curvature perturbations are illustrated at $\tau_{\rm dec}$ but as a function of $k$, i.e. the comoving wave-number. For wavelengths 
larger than the Hubble radius, the flat plateau appearing in both plots of Fig. \ref{F6a} is accurately predicted by Eqs. (\ref{MAT1}) and (\ref{MAT2}).

The techniques presented here may have useful implications. For instance, 
in \cite{brown} it has been correctly argued that magnetic fields may be 
a source of non-Gaussian signatures in CMB maps. It can be easily argued 
that if the magnetic fields are Gaussianly distributed, the magnetic 
energy density will be non-Gaussian (being the square of a Gaussian variable).
This argument is usually made by looking at the magnetic energy-momentum 
tensor. Here it has been shown that , in the long wavelength limit, the 
magnetic contribution to curvature perturbations is exactly proportional to the 
magnetic energy density with a numerical coefficient that is dynamically determined by the features of the radiation-matter transition. The results 
of \cite{brown} on higher order correlators must then be complemented 
by a consistent framework where adiabatic and non-adiabatic initial 
conditions are properly set.

It would be interesting to go to higher order  in the tight-coupling expansion 
and to adapt the semi-analytical techniques already developed 
in the standard case \cite{jorgensen,HS3}.  This will allow a semi-aanalytical 
treatment of the implications of magnetized initial conditions on polarization 
power spectra.

\renewcommand{\theequation}{5.\arabic{equation}}
\section{The Sachs-Wolfe plateau for magnetized initial conditions}
\setcounter{equation}{0}

Magnetized curvature and entropy perturbations,
initially present for $\tau \ll \tau_{\rm eq}$, have been 
followed through the radiation matter transition. This analysis allows 
the calculation  of the observables that determine 
the large-scale temperature fluctuations when the relevant modes are outside the horizon after equality but before decoupling.  
While this calculation is much simpler 
in the absence of magnetic fields, it has been shown that the inclusion of magnetic fields  well defined since the regular modes 
present deep in the radiation epoch propagate regularly 
through $\tau_{\rm eq}$. It is essential 
to treat carefully not only the magnetic anisotropic stress but also the anisotropic stress of the fluid (mainly induced by collisionless neutrinos). Therefore, a posteriori, it is also clear why, in the absence of magnetic fields, the analytical calculation is much simpler: in 
the standard case the neutrino anisotropic stress can be neglected and its inclusion only produces a weak dependence on the neutrino fractional contribution that enters, for instance, in the relation between the curvature perturbations ${\cal R}$ and the Bardeen potential.
It is then appropriate to proceed further and to evaluate the Sachs-Wolfe (SW) 
plateau induced by magnetized curvature perturbations. This analysis may serve as a starting point for future numerical investigations aimed at the inclusion of regular magnetized modes in the discussion of parameter extraction from present and forthcoming CMB data.
The temperature fluctuations induced by the scalar modes, i.e. the brightness 
perturbations of the intensity of the radiation field, can be written 
as (see, for instance, \cite{g3}) 
\begin{equation}
\Delta_{\rm I} (\vec{x}, \hat{n}, \tau) = \frac{1}{(2\pi)^{3/2}} 
\int d^{3} k e^{-i \vec{k}\cdot \vec{x}} \Delta_{\rm I} (\vec{k}, \hat{n}, \tau),
\label{FT}
\end{equation}
where $\hat{n}$ represents the direction of the photon momentum and,
for convenience we also introduce $\mu = \hat{n}\cdot \hat{k}$, i.e. 
the projection of the photon momentum on the direction 
of a given Fourier mode. Notice that the subscript I appearing in Eq.  (\ref{FT}) 
refers to the first Stokes parameter measuring the intensity of the radiation field.
The Fourier amplitude appearing in Eq. (\ref{FT}) can then be expanded 
in a series of Legendre polynomials as \footnote{We report the following formula 
in order to set accurately the notations. In fact different authors 
choose to perform the deceomposition in terms of Legendre polynomials 
following different conventions. For instance, some authors include 
the factors $(-i)^{\ell}$ (or even $(2 \ell + 1)$) in the definition of 
$\Delta_{{\rm I}\ell}$. Of course these differences are immaterial once 
the calculation is performed consistently in the same framework.}
\begin{equation}
\Delta_{\rm I} (\vec{k}, \hat{n}, \tau) = \sum_{\ell} (-i )^{\ell}
( 2 \ell + 1 ) \Delta_{{\rm I}\ell}(\vec{k},\tau) P_{\ell}(\mu).
\label{LT}
\end{equation}
Assuming then that the receiver is located at $\vec{x}=0$,
the temperature anisotropies can be expanded in spherical harmonics as 
\begin{equation}
\Delta_{\rm I}(\hat{n},\tau) = \sum_{\ell\,\,m} a_{\ell\,\,m} Y_{\ell\,\,m}(\hat{n}).
\label{SH}
\end{equation}
Hence the coefficients $a_{\ell\,\,m}$ can be determined as 
\begin{equation}
a_{\ell\,\,  m} = \frac{4\pi}{ (2\pi)^{3/2}} (-i)^{\ell} \int d^{3}k  Y_{\ell\,\,m} \Delta_{{\rm I}\,\ell}(\vec{k},\tau), 
\label{alm}
\end{equation}
where the theorem of addition of spherical harmonics \cite{abr,tric} has been used.

Since the magnetic field is fully inhomogeneous the isotropy of space-time 
is preserved and, therefore, the angular power spectrum will not depend on the 
azimuthal angle.  In particular the magnetic field is described in terms of a two point correlation function (see Eqs. (\ref{defFT2}) and (\ref{Bcorr})) 
that, therefore, does 
not imply any specific preferred direction. This means that the ensemble average 
of the $a_{\ell\,\,m}$ can be expressed as 
\begin{equation}
\langle a_{\ell\,m} a_{\ell ' \, m'}^{*} \rangle = C_{\ell} \delta_{\ell\,\ell'} \delta_{m\,m'}.
\label{alens}
\end{equation}
Using again the addition theorem for spherical harmonics \cite{abr,tric} the 
angular power spectrum will then be expressed directly in terms 
of the $C_{\ell} $, i.e. 
\begin{equation}
C(\theta) = \langle \Delta_{\rm I} (\hat{n}_{1},\tau_{0}) \Delta_{\rm I} (\hat{n}_{2},\tau_{0}) \rangle = \frac{1}{4\pi} \sum_{\ell} ( 2 \ell + 1) C_{\ell} P_{\ell}(\hat{n}_{1} \cdot 
\hat{n}_{2}).
\end{equation}
We have now to compute $C_{\ell}$ in terms of the curvature fluctuations. 
In order to do so we solve the heat transfer equation in the sudden decoupling limit.  The large-scale contribution \footnote{We neglect here the
integrated Sachs-Wolfe contribution as well as, for the moment,  the Doppler contribution that is relevant for smaller angular scales} to $\Delta_{\rm I}(\vec{k},\tau)$ at the present time  $\tau_{0}$ can then be written as 
\begin{equation}
\Delta_{\rm I} (\vec{k},\tau_{0}) = e^{- i k \mu(\tau_{0} - \tau_{\rm dec}) } \biggl[ - 
\frac{R_{*}(\vec{k})}{5} +  \frac{2}{5} {\cal S}_{*}(\vec{k}) + \frac{3 R_{\gamma}}{10} \Omega_{\rm B}(\vec{k}) \biggr],
\label{DELex}
\end{equation}
where $\tau_{\rm dec} \ll \tau_{0}$ and its presence can be neglected.
The terms at the right-hand side of Eq. (\ref{DELex}) represent 
the contribution of a general perturbation to the temperature anisotropy. 

If ${\cal S}_{*}= \Omega_{\rm B}=0$ the only contribution to the angular power 
spectrum  comes from the usual adiabatic mode. The dependence upon the 
primordial perturbation is also canonical. For instance, recalling that, in the 
longitudinal gauge, ${\cal R}_{*} = - (5/3) \psi_{\rm m}$ we have that the contribution is $\psi^{\rm ad}_{\rm m}/3$ where $\psi^{\rm ad}_{\rm m}$ is the constant adiabatic 
mode during matter and in the conformally Newtonian gauge. Similarly, if the the only non-vanishing contribution is of isocurvature type we will have that 
$\psi^{\rm iso}_{\rm m} = {\cal S}_{*}/5$ reproducing, therefore, the 
isocurvature contribution to the SW plateau that goes, in the conformally 
Newtonian gauge, as $ 2 \psi^{\rm iso}_{\rm m}$. 
We can then expand the plane wave appearing in Eq. (\ref{DELex}) in Legendre 
polynomials, with the result, that $\Delta_{{\rm I}\ell}$ is determined to be:
\begin{equation}
\Delta_{{\rm I}\ell}(\vec{k},\tau_{0}) = \biggl[ - \frac{{\cal R}_{*}(\vec{k})}{5} + \frac{2}{5} {\cal S}_{*}(\vec{k}) + \frac{3 R_{\gamma}}{10} \Omega_{\rm B}(\vec{k}) \biggr] j_{\ell} ( k\tau_{0}),
\label{DELLex}
\end{equation}
where $j_{\ell}(k \tau_{0})$ are the spherical Bessel functions i.e. \cite{abr,tric}
\begin{equation}
j_{\ell} (k \tau_{0}) = \sqrt{\frac{\pi}{2\,k\tau_{0}} }J_{\ell + 1/2}( k\tau_{0}).
\label{BESSex}
\end{equation}
Now the calculation goes as follows. Equation (\ref{DELLex}) has to be inserted 
into Eq. (\ref{alm}). Then, the obtained result has to be inserted into 
Eq. (\ref{alens}) and the $C_{\ell}$ can be determined from the 
explicit expression of the ensemble average of the $a_{\ell\,m}$. To 
compute explicitly $\langle a_{\ell\,m}^{*} a_{\ell'\,m'}\rangle$ we have to know 
the ensemble averages of the various random fields appearing in Eq. (\ref{DELLex}).
In particular, inserting Eq. (\ref{DELLex}) into Eq. (\ref{alens}) and trying to evaluate 
the expectation value $\langle a_{\ell\,m} a_{\ell'\, m}\rangle$ the following 
expectation value has to be computed
\begin{eqnarray}
&&\langle \Delta_{{\rm I}\ell}(\vec{k},\tau_{0}) \Delta_{{\rm I}\ell'}(\vec{p},\tau_{0})^{*}\rangle
 = j_{\ell}(k\tau_{0}) j_{\ell'} (p\tau_{0}) \biggl[
  \frac{1}{25} \langle {\cal R}_{*}(\vec{k},\tau_{0}) {\cal R}_{*}(\vec{p},\tau_{0})^{*}\rangle 
  + \frac{4}{25} \langle {\cal S}_{*}(\vec{k},\tau_{0}) {\cal S}_{*}(\vec{p},\tau_{0})^{*}\rangle 
\nonumber\\
&& + \frac{9}{100} \langle \Omega_{\rm B}(\vec{k},\tau_{0}) \Omega_{\rm B}(\vec{p},\tau_{0})^{*} \rangle 
 - \frac{4}{25}  \langle {\cal R}_{*}(\vec{k},\tau_{0}) {\cal S}_{*}(\vec{p},\tau_{0})^{*}\rangle  
\nonumber\\
&& -\frac{3}{25} \langle \Omega_{\rm B}(\vec{k},\tau_{0}) {\cal R}_{*}(\vec{p},\tau_{0})^{*} \rangle 
+ \frac{6}{25}  \langle {\cal S}_{*}(\vec{k},\tau_{0}) \Omega_{\rm B}(\vec{p},\tau_{0})^{*} \rangle \biggr]
\label{alens2}
\end{eqnarray}
The stochastic averages appearing in Eq. (\ref{alens2}) can be obtained by using 
the expressions appearing in Eqs. (\ref{SP1})--(\ref{corrRS}), (\ref{autoc}) and (\ref{crossc}).
Recalling then Eq. (\ref{alens})  the expression for the $C_{\ell}$ is
\begin{equation}
C_{\ell } = 4 \pi \int d\ln{k}\, {\cal G}(k)\, j_{\ell}^2(k\tau_{0}),
\label{CLEex}
\end{equation}
where 
\begin{eqnarray}
&&{\cal G}(k) = \frac{{\cal P}_{{\cal R}}(k,\tau_{0})}{25} + \frac{4}{25}{\cal P}_{{\cal S}}(k,\tau_{0}) + 
\frac{9 }{100} R_{\gamma}^2\, {\cal P}_{\rm B}(k,\tau_{0}) -  \frac{4}{25} \sqrt{{\cal P}_{{\cal R}}(k,\tau_{0}) {\cal P}_{\cal S}(k,\tau_{0})} \cos{\gamma_{rs}}
\nonumber\\
&& - 
\frac{3}{25} R_{\gamma}\sqrt{{\cal P}_{\rm B}(k,\tau_{0}) {\cal P}_{{\cal R}}(k,\tau_{0})} \cos{\gamma_{br}}  + 
\frac{6}{25} R_{\gamma} \sqrt{{\cal P}_{\cal S}(k, \tau_{0}) {\cal P}_{\rm B}(k,\tau_{0})}   \cos{\gamma_{sb}}.
\label{G}
\end{eqnarray}
Concerning Eq. (\ref{G}) the following comments are in order:
\begin{itemize}
\item{} if the adiabatic, non-adiabatic and magnetic contributions are all uncorrelated (i.e. 
$\gamma_{rs}=\gamma_{br}= \gamma_{sb} = \pi/2$), 
the Sachs-Wolfe contribution will be enhanced if compared with the case where only the adiabatic mode is present;
\item{} if the adiabatic and the non-adiabatic modes are positively correlated (i.e. $\cos{\gamma_{rs}}>0$)
but uncorrelated with the magnetic contribution (i.e. $\gamma_{br} = \gamma_{bs}= \pi/2$)
the correlation reduces the contribution to the SW plateau;
\item{} if the adiabatic and non-adiabatic contribution are both uncorrelated (i.e. $\gamma_{rs} = \pi/2$) 
but the adiabatic and magnetic components are positively correlated (i.e. $\cos{\gamma_{rs}} >0$)
again the amplitude of the SW plateau is reduced;  on the contrary, if the non-adiabatic and magnetic 
components are positively correlated  (i.e. $\cos{\gamma_{sb}} >0$) the amplitude of the SW plateau 
is enhanced;
\item{} the previous cases are of course reversed when the various modes instead of being 
correlated are anti-correlated (i.e. $\cos{\gamma} <0$ where $\gamma$ stands for one or more 
corrrelation angles).
\end{itemize}
It is then clear that the presence of a regular magnetized mode during the radiation dominated epoch induces, after $\tau_{\rm eq}$  super-horizon fluctuations 
whose effect on the SW plateau depend on the mutual correlation 
of the various components. This situation is, in some way, similar to the case where 
abiabatic and isocurvature modes are correlated (or anti-correlated). The present situation is, however, 
phenomenologically more rich due to the presence of the magnetic field.

By performing the integral over $k$ and by recalling Eq. (\ref{BESSex}), 
Eq. (\ref{G}) allows the explicit evaluation of the $C_{\ell}$ once the form of the power spectrum is specified.  
Each of the power spectra appearing in Eq. (\ref{G}) will have a different spectral slope. So the 
general notation adopted here will be the one summarized in Eqs. (\ref{PSRS}) and (\ref{corrRS}) (for the 
autocorrelation of the adiabatic and isocurvature modes and their correlation), in Eqs. (\ref{autoc}) and (\ref{crossc}) 
(for the autocorrelation of the magnetic component and for the cross correlations between the magnetic sector and the 
adiabatic or non-adiabatic contributions).  As discussed in the appendix, the power spectra of the various contributions 
are all referred to the same (pivot) wave-number $k_{\rm p}$ whose specific value may range, according to present conventional choices between $0.05$ ${\rm Mpc}^{-1}$ and $0.002$ ${\rm Mpc}^{-1}$.

With these necessary specifications the integration over $k$ in Eq. (\ref{G}) gives
\footnote{The integrals in Eq. (\ref{G}) converge provided $-3 < n_{s} < 3$, $-3 < n_{r} < 3$ and 
$-3 < n_{rs} < 3$.}
\begin{eqnarray}
C_{\ell} &=& \biggl[ \frac{{\cal A}_{\cal R}}{25} \,{\cal Z}_{1}(n_{r},\ell) + \frac{4}{25} \,{\cal A}_{{\cal S}} \,
{\cal Z}_{1}(n_{s},\ell) +
\frac{9}{100} \, R_{\gamma}^2  \overline{\Omega}^{2}_{{\rm B}\,L} {\cal Z}_{2}(\epsilon,\ell) - 
\frac{4}{25} \sqrt{{\cal A}_{{\cal R}} {\cal A}_{{\cal S}}} {\cal Z}_{1}(n_{r\,s},\ell) \cos{\gamma_{rs}}
\nonumber\\
&-& 
\frac{3}{25} \sqrt{ {\cal A}_{{\cal R}}} \, R_{\gamma} \,\overline{\Omega}_{{\rm B}\, L}\,{\cal Z}_{3} (n_{r},\varepsilon, \ell) \cos{\gamma_{br}} 
+ \frac{6}{25} \sqrt{ {\cal A}_{{\cal S}}}\,R_{\gamma} \overline{\Omega}_{{\rm B}\,L}\, {\cal Z}_{3}(n_{s},\varepsilon, \ell)\cos{\gamma_{b s}} \biggr],
\label{CELL2}
\end{eqnarray}
where the functions ${\cal Z}_{1}(m,\ell)$, ${\cal Z}_{2}(\epsilon,\ell)$ and ${\cal Z}_{3}(m, \epsilon,\ell)$ 
are defined, for a generic spectral index $n$
\begin{eqnarray}
{\cal Z}_{1}(n,\ell) &=& \frac{\pi^2}{4} \biggl(\frac{k_0}{k_{\rm p}}\biggl)^{n-1} 2^{n} \frac{\Gamma( 3 - n) \Gamma\biggl(\ell + 
\frac{ n -1}{2}\biggr)}{\Gamma^2\biggl( 2 - \frac{n}{2}\biggr) \Gamma\biggl( \ell + \frac{5}{2} - \frac{n}{2}\biggr)},
\label{Z1}\\
{\cal Z}_{2}(\varepsilon,\ell) &=& \frac{\pi^2}{2} 2^{2\varepsilon} {\cal F}(\varepsilon) \biggl( \frac{k_{0}}{k_{L}}\biggr)^{ 2 \varepsilon} \frac{ \Gamma( 2 - 2 \varepsilon) 
\Gamma(\ell + \varepsilon)}{\Gamma^2\biggl(\frac{3}{2} - \varepsilon\biggr) \Gamma(\ell + 2 -\varepsilon)},
\label{Z2}\\
{\cal Z}_{3}(n,\varepsilon,\ell) &=&\frac{\pi^2}{4} 2^{\varepsilon} 2^{\frac{n +1}{2}} \,\sqrt{{\cal F}(\varepsilon)}\, \biggl(\frac{k_{0}}{k_{L}}\biggr)^{\varepsilon} \biggl(\frac{k_{0}}{k_{\rm p}}\biggr)^{\frac{n + 1}{2}} \frac{ \Gamma\biggl(\frac{5}{2} - \varepsilon - \frac{n}{2}\biggr) \Gamma\biggl( \ell + 
\frac{\varepsilon}{2} + \frac{n}{4} - \frac{1}{4}\biggr)}{\Gamma^2\biggl(\frac{7}{4} - \frac{{\varepsilon}}{2} - \frac{n}{4}\biggr)
\Gamma\biggl( \frac{9}{4} + \ell - \frac{\varepsilon}{2} - \frac{n}{4} \biggr)}.
\label{Z3}
\end{eqnarray}
The generic spectral index $n$ appearing in Eqs. (\ref{Z1}), (\ref{Z2}) and (\ref{Z3}) may correspond 
either to $n_{r}$, or to $n_{s}$ or even to $n_{r\,s} = (n_{r} + n_{s})/2$. Furthermore, the function ${\cal F}(\varepsilon)$ 
appearing in Eqs. (\ref{Z1}) and (\ref{Z2}) is given in Eq. (\ref{Feps}) of the appendix.

As explained after Eq. (\ref{G}), the phenomenological situation 
implied by the magnetized adiabatic and isocurvature modes 
is rather rich since we essentially have three spectral indices 
(i.e. $n_{r}$, $n_{s}$ and $\epsilon$), three amplitudes (i.e. ${\cal A}_{\cal R}$,
${\cal A}_{\cal S}$ and $\overline{\Omega}_{B\, L}$), and three correlation 
angles\footnote{In principle the spectral slope of the cross-correlations may 
not be simply related to the spectral indices of the autocorrelation. In this 
case the formulae given in the present section are slightly modified. The modifications can be easily derived from the results of the appendix 
by assuming that, for instance, $n_{r\,s}$ is not given, as assumed here, 
by $(n_{r} + n_{s})/2$.}.

\begin{figure}
\begin{center}
\begin{tabular}{|c|c|}
      \hline
      \hbox{\epsfxsize = 6.7 cm  \epsffile{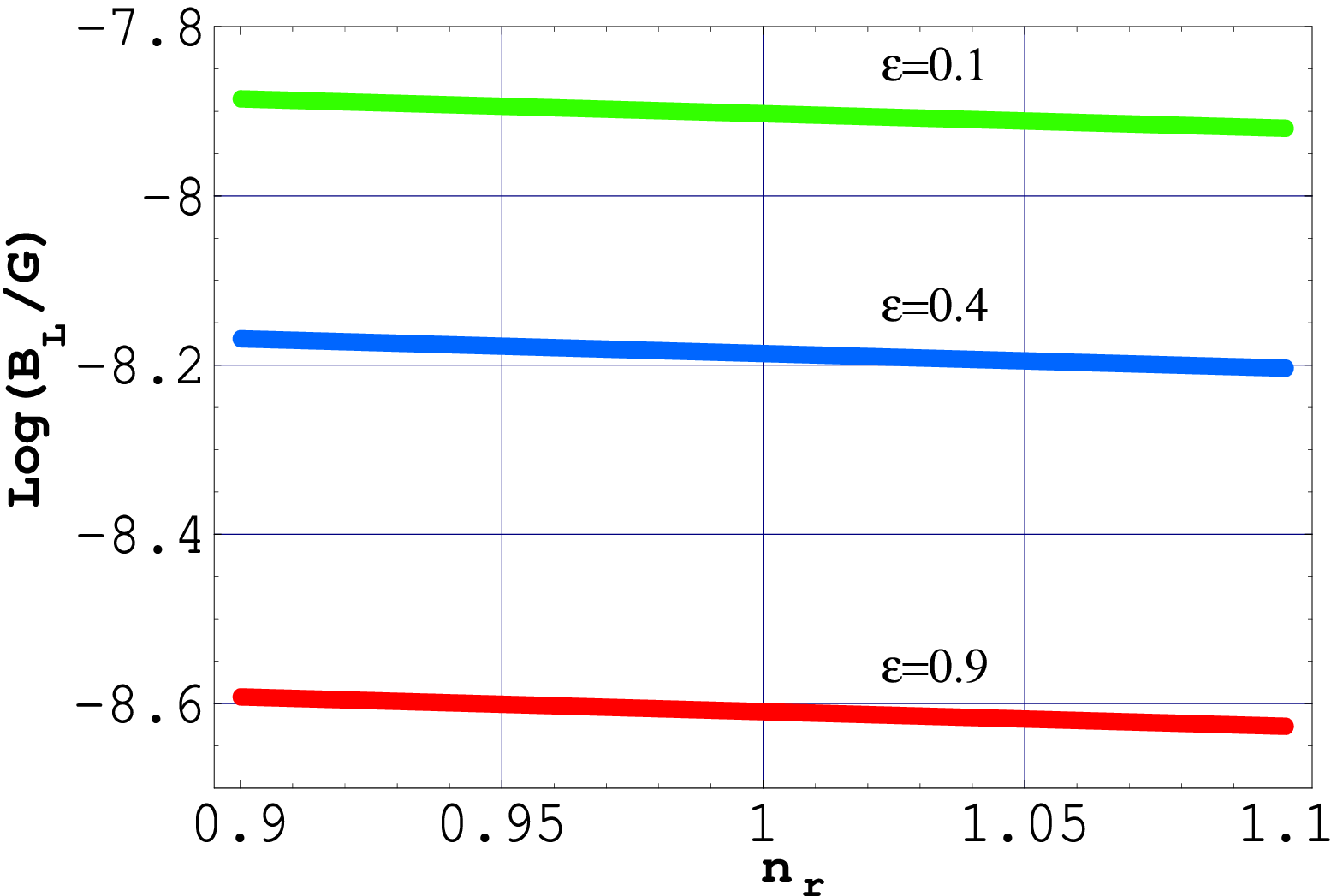}} &
      \hbox{\epsfxsize = 7 cm  \epsffile{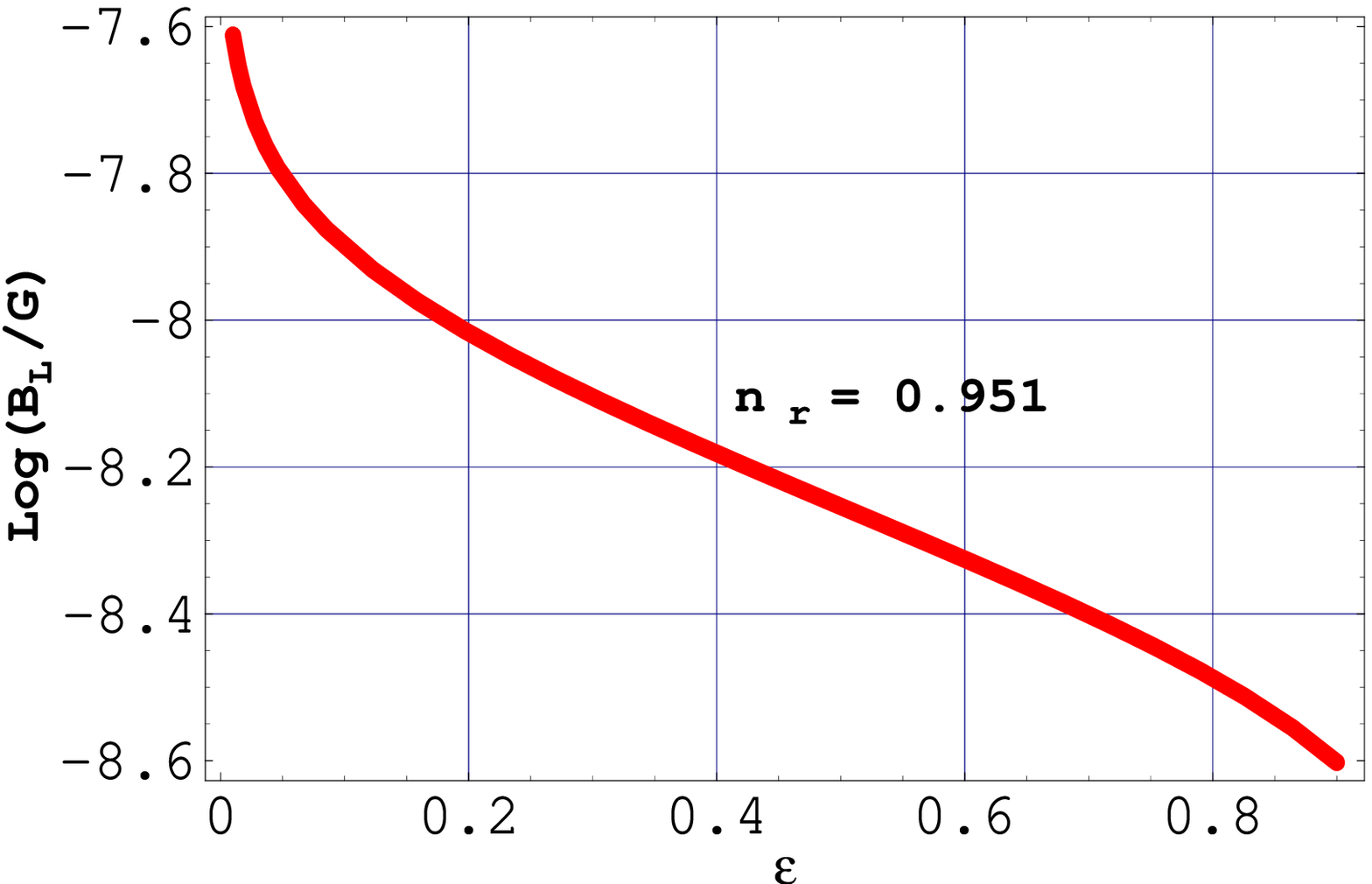}}\\
      \hline
\end{tabular}
\end{center}
\caption[a]{The bounds on the magnetic field intensity derived from 
Eq. (\ref{CELL3}) assuming maximal anticorrelation between the magnetic and adiabatic components.}
\label{Fb1}
\end{figure}
In order to discuss this situation in fully consistent terms it is 
necessary to perform a numerical analysis of parameters extraction
similar to the one reported in \cite{hannu2} where, however, the magnetized 
contribution is taken into account both at the level of the initial conditions 
and at the level of the normalization. 

To get an estimate of the possible bounds obtainable in the 
present framework on the intensity of the magnetic field the minimal 
situation will be considered. Suppose, indeed, that the SW plateau 
is just produced by a magnetized adiabatic mode.  In this case Eq. 
(\ref{CELL2}) reduces to 
\begin{equation}
C_{\ell} = \biggl[ \frac{{\cal A}_{\cal R}}{25} \,{\cal Z}_{1}(n_{r},\ell) 
+\frac{9}{100} \, R_{\gamma}^2  \overline{\Omega}^{2}_{{\rm B}\,L} {\cal Z}_{2}(\epsilon,\ell) - \frac{3}{25} \sqrt{ {\cal A}_{{\cal R}}} \, R_{\gamma} \,\overline{\Omega}_{{\rm B}\, L}\,{\cal Z}_{3} (n_{r},\varepsilon, \ell) \cos{\gamma}\biggr]. 
\label{CELL3}
\end{equation}
where the correlation angle has been simply denoted by $\gamma$, since it is, in  this case, the only correlation angle. Since the contribution to the SW 
plateau must be predominantly adiabatic we do know that the scalar spectral 
index must be nearly scale invariant. In the case where $n_{r} =1$ and in the 
absence of magnetic field, Eq. (\ref{CELL3}) implies 
that $\ell(\ell +1) C_{\ell}/2\pi = {\cal A}_{{\cal R}}/25$. The WMAP determination 
will then impose ${\cal A}_{{\cal R}} \simeq 2.65 \times 10^{-9}$. 
The recent release of WMAP data \cite{map3} seems to prefer slightly red spectra 
with preferred value (taking WMAP alone) $n_{r} = 0.951$. 
The spectrum of the magnetized contribution is controlled now by 
$\varepsilon$. If $0< \varepsilon < 1$ we are in the interesting situation 
where also the magnetic field spectrum is nearly scale-invariant (though with a 
blue tilt since $\varepsilon$ is directly related with the tilt of the magnetic energy 
spectrum ${\cal P}_{\Omega}$, in the notations of the appendix). 
The range $0<\varepsilon< 1$ is phenomenologically interesting not only 
because, in this range, the magnetic energy spectrum is nearly scale-invariant 
but also because the diffusive effects (i.e. finite thermal and magnetic
diffusivity scales) do not affect the properties of the large-scale 
magnetic field (see, for instance, \cite{kan1}).

If the contribution to the SW plateau is predominantly adiabatic, then, the 
limit on the magnetic field intensity for a given range of $n_{r}$ and $\epsilon$ 
are illustrated in Fig. \ref{Fb1} assuming the validity of Eq. (\ref{CELL3}) and 
enforcing the WMAP (i.e. experimental) determination of the SW plateau.  
In the left plot of Fig. \ref{Fb1} the maximal 
 magnetic field intensity is illustrated as a 
function of the scalar spectral index $n_{r}$ for different values of $\varepsilon$.
For the selected values of the parameters the magnetic field has to be always 
below each of the curves.
As discussed above (and also in the appendix), $B_{L}$ is the magnetic field 
regularized over a typical comoving wave-number $k_{L}$ that we take to be, according to the conventions, $1$ ${\rm Mpc}^{-1}$. The pivot scale 
has been fixed at $k_{\rm p} = 0.05\, {\rm Mpc}^{-1}$. To obtain the estimates 
illustrated in Fig. \ref{Fb1} we assumed, in Eq. (\ref{CELL3}) the less
favourable situation, i.e. the one where $\gamma = \pi$. In this case 
(maximal antcorrelation) the two autocorrelations and the cross-correlation 
term all add-up to the plateau and there is no possibility of (artificial) 
reduction of the amplitude. From the left 
plot of Fig. \ref{Fb1} it is clear that the present value of the magnetic
field is bounded to be $B_{L} \leq 10^{-8}-10^{-9}\, {\rm G}$. Similar conclusion 
can be drawn from the plot at the right hand side of Fig. \ref{Fb1} where 
the allowed region is illustrated for a specific value $n_{r}$ (preferred by WMAP
data alone) as a function of $\varepsilon$.
It is clear that by allowing a positive correlation (i.e. $\cos{\gamma}>0$) 
instead of a negative one (as previously assumed) the bounds may be relaxed.
It could be argued that the limit on magnetic fields arising from this 
analysis is competitive with the ones obtainable, for instance, from Faraday 
rotation of the E-modes into B-modes. The main conclusion 
is however, that it is now possible to include the magnetic fields in 
a systematic strategy of parameter extraction from available CMB data 
possibly combined with large scale structure data.
 
\renewcommand{\theequation}{6.\arabic{equation}}
\section{Discussion}
\setcounter{equation}{0}

There are no compelling reasons why the scalar modes induced by large-scale 
magnetic fields should not be considered as a possible source of CMB 
anisotropies. In this paper a new framework for this analysis 
has been proposed. The main findings can be summarized as follows:
\begin{itemize}
\item{} the presence of large-scale magnetic fields (with nearly scale-invariant 
magnetic energy spectrum) affects both the adiabatic and 
the entropy modes;
\item{} a general gauge-invariant system for the analysis of the initial
conditions has been proposed: in such a system the gravitational 
sector is described in terms of the curvature perturbations on comoving 
orthogonal hypersurfaces, the density constrast on uniform density 
hyepersurfaces and the Bardeen potential;
\item{} this gauge-invariant system has been analytically 
 solved deep in the radiation 
dominated epoch (well before equality) and then numerically followed 
after equality (but before recombinantion) in the tight coupling approximation;
\item{} the numerical estimates (corroborated by the analytical results)
allow to compute the precise value of the magnetized adiabatic and non-adiabatic 
curvature perturbations when the relevant modes are outside the horizon before 
decoupling;
\item{} as an example the effect of magnetic fields on the SW plateau has 
been discussed.
\end{itemize}

The magnetized curvature and entropy perturbations propagate then smoothly 
through equality. This result seems particularly useful since 
the common lore was that the presence of magnetic fields induce a 
singular modes. As discussed, this presumption was a simple consequence 
of the fact that, in most studies, the anisotropic stress of the neutrinos 
was simply neglected. In the present analysis, on the contrary, the four-fluid 
system describing the most general initial conditions for CMB anisotropies 
has been consistently studied and solved. 

The present paper opens various possible developments. First of all 
the analysis reported in the present paper involves the lowest 
order in the tight coupling expansion. It will be interesting 
to go beyond and study, within the same framework, 
the effects arising to first-order in the tight coupling expansion. This 
will allow the treatment of various phenomena (including Faraday effect) 
that have been discussed, up to now, just assuming the standard 
adiabatic mode that is, however, itself modified by the presence of the 
magnetic field as demonstrated in the present paper.
Numerical studies on the relevance of magnetized fluctuations 
correlated (or anticorrelated) with adiabatic and non-adiabatic modes are now
conceptually possible since the super-horizon curvature perturbations 
have been carefully computed (after equality) starting from the solution 
of the system of perturbations after neutrino decoupling. Possible 
strategies of parameter extraction adapted to this new situation should be 
discussed in future investigations.

\newpage

\begin{appendix}
\renewcommand{\theequation}{A.\arabic{equation}}
\setcounter{equation}{0}
\section{Power spectra}
In the present appendix the main conventions for the evaluation of the power spectra will be summarized. This technical discussion is relevant since 
there are conventions that are used in the magnetic field literature  (for instance 
concerning the relative normalization of the power spectra) and that differ 
from the conventions used in CMB physics.
According to the  conventions adopted in this paper, the Fourier components of the curvature and entropy fluctuations 
are defined as 
\begin{equation}
{\cal R}(\vec{x},\tau) = \frac{1}{(2\pi)^{3/2}} \int d^{3} k {\cal R}(\vec{k}, \tau) e^{-i \vec{k}\cdot\vec{x}}, \qquad 
{\cal S}(\vec{x},\tau) = \frac{1}{(2\pi)^{3/2}} \int d^{3} k {\cal S}(\vec{k},\tau) e^{-i \vec{k}\cdot\vec{x}},
\label{defFT}
\end{equation}
where ${\cal R}^{*}(\vec{k},\tau) = {\cal R}(-\vec{k},\tau)$ and  ${\cal S}^{*}(\vec{k},\tau) = {\cal S}(-\vec{k},\tau)$.
The correlation functions of the primordial adiabatic and isocurvature modes are then 
defined as
\begin{equation}
 \langle {\cal R}_{*}(\vec{k},\tau) {\cal R}_{*}(\vec{p},\tau) \rangle = \frac{2\pi^2 }{k^3} 
{\cal P}_{\cal R}(k,\tau) \delta^{(3)}(\vec{k} + \vec{p}),\qquad \langle {\cal S}_{*}(\vec{k},\tau) {\cal S}_{*}(\vec{p},\tau) \rangle = \frac{2\pi^2 }{k^3} {\cal P}_{\cal S}(k,\tau) \delta^{(3)}(\vec{k} + \vec{p}),
\label{SP1}
\end{equation}
where $k = |\vec{k}|$  and where the power spectra can be parametrized as
\begin{equation}
{\cal P}_{{\cal R}}(k,\tau) = A_{{\cal R}} \biggl(\frac{k}{k_{\rm p}}\biggr)^{n_{r} -1}, \qquad
{\cal P}_{{\cal S}}(k,\tau) = A_{{\cal S}} \biggl(\frac{k}{k_{\rm p}}\biggr)^{n_{s} -1}.
\label{PSRS}
\end{equation}
In Eq. (\ref{PSRS}),  $k_{\rm p}$ is the ``pivot" scale at which the spectra are normalized. A typical choice\footnote{Another 
possible choice for $k_{\rm p}$ is $0.002$ ${\rm Mpc}^{-1}$. See \cite{map3} and also \cite{hannu2} for 
a discussion of the sensitivity of the parameter estimation on $k_{\rm p}$.} 
for $k_{\rm p}$ is $0.05 \, {\rm Mpc}^{-1}$.
If we allow a correlation between adiabatic and isocurvature modes, the initial value problem may require 
a further primordial correlation function, namely:
\begin{equation}
\langle {\cal R}_{*}(\vec{k},\tau) {\cal S}_{*}(\vec{p},\tau) \rangle= \frac{2\pi^2 }{k^3} 
\sqrt{{\cal P}_{\cal R}(k,\tau)\,\,{\cal P}_{\cal S}(k,\tau)} \cos{\gamma_{ r\,s}} \delta^{(3)}(\vec{k} + \vec{p}),
\label{corrRS}
\end{equation}
where $\gamma_{r\,s}$ is the correlation angle. If $\cos{\gamma_{r\,s}}=0$ the primordial modes 
are not correlated.   

With similar notations, the Fourier modes for the magnetic field intensity are
\begin{equation}
B_{i}(\vec{x},\tau) = \frac{1}{(2\pi)^{3/2}} \int d^{3} k B_{i}(\vec{k},\tau) e^{ - i \vec{k} \cdot \vec{x}},
\label{defFT2}
\end{equation}
again with $B_{i}^{*}(\vec{k},\tau) = B_{i}(-\vec{k},\tau)$. 
The magnetic fields studied in the present paper are fully inhomogeneous and characterized by their two-point 
function, i.e. 
\begin{equation}
\langle B_{i}(\vec{k},\tau) B^{j}(\vec{p},\tau) \rangle =  \frac{2\pi^2}{k^3} \,P_{i}^{j}(k)\, P_{\rm B}(k,\tau)\, \delta^{(3)}(\vec{k} + \vec{p}),
\label{Bcorr}
\end{equation}
where 
\begin{equation}
P_{i}^{j}(k) = \biggl(\delta_{i}^{j} - \frac{k_{i}k^{j}}{k^2} \biggr),\qquad P_{\rm B}(k,\tau) = A_{\rm B} 
\biggl(\frac{k}{k_{\rm p}}\biggr)^{m-1}.
\label{MPS}
\end{equation}
The amplitude of the magnetic power spectrum, $A_{\rm B}$, can be traded for the magnetic energy density regularized 
over a typical scale of size $L$. In fact, by using a Gaussian window function $e^{ - k^2 L^2/2}$ for each 
magnetic field intensity, the (real space) magnetic autocorrelation 
function at two coincident spatial points is 
\begin{equation}
B_{L}^2 = \langle B_{i}(\vec{x},\tau) B^{i}(\vec{x},\tau) \rangle =  
A_{\rm B} \biggl(\frac{k_{L}}{k_{\rm p}}\biggr)^{m -1} \Gamma\biggl(\frac{m -1}{2}\biggr)
\label{AB}
\end{equation}
where $k_{L} = 2\pi/L$.  Equation (\ref{AB}) implies that the amplitude of the magnetic power spectrum 
appearing in Eqs. (\ref{Bcorr}) and (\ref{MPS}) can be expressed in terms of $B_{L}^2$ as 
\begin{equation}
A_{\rm B} = (2\pi)^{\varepsilon} \frac{B_{L}^2}{\Gamma(\varepsilon/2)} \biggl(\frac{k_{\rm p}}{k_{L}}\biggr)^{\varepsilon},
\label{AtoB}
\end{equation}
where, for convenience, the magnetic spectral tilt $\varepsilon = (m -1)$ has been introduced. Different authors (see, for instance, \cite{BB1,bs2,bs3,bs4}) employ slightly 
different conventions both for  Eqs. (\ref{MPS})--(\ref{AtoB}). In particular, sometimes the factor $2\pi^2/k^3$ appearing 
in Eq. (\ref{Bcorr}) is dropped and reabsorbed in the definition of $P_{B}(k)$ which is, in turn, often parametrized as 
$P_{B} = \tilde{A} k^{\tilde{m}}$ (without any typical pivot scale). If these two choices are consistently adopted, a nearly 
scale invariant magnetic energy spectrum (see below)  occurs for $-3 < \tilde{m} \
< - 2$. Within the present conventions, that are homogeneous with the ones customarily adopted in the analysis of primordial adiabatic and isocurvature modes, 
the nearly scale-invariant magnetic energy spectrum is obtained in the range   $ 0 <\varepsilon < 1$.

The nearly scale-invariant spectrum is particularly interesting for the purposes of the present investigation. 
In this limit, in fact, the magnetic field energy spectrum (as well as the spectrum of the Lorentz force) is practically 
unaffected by diffusive damping provided by the effective Alfv\'en damping scale \cite{bs1,bs2}. The Alfv\'en
damping occcurs for typical wavelengths that are of the order of $L_{\rm A} = v_{\rm eff} L_{\rm S}$ where 
$v_{\rm eff}$ is the effctive Alfv\'en velocity and $L_{\rm S}$  is the Silk damping scale. Since $v_{\rm eff}$ will be proportional, in the present notations, to $B_{L}^2/\overline{\rho}_{\gamma}$, we can also appreciate that $L_{\rm A}$ will 
simply be a fraction of the Silk scale. In the case covered by the present investigation, on the contrary, we are not interested in setting 
bounds on very steep (violet) spectra of the magnetic energy density that are, in any case, already strongly 
constrained by the analysis of the vector and tensor modes. On the contrary, since in the 
present analysis, the magnetic fields induce some correlated isocurvature modes, it is also reasonable 
to deepen the analysis in the nearly scale-invariant case. With these 
specifications, the typical scale where diffusion becomes important will not be directly related with 
the Alfv\'en damping but rather with the typical scale at which the wavelength of the fluctuations becomes 
comparable with the photon mean free path corresponding to a comoving wave-number 
\begin{equation}
k_{\rm d} \simeq 0.3 \, \biggl( \frac{h^2 \Omega_{\rm m}}{ 0.134}\biggr)^{1/4} \, \biggl(\frac{h^2 \Omega_{\rm b}}{0.023 } 
\biggr)^{1/2} \biggl(\frac{a_{\rm dec}}{a}\biggr)^{5/4} \, {\rm Mpc}^{-1}.
\end{equation}

From Eq. (\ref{Bcorr}) it is possible to compute, by convolution, the appropriate quadratic observables appearing 
in the treatment of the scalar problem. In particular,
as discussed in the bulk of the paper, the effect of the magnetic fields on scalar modes 
depends upon the magnetic energy and pressure densities, (i.e. $\delta\rho_{\rm B}$ and $\delta p_{\rm B}$) as 
well as on the divergence of the Lorentz force. This last quantity, in MHD, is proportional to 
$\vec{\nabla}\cdot[ (\vec{\nabla} \times \vec{B})\times \vec{B}]$. 
Recalling, in fact, the notations established in  Eqs. (\ref{anis2}) and (\ref{OMB}) 
we have that, in real space 
\begin{equation}
\Omega_{\rm B}(\vec{x}, \tau) = \frac{\delta \rho_{\rm B}( \vec{x},\tau)}{\rho_{\gamma}(\tau)} = \frac{1}{(2\pi)^{3/2}}
\int d^{3} q \,\,\Omega_{\rm B} (\vec{q},\tau) \,e^{- i\vec{q}\cdot \vec{x}}, \qquad \delta p_{\rm B} = \frac{\delta \rho_{\rm B}}{3},
\label{OMFT}
\end{equation}
where 
\begin{equation}
\Omega_{\rm B}(\vec{q},\tau) = \frac{1}{(2\pi)^{3/2}}  \frac{1}{8\pi \overline{\rho}_{\gamma}}  \int d^{3} p B_{i}(\vec{p}) 
B^{i} (\vec{q} - \vec{p}).
\label{OMEGA}
\end{equation}
In Eq. (\ref{OMEGA}) the quantity $\overline{\rho}_{\gamma} = \rho_{\gamma}(\tau) a^4(\tau)$ has been introduced for 
practical reasons.
Consistently with Eq. (\ref{defFT2}), the correlation 
function of the magnetic energy density is then:
\begin{equation}
\langle \Omega_{\rm B}(\vec{k},\tau ) \Omega_{\rm B}(\vec{p},\tau)\rangle = \frac{2\pi^2 }{k^3} 
{\cal P}_{\Omega}(k,\tau) \delta^{(3)}(\vec{k} + \vec{p}).
\label{autoc}
\end{equation}
where 
\begin{eqnarray}
&&{\cal P}_{\Omega}(k) = {\cal F}(\varepsilon) \overline{\Omega}_{{\rm B}\, L}^2 \biggl(\frac{k}{k_{L}}\biggr)^{2 \varepsilon},
\label{POM}\\
&& {\cal F}(\varepsilon) = \frac{(6 - \varepsilon) ( 2 \pi)^{ 2 \varepsilon}}{\varepsilon ( 3 - 2 \varepsilon)
 \Gamma^2(\varepsilon/2)},
\label{Feps}\\
&&  \overline{\Omega}_{{\rm B}\,\, L} = \frac{\rho_{{\rm B}\,L}}{\overline{\rho}_{\gamma}}, 
\qquad \rho_{{\rm B}\,\, L}=\frac{ B_{L}^2}{8\pi}.
\label{OMC2}
\end{eqnarray}
Notice that the obtained expression for ${\cal P}_{\Omega}$ follows from the mode-coupling integral 
\begin{equation}
{\cal P}_{\rm B}(q) = \frac{1}{4 (2\pi)^{7}} \frac{q^{3}}{\rho_{\gamma} a^4} \int 
d^{3} k \biggl\{ 1 + \frac{[ \vec{k} \cdot (\vec{q} + \vec{k})]^2}{k^2 (\vec{q} + \vec{k})^2} \biggr\} P_{\rm B} ( \vec{q}
+ \vec{k}) P_{\rm B}( \vec{k}).
\label{OMCORR}
\end{equation}
in the limit $0< \epsilon \leq 0.3$.
According to Eq. (\ref{anis2}), 
the correlator of $\sigma_{\rm B}$ determines, together with $\Omega_{\rm B}$ the Lorentz force.
As far as the two-point function of $\sigma_{\rm B}$ is concerned, we can start by noticing that 
\begin{equation}
\sigma_{\rm B}(\vec{x}) = \frac{1}{(2\pi)^{3/2}} \int d^{3} q  e^{ - i \vec{q} \cdot\vec{x}} \sigma_{\rm B}(q),
\end{equation}
where
\begin{equation}
\sigma_{\rm B}(q) = \frac{1}{(2\pi)^{3/2}} \frac{1}{16\pi a^{4} \rho_{\gamma}} \frac{1}{q^2} \int d^{3} p \biggl[ B_{i}(\vec{q} - \vec{p}) B^{i}(\vec{p}) + 3 ( q^{j} - p^{j}) p^{i} B_{j}(\vec{p}) B_{i}(\vec{q} -\vec{p})\biggr].
\end{equation}
Consequently, 
\begin{equation}
\langle \sigma_{\rm B} (\vec{q}) \sigma_{\rm B}(\vec{q}') \rangle = 
\frac{2 \pi^2}{q^3} {\cal P}_{\sigma}(q) \delta^{(3)}(\vec{q} +\vec{q}').
\end{equation}
where, for $\epsilon < 0.2$, 
\begin{equation}
{\cal P}_{\sigma}(q)\simeq \frac{( 3 - 2\varepsilon) (188 -4 \varepsilon^2 - 66 \varepsilon )}{3 (6 - \varepsilon) (3 -\varepsilon  ) ( 2 \varepsilon + 1)} {\cal P}_{\Omega}(q).
\end{equation}

In the general phenomenological situation, the magnetic field may also be correlated 
with the primordial adiabatic and isocurvature modes. In this case, two new cross-correlations 
have to be taken into account when computing, for instance, the Sachs-Wolfe plateau:
\begin{eqnarray}
&& \langle {\cal R}_{*}(\vec{k}) \Omega_{\rm B} (\vec{p}) \rangle= \frac{2\pi^2 }{k^3} 
\sqrt{{\cal P}_{\cal R}(k)\,\,{\cal P}_{\Omega}(k)} \cos{\gamma_{ r\,b}} \delta^{(3)}(\vec{k} + \vec{p}), 
\nonumber\\
&& \langle {\cal S}_{*}(\vec{k}) \Omega_{\rm B} (\vec{p})\rangle = \frac{2\pi^2 }{k^3} 
\sqrt{{\cal P}_{\cal S}(k)\,\,{\cal P}_{\Omega}(k)} \cos{\gamma_{ s\,b}} \delta^{(3)}(\vec{k} + \vec{p}).
\label{crossc}
\end{eqnarray}
Again, in Eq. (\ref{crossc}) the various ``angles" parametrize the correlation between
different modes. If $\gamma_{r\,s}= \gamma_{r\,b} = \gamma_{b\,s}= \pi/2$ the modes 
are not correlated. 
\end{appendix}

\end{document}